\documentclass[useAMS,usenatbib]{mnras}
\usepackage{longtable}
\usepackage{pdflscape}
\usepackage{rotating}
\usepackage{graphicx}
\usepackage{grffile}
\usepackage{courier}
\usepackage{amsmath} 
\usepackage{amssymb} 
\usepackage{lscape}
\usepackage{color}
\usepackage{longtable}
\usepackage[normalem]{ulem}   

\usepackage{xspace}

\def\kms{\ifmmode{{\rm km \, s^{-1}}}\else{${\rm km \, s^{-1}}$}\xspace\fi}
\def\Mchar{\ifmmode{{M_{1/2}}}\else{$M_{1/2}$}\xspace\fi}
\def\vchar{\ifmmode{{v_{1/2}}}\else{$v_{1/2}$}\xspace\fi}
\def\vmax{\ifmmode{{v_{\rm max}}}\else{$v_{\rm max}$}\xspace\fi}
\def\vcirc{\ifmmode{{v_{\rm circ}}}\else{$v_{\rm circ}$}\xspace\fi}

\def\hMpc{\ifmmode{\:h^{-1}\,{\rm Mpc}}\else{$h^{-1}\,{\rm Mpc}$}\fi}
\def\hGpc{\ifmmode{\:h^{-1}\,{\rm Gpc}}\else{$h^{-1}\,{\rm Gpc}$}\fi}
\def\hGpcc{\ifmmode{\:h^{-3}\,{\rm Gpc^3}}\else{$h^{-3}\,{\rm Gpc}^3$}\fi}
\def\hkpc{\ifmmode{\:h^{-1} \,{\rm kpc}}\else{$h^{-1}\,{\rm kpc}$}\fi}
\def\hMsun{\ifmmode{\:h^{-1}\,{\rm M}_\odot}\else{$h^{-1}\,{\rm M}_\odot$}\fi}
\def\uG{\ifmmode{\:\mu{\rm G}}\else{$\mu G$}\fi}
\def\mjyb{\ifmmode{\:{\rm mJy}\,{\rm beam}^{-1}}\else{\,mJy$\,$beam$^{-1}$}\fi}
\def\ffergs{\ifmmode{10^{44}\:{\rm erg \, s^{-1}}}\else{$10^{44}\:{\rm erg\,s^{-1}}$}\fi}

\title[Merger shocks and radio relics]{Can cluster merger shocks reproduce the luminosity and shape distribution of radio relics?}

\author[Nuza et al.]{
	Sebasti\'an\,E.\,Nuza$^{1,2,3}$\thanks{E-mail: snuza@aip.de},
	Jakob\,Gelszinnis$^{4}$, 
	Matthias\,Hoeft$^{4}$, 
	and Gustavo\,Yepes$^{5}$ \\
	$^1$Leibniz-Institut f\"ur Astrophysik Potsdam, An der Sternwarte 16, 14482 Potsdam, Germany\\
	$^2$Instituto de Astronom\'{\i}a y F\'{\i}sica del Espacio (IAFE, CONICET-UBA), CC 67, Suc. 28, 1428 Buenos Aires, Argentina\\
	$^3$Facultad de Ciencias Exactas y Naturales (FCEyN), Universidad de Buenos Aires (UBA), Buenos Aires, Argentina\\
	$^4$Th\"uringer Landessternwarte, Sternwarte 5, 07778 Tautenburg, Germany\\
	$^5$Grupo de Astrof\'{\i}sica, Universidad Aut\'onoma de Madrid, Cantoblanco, 28039 Madrid, Spain
	}

\date{}

\begin{document}

\maketitle

\begin{abstract}
	 Radio relics in galaxy clusters are believed to trace merger shock fronts. If cosmological structure formation 
	 determines the luminosity, size and shape distributions of radio relics then merger shocks need to be lighted up in 
	 a homogeneous way. We investigate if a mock relic sample, obtained from zoomed galaxy cluster simulations, 
	 is able to match the properties of relics measured in the NRAO VLA Sky Survey (NVSS). We compile a list of all 
	 radio relics known to date and homogeneously measure their parameters in all NVSS images and apply the same 
	 procedure to relics in our simulations. Number counts in the mock relic sample increase 
	 more steeply towards lower relic flux densities, suggesting an incompleteness of NVSS in this regime. 
	 Overall, we find that NVSS and mock samples show similar properties. However, large 
	 simulated relics tend to be somewhat smaller and closer to the cluster centre than observed ones. 
	 Besides this, the mock sample reproduces very well-known correlations for radio relics, in particular those 
	 relating the radio luminosity with the largest linear size and the X-ray luminosity. 
	 We show that these correlations are largely governed by the sensitivity of the NVSS observations. 
	 Mock relics show a similar orientation with respect to the direction to the cluster 
	 centre as the NVSS sample. Moreover, we find that their {\it maximum} radio luminosity roughly correlates 
	 with cluster mass, although displaying a large scatter. The overall good agreement 
	 between NVSS and the mock sample suggests that properties of radio relics are indeed governed by merger shock 
	 fronts, emitting in a homogeneous fashion. Our study demonstrates that the combination of mock observations and data from 
	 upcoming radio surveys will allow to shed light on both the origin of radio relics and the nature of the intracluster medium.  
\end{abstract}


\section{Introduction}
\label{sec:intro}


	%
	Diffuse extended radio sources in galaxy clusters with a steep power-law spectrum and without optical 
	counterparts are known since many decades \mbox{\citep{1996IAUS..175..333F}}. The sources found in the periphery of 
	clusters were initially attributed to radio galaxies in an inactive phase, i.e. {\em relics} of radio galaxies.
	In such a scenario, the radio spectrum should show a clear break towards high frequencies as a result of the ageing 
	of the electron population. 
	Instead, relics show spectra close to a power-law at least up to frequencies of about $10\,$GHz, suggesting the existence of recently
	accelerated electrons. \mbox{\citet{1998A&A...332..395E}} proposed a link between radio relics and the shock fronts
	induced by galaxy cluster mergers. The study of several spectacular relics seems to support this scenario, as suggested by the 
	following facts: 
	(i) X-ray surface brightness jumps attributed to shock fronts have been found in these systems; 
	(ii) spectral index maps of the diffuse radio emission indicate that electron populations are ageing towards the 
	cluster centre (i.e., opposite to the propagation direction of the shock), 
	and (iii) relics display a significant degree of polarization with B-vectors 
	aligned with the shock surface, as can be seen, for instance, in the relic in Abell\,3667 \citep{1997MNRAS.290..577R,2010ApJ...715.1143F}, 
	in the `Sausage' relic in CIZA\,J2242$+$53 \citep{2010Sci...330..347V,2013A&A...555A.110S,2015A&A...582A..87A}, and in the 
	`Toothbrush' relic in 1RXS\,J0603$+$42 \citep{2012A&A...546A.124V,2016ApJ...818..204V}.

	%
	Systematic searches for relics have been carried out, e.g. by investigating both 
	X-ray bright Abell-type clusters \citep{giovannini:99} in the NRAO VLA Sky Survey 
	(NVSS; \citealt{1998AJ....115.1693C}) and Abell clusters \citep{2001ApJ...548..639K} 
	accessible to the Westerbork Northern Sky Survey (WENSS; \citealt{1997A&AS..124..259R}), 
	by searching for steep spectrum sources in the NVSS \citep{2011A&A...533A..35V}, 
	or --as a by-product-- in the Giant Metrewave Radio Telescope (GMRT) radio halo survey, 
	which is based on a redshift- and flux-limited cluster sample selected from the REFLEX and eBCS 
	catalogues \citep{2015A&A...579A..92K}. Recently, the signature of the 
	Sunyaev-Zel'dovich (SZ) effect in the intracluster medium (ICM) has also been used to identify galaxy 
	clusters (see e.g., \citealt{2015A&A...581A..14P}) that can be later chosen as targets for
	the relic searches \citep{2011ApJ...736L...8B,2015MNRAS.453.3483D}. 
	All of these works, with the exception of the search performed 
	by \citet{2011A&A...533A..35V}, have been done by observing at the location of rather 
	massive galaxy clusters. Therefore, it is not surprising that only very few relic 
	candidates are known to be present in poor clusters.

	%
	The sample of known radio relics shows a plethora of correlations between relic and 
	cluster properties. Some of them include: the relation between radio emission 
	and cluster X-ray luminosity or mass \citep{2012A&ARv..20...54F,2014MNRAS.444.3130D}, 
	between radio relic luminosity and redshift \citep{2014MNRAS.444.3130D} and 
	largest linear size (LLS; \citealt{2012A&ARv..20...54F}). 
	Moreover, radio relics tend to be tangentially oriented with respect to the isocontours 
	of the X-ray surface brightness distribution of clusters \citep{2011A&A...533A..35V}. 
	Despite of the existence of these intriguing correlations, their origin, and the systematic 
	effects shaping them, is up to now little studied.

	%
	As aforementioned, radio relics most likely originate from cluster merger shock fronts. 
	For several relics, the jump in X-ray surface brightness and/or temperature maps has 
	been clearly measured \citep[e.g.,][]{2014xru..confE.181S,2015A&A...582A..87A,2016MNRAS.460L..84B,2016ApJ...818..204V}. 
	In addition to the shocks found in several clusters, the disturbance of the cluster X-ray 
	surface brightness, as measured by power ratios, centroid shifts and concentration 
	parameters, shows that relics are, in general, related to ongoing mergers 
	\citep{2015ApJ...813...77Y}. 
	
	%
	The high temperatures of the ICM result from shock dissipation, 
	hence, the latter must be included when simulating its evolution. 
	In a pioneering work, \citet{2000ApJ...542..608M} studied the properties of shock 
	fronts in a cosmological simulation. These authors distinguished between {\it accretion} shocks, 
	responsible for heating the primordial intergalactic medium, and {\it merger} shocks, which show 
	a complex three-dimensional structure and provide additional gas heating. 
	The properties and distribution of cosmological shock waves have been studied in 
	several subsequent works 
	\citep[e.g.,][]{2003ApJ...593..599R,2006MNRAS.367..113P,2008ApJ...689.1063S,2009MNRAS.395.1333V}. 
	It has been shown that the distribution and strength of the shocks vary significantly with 
	resolution and numerical scheme \citep{2011MNRAS.418..960V}. However, different simulations 
	basically agree that accretion shocks tend to show Mach numbers of roughly $100-1000$, 
	and that the ICM is filled with a large number of lower Mach number 
	shocks in the range from 1 to 10. 
	Recent works using improved numerical schemes and increased resolution confirm 
	the abundance and complexity of shock fronts in the ICM 
	\citep{2014ApJ...782...21M,2016MNRAS.461.4441S}.

	%
	It has been known for decades that strong shocks in supernova remnants serve as efficient 
	particle accelerators (see \citealt{2008ARA&A..46...89R} for a review). 
	Even if the physical conditions of intergalactic and intracluster media are quite 
	different from those of the interstellar medium, it appears likely that accretion 
	and merger shocks are also capable of accelerating electrons, protons and nuclei. 
	In this respect, radio relics may be regarded as immediate proof for electron 
	acceleration. However, evidence for proton and nuclei acceleration 
	at these locations is still lacking. A plausible acceleration mechanism is the so-called 
	diffusive shock acceleration 
	\citep[DSA;][]{1983SSRv...36...57D,1987PhR...154....1B,2012JCAP...07..038C}. 
	It has been shown that merger shocks may accelerate electrons and protons via DSA
	\citep{2012ApJ...756...97K,2013ApJ...764...95K}. However, a simple scenario for the 
	origin of radio emission in which relativistic electrons originate 
	from the thermal pool by acceleration at the shock front via DSA in the test-particle 
	regime is challenged by several observational findings, namely: 
	(i) Mach numbers derived from the X-ray surface brightness and/or temperature jumps 
	are often too low compared to those inferred from the radio luminosity and spectral index 
	(see e.g., \citealt{2014MNRAS.443.2463O,2016ApJ...818..204V}); (ii) the spectra of radio relics are 
	possibly curved at frequencies above 10\,GHz \citep{2016MNRAS.455.2402S}, 
	and (iii) upper limits for the $\gamma$-ray emission from clusters are inconsistent 
	with standard assumptions in DSA \citep{2014MNRAS.437.2291V,2016MNRAS.459...70V}. 
	Improved modelling including, for instance, re-acceleration of mildly relativistic 
	electrons \citep{2013MNRAS.435.1061P,2015ApJ...809..186K,2016ApJ...823...13K}, 
	and/or different acceleration mechanisms such as shock drift acceleration 
	\citep{2014ApJ...794..153G}, may alleviate the tension between models and observations. 
	Similarly, the recent discovery that fossil relativistic electrons from active galactic 
	nuclei (AGN) can be re-accelerated at cluster shocks could provide an explanation for 
	these discrepancies, at least for some relics \citep{2017NatAs...1E...5V}. 
	We stress that the origin of the relativistic electrons responsible for the non-thermal 
	radio emission in the general relic population is still a matter of ongoing debate.

	\begin{figure}
		\includegraphics[width=0.48\textwidth]{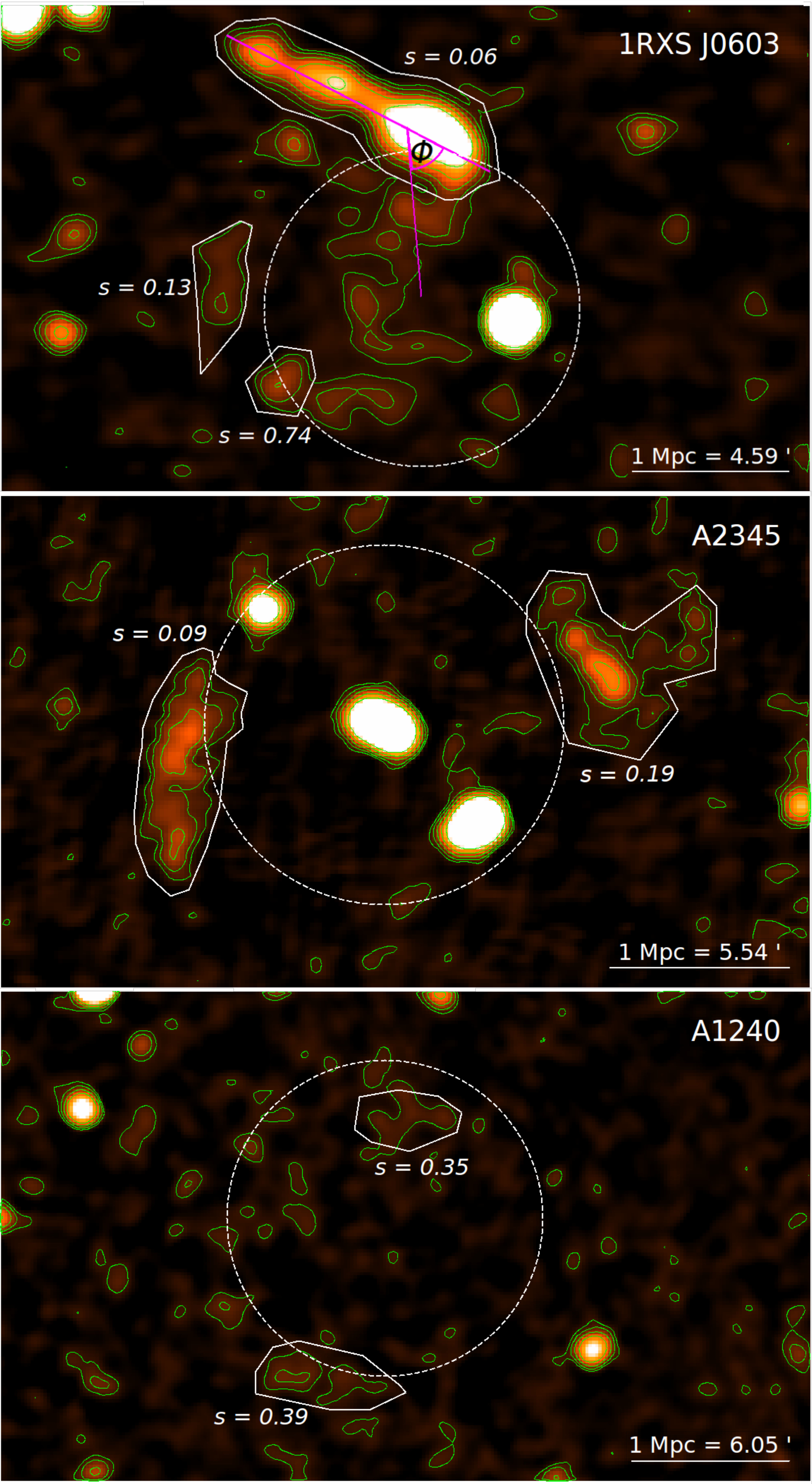} 
		\caption{
			Examples of relic image analysis within NVSS.  
			The colour scale indicates the surface brightness distribution. 
			Contours (green solid lines) are drawn at 
			$[2,4,8,16,32]\times\sigma_\text{NVSS}$, where $\sigma_\text{NVSS}$ 
			is the average rms brightness fluctuations in the NVSS images. 
			The manually defined relic regions are also shown (white solid lines). 
			Cluster positions are indicated by a circle (white dashed lines) with 
			1\,Mpc radius centred on the coordinates given by SIMBAD. The angle $\phi$ 
			between purple solid lines measures relic orientation. 
			The prominent elongated structure in the top panel is usually referred to as the {\it Toothbrush} relic. 
			For Abell\,2345 the brightness distribution is shown after subtracting several bright sources 
			in the relic regions.
		}
		\label{fig:NVSSimage}
	\end{figure}

	%
	The strength, structure and origin of the magnetic fields in relics are also loosely 
	constrained. No direct measurement of their field strength has been possible 
	so far. The most stringent lower limit for the magnitude of the field has been derived 
	for the northern relic in Abell\,3667 via upper limits of the inverse Compton (IC) emission, 
	namely $>3 \,\rm \mu G$ \citep{2010ApJ...715.1143F}. 
	For the relic in the Coma cluster, \citet{2013MNRAS.433.3208B} inferred a field 
	strength of $\sim 3\,\rm \mu G$ using Faraday rotation measurements. 
	The `Sausage' relic has a small intrinsic width, which suggests a lower limit 
	for the magnetic field strength of $\gtrsim 5\,\rm \mu G$ \citep{2010Sci...330..347V}. 
	Magnetic fields in relics may reflect those present in the ICM, 
	however, this underlying field might be also amplified at the shock front, e.g. 
	by the firehose instability \citep{2014ApJ...797...47G}, which illustrates 
	the complex non-linear nature of the problem.

	%
	For simulating radio relics in cosmological simulations assumptions about 
	both the acceleration of electrons and the magnetic fields in relics have to be made. 
	\citet{2007MNRAS.375...77H} introduced a semi-analytic formula relating the 
	non-thermal radio emission, the physical conditions of the downstream medium 
	and the Mach number of the shocks. By grafting this prescription on to simulations 
	it has been shown that cluster mergers may reproduce typical radio relic morphologies 
	\citep{2008MNRAS.391.1511H,2011JApA...32..509H,2011ApJ...735...96S}. 
	Using higher resolution simulations including the same or similar prescriptions 
	for the radio emission it has been shown that typical relic distances range 
	from a few hundred kpc to about one Mpc from the cluster centre \citep{2012MNRAS.421.1868V} 
	in agreement with observations, and that 
	the spectral index gradient may partly reflect the Mach number distribution 
	of the shock front \citep{2013ApJ...765...21S}. In contrast, \citet{2015ApJ...812...49H} 
	found that radio emission associated with a set of simulated relic regions was located 
	at distances larger than observed, suggesting that a more efficient electron 
	acceleration at low Mach number shocks could alleviate the discrepancy.

	%
	As already mentioned, powerful radio relics are preferentially found in massive galaxy clusters 
	\citep{2014MNRAS.444.3130D}. In order to mimic the entire sample of observed 
	radio relics using simulations, the simulated volume has to be large 
	enough to contain a sufficiently large number of massive clusters. 
	On the other hand, shock fronts need to be resolved with enough resolution to properly 
	describe their morphology. 
	These two conditions can be achieved by means of the so-called `zoom-in' technique, 
	where a set of cluster regions, previously selected from a large cosmological box, 
	can be re-simulated at high resolution \citep{2001ApJ...554..903K}. 
	This technique has been applied to simulate 282 cluster regions drawn 
	from a $1\,h^{-3}\,\text{Gpc}^3$ cosmological simulation ($h$ denoting 
	the reduced Hubble constant) dubbed as the MUSIC-2 galaxy cluster sample \citep{2013MNRAS.429..323S}.

	%
	In a simplified view, merger shock fronts get `illuminated' via synchrotron emission of relativistic 
	electrons and magnetic fields in the downstream region. To some extent, this resembles the non-related 
	process of stars in galaxies revealing the presence of the dark matter haloes hidden in the cosmic web. 
	However, little is known about the relation between radio luminosity and shock properties such as the Mach number. 
	The fact that only about 50 radio relics have been discovered so far may suggests that only the strongest shocks 
	become radio luminous or, alternatively, that additional factors could play a crucial role when forming these objects. 
	In \citet{2012MNRAS.420.2006N}, to avoid translating cluster merger properties into shock parameters and their corresponding 
	radio luminosities, we derived predictions for the abundance of radio relics simply by introducing the probability 
	of finding radio-luminous shocks with a given power, $P_{1.4}$, as a function of cluster mass.

	%
	In this work we pursue the question if cosmological shock fronts illuminated 
	with a uniform prescription result in a sample 
	of radio relics with a similar flux density distribution, morphology and location 
	within galaxy clusters as the observational one. To this end, we 
	homogeneously measure flux densities, shapes and positions for all known relics and 
	perform a comparison to a mock relic sample. 
	If the properties of the synthetic relics match observations 
	then, any additional requirement for electron acceleration such as, 
	for instance, a pre-existing electron population, must be fulfilled at any 
	other merger shock as well.

	%
	The paper is organized as follows. In Section~\ref{sec:NVSS}, we compile a list 
	of all radio relics known to date and consistently measure their flux densities, 
	shapes, and positions within galaxy clusters found in the NVSS. 
	In Section~\ref{sec:MockRelics} we describe our mock observations of relics  
	belonging to the simulated MUSIC-2 cluster sample. 
	In Section~\ref{sec:Compare} we perform the comparison between the two samples 
	and, in Section~\ref{sec:Discussion}, we discuss the results. 
	Finally, in Section~\ref{sec:Summary}, we summarize our findings.

	%
	Throughout this paper, we assume a $\Lambda$ cold dark matter cosmology 
	with $\Omega_{\rm M}=0.27$, $\Omega_{\Lambda}=0.73$ 
	and $H_0=70\,\rm km\,s^{-1}\,Mpc^{-1}$.

\section{Radio relics in the NVSS} 
\label{sec:NVSS}

	\begin{figure}
		\hspace*{-0.2cm}
		\includegraphics[width=0.47\textwidth]{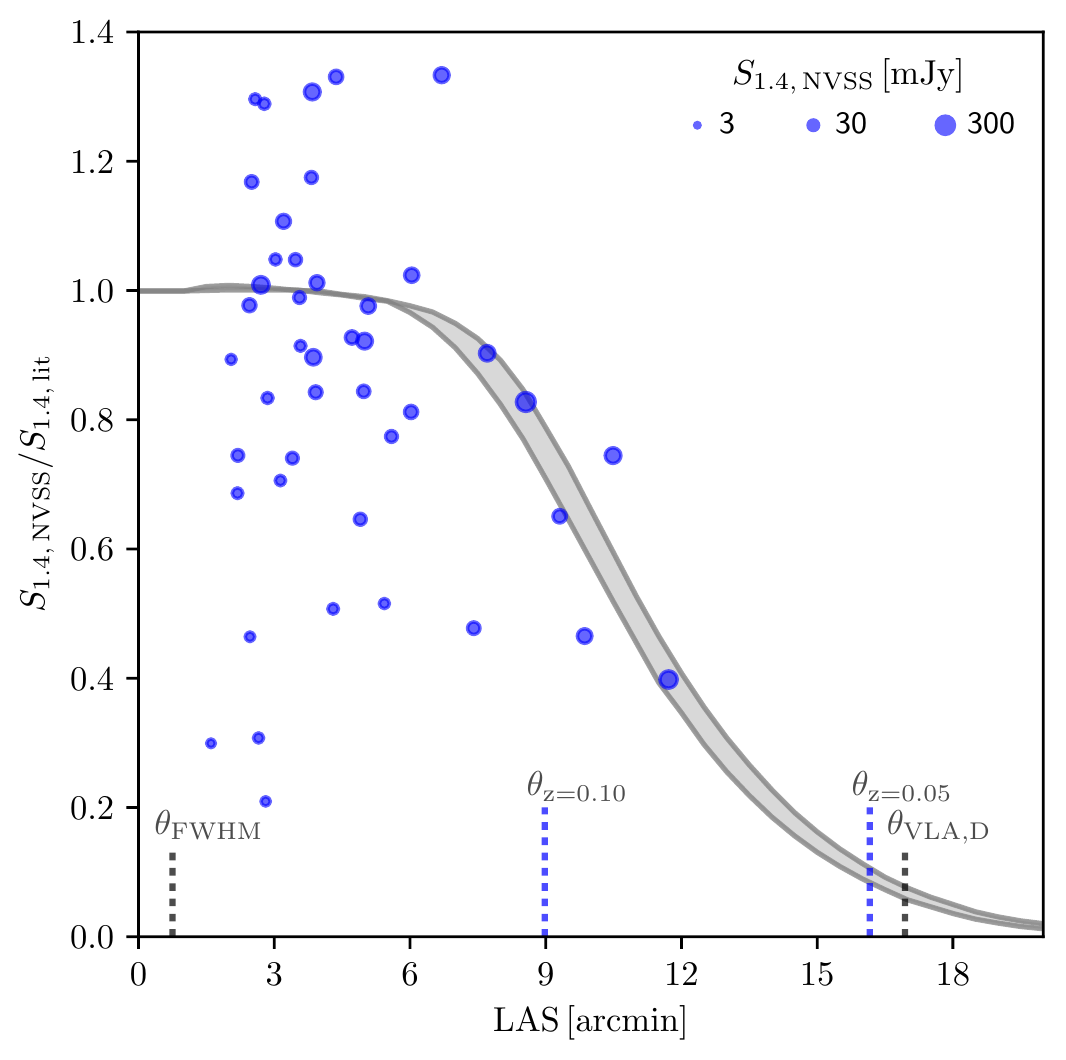}
		\caption{
			Ratio of flux densities measured in the NVSS images and 
			those reported in the literature versus the largest angular 
			size, LAS, of the largest radio-emitting region in each cluster. 
			Symbol size scales with NVSS-measured flux. The shaded grey region 
			shows the expectation for the NVSS flux recovery as derived 
			by \citet{2013ApJ...779..189F}. Also shown are relevant 
			angular scales.
			} 
		\label{fig:FluxesLAS}
	\end{figure}

\subsection{Compilation of known relics to date}
\label{sec:relicCompile}

	%
	Our aim is to compile a list of radio relics as complete as possible for later 
	comparison with simulations. \citet{2012A&ARv..20...54F} inventoried diffuse radio emission 
	in galaxy clusters and their observational properties. Since then, several 
	new relics have been reported in the literature. In this section, we present an up-to-date 
	collection of relics based on the compilations of \citet{2012MNRAS.420.2006N}, 
	\citet{2012A&ARv..20...54F}, and \citet{2015ApJ...813...77Y}, complemented by 
	recently reported objects.

	%
	In observations, extended diffuse radio emission without optical counterpart that is located in the cluster periphery is 
	usually classified as a relic. In our simulations, we model radio relics assuming Fermi-I acceleration of electrons 
	at merger shocks. Therefore, the relic sample compiled from observations may systematically differ from the one derived 
	in simulations as the former naturally misses some of the emission that is sufficiently extended 
	or too close to the cluster centre. On the other hand, the observed sample may also comprise diffuse emission 
	without optical counterparts but which does not originate from electron acceleration at merger shocks. Such objects 
	have been identified in observations. A plausible explanation for their origin is given by shock compression of fossil 
	AGN plasma \citep{2001A&A...366...26E}. 
	Following the convention of \citet{2004rcfg.proc..335K}, we use the names {\it gischt} and {\it phoenix} to label 
	diffuse radio emission if it is attributed to Fermi-I acceleration or shock compression, respectively. 
	Since in our simulations we only model radio gischts, a careful inspection of the classification is needed 
	before proceeding with the comparison. In this regard, the spectral properties, and their associated 
	index maps, provide the most clear evidence to perform the classification: an overall power-law spectrum with spectral 
	index steeper than $-1$ and spectral steeping towards the cluster centre strongly suggest the presence of a 
	gischt \citep{2007MNRAS.375...77H}, while a break in the overall spectrum and an irregular distribution of spectral 
	indices indicate that we are dealing with a phoenix.


\begin{table*}
	\caption{Radio relic collection. {Column 1:} cluster name (common names for the cluster or relic are indicated in italic). 
		{Column 2:} cluster redshift. 
		{Column 3:} cluster X-ray luminosity in the $0.1-2.4\,$keV energy band.
		{Column 4:} flux density of radio relics as reported in literature. 
		{Column 5:} flux density measured in NVSS images. 
		  Relics are excluded when: 
			  (i) the origin is most likely not shock acceleration, i.e. not gischt or gischtlet;
			  (ii) the source position is not covered by NVSS, i.e. the declination is too low;
			  (iii) the emission is too faint;
			  (iv) the source is too extended, hence NVSS cannot recover the flux density, and
			  (v) the contamination by other sources is too strong.
		{Column 6:} Classification (`can.' stands for candidate).  
		{Column 7:} References for cluster X-ray luminosity and radio flux density.}
	\begin{scriptsize}
	\hspace*{-0.6cm}
	\begin{tabular}{llr@{}lr@{}lr@{}llr@{}l}
		\hline
		Cluster                           & \multicolumn{1}{c}{$z$} & \multicolumn{2}{c}{$L_{\rm X}$} & \multicolumn{2}{c}{$S_{1.4,\text{lit}}$} & \multicolumn{2}{c}{$S_{1.4,\text{NVSS}}$} & Classification                & \multicolumn{2}{c}{References} \\
	 	                                  &                         & \multicolumn{2}{c}{(\ffergs)}     & \multicolumn{2}{c}{(mJy)}                  & \multicolumn{2}{c}{(mJy)}                   &                               &              &                 \\
  	(1)                               & \multicolumn{1}{c}{(2)} & \multicolumn{2}{c}{(3)}         & \multicolumn{2}{c}{(4)}                  & \multicolumn{2}{c}{(5)}                   & (6)                           & \multicolumn{2}{c}{(7)}        \\
	 	\hline                                                                                                                                                                                                
 	 	Abell\,13                         & 0.094                   &  \rule{0.38cm}{0cm} 1&.18       &                      31&.0               & \multicolumn{2}{c}{not gischt}            & phoenix can.                  &     Fer12 \,&$\cdot$ Sle01     \\
	 	Abell\,85                         & 0.055                   &                     4&.10       &                      43&.0               & \multicolumn{2}{c}{not gischt}            & phoenix                       &     Fer12 \,&$\cdot$ Sle01     \\
	 	Abell\,115                        & 0.197                   &                     8&.91       &                     147&.0               & \rule{0.50cm}{0cm}    68&.4               & gischt                        &     Fer12 \,&$\cdot$ Gov01     \\
	 	Abell\,133                        & 0.057                   &                     1&.93$^a$   &                     137&.0               & \multicolumn{2}{c}{not gischt}            & phoenix can.                  &     Ebe96 \,&$\cdot$ Ran10     \\
	 	Abell\,209                        & 0.206                   &                     6&.17       &                       1&.0               & \multicolumn{2}{c}{too faint}             & gischt can. ($+$halo)         &     Gio09 \,&$\cdot$ Gio09     \\
	 	Abell\,521                        & 0.247                   &                     8&.47       &                      15&.0               &                       13&.7               & gischt ($+$ halo)             &     Fer12 \,&$\cdot$ Gia06     \\
	 	Abell\,523                        & 0.104                   &                     1&.07       &                      61&.0               &                       29&.1               & gischtlet can.                 &     Fer12 \,&$\cdot$ Wee11c    \\              
	 	Abell\,548b                       & 0.045                   &                     0&.15       &                     121&.0               &   \multicolumn{2}{c}{too extended}       & gischtlet can.                 &     Fer12 \,&$\cdot$ Fer06     \\
	 	Abell\,610                        & 0.095                   &                     0&.8$^b$    &                      17&.3               &                       15&.5               & gischt can.                   &     Bos08 \,&$\cdot$ Gio00     \\ 
		Abell\,725                        & 0.092                   &                     0&.45$^a$   &                       6&.0               & \multicolumn{2}{c}{not gischt}            & phoenix can.                   & B{\"o}h00 \,&$\cdot$ Kem01     \\ 
	 	Abell\,746                        & 0.232                   &                     3&.68       &                      24&.5               &                       12&.4               & gischt ($+$ halo)             &     Fer12 \,&$\cdot$ Wee11c    \\ 
	 	Abell\,S753                       & 0.014                   &                     0&.5$^c$    &                     460&.0               & \multicolumn{2}{c}{not gischt}            & phoenix can.                  &     Sub03 \,&$\cdot$ Sub03     \\  
	 	Abell\,754                        & 0.054                   &                     2&.21       &                       6&.0               & \multicolumn{2}{c}{too faint}             & gischtlet can. ($+$ halo)      &     Fer12 \,&$\cdot$ Mac11     \\
	 	Abell\,781                        & 0.300                   &                     4&.6        &                      15&.5               &                       16&.2               & gischt ($+$ halo)             &     Fer12 \,&$\cdot$ Gov11     \\
	 	Abell\,910                        & 0.206                   &                     3&.39$^a$   &                      12&.1               & \multicolumn{2}{c}{contam.}               & gischtlet can.                 &     Ebe00 \,&$\cdot$ Gov12     \\
	 	Abell\,1033                       & 0.130                   &                     4&.43$^d$   &                      46&.9               & \multicolumn{2}{c}{not gischt}            & phoenix can.                  &    Gas15a \,&$\cdot$ Gas15a    \\ 
	 	Abell\,1240                       & 0.159                   &                     1&.0        &                      16&.1               &                       11&.4               & gischt double                 &     Fer12 \,&$\cdot$ Bon09     \\
	 	Abell\,1300                       & 0.307                   &                    13&.73       &                      20&.0               &                       19&.8               & gischt can. ($+$ halo)        &     Fer12 \,&$\cdot$ Ven13     \\
 	 	Abell\,1351                       & 0.322                   &                     8&.4        &                      13&.0               &                       16&.8               & gischtlet can. ($+$ halo)      &     Gia09 \,&$\cdot$ Gia09     \\
	 	Abell\,1367                       & 0.022                   &                     0&.82       &                     232&.0               & \multicolumn{2}{c}{too extended}          & gischt can. ($+$ halo)        &     Far13 \,&$\cdot$ Far13     \\
	 	Abell\,1443                       & 0.270                   &                     6&.2        &                      9&.8$^A$           &                        8&.8               & gischt can. ($+$ halo)        &     Bon15 \,&$\cdot$ Bon15     \\
	 	Abell\,1612                       & 0.179                   &                     2&.41       &                      62&.8               &                       61&.3               & gischt can.                   &     Fer12 \,&$\cdot$ Wee11c    \\
	 	Abell\,1656: {\it Coma}           & 0.023                   &                     3&.99       &                     260&.0               & \multicolumn{2}{c}{too extended}          & gischt can. ($+$ halo)        &     Fer12 \,&$\cdot$ Fer12     \\
	 	Abell\,1664                       & 0.128                   &                     3&.09       &                     107&.0               &                       96&.6               & gischt can.                   &     Fer12 \,&$\cdot$ Gov01     \\
		Abell\,1682                       & 0.226                   &                     7&.0        &                      26&.9$^B$           &                       20&.0               & gischtlet can. ($+$ halo)      &     Ven13 \,&$\cdot$ Ven13     \\
	 	Abell\,1758N                      & 0.280                   &                     7&.6$^a$    &                      12&.8               &                       22&.1               & gischt double can. ($+$ halo) &     Ebe98 \,&$\cdot$ Gio09     \\
		Abell\,2034                       & 0.113                   &                     3&.90       &                      24&.0               &                       28&.0               & gischt can. ($+$ halo)        &     Ebe98 \,&$\cdot$ Wee11c    \\
	 	Abell\,2048                       & 0.097                   &                     1&.91       &                      19&.0               & \multicolumn{2}{c}{not gischt}            & phoenix                       &    Wee11a \,&$\cdot$ Wee11a    \\
	 	Abell\,2061                       & 0.078                   &                     3&.95       &                      27&.6               &                        5&.8               & gischt                        &    Wee11c \,&$\cdot$ Wee11c    \\   
	 	Abell\,2063                       & 0.035                   &                     0&.98       &                      81&.5$^C$           & \multicolumn{2}{c}{not gischt}            & phoenix can.                  &     Fer12 \,&$\cdot$ Kom94     \\
	 	Abell\,2163                       & 0.203                   &                    22&.73       &                      18&.7               &                       12&.8               & gischt can.                   &     Fer12 \,&$\cdot$ Fer01     \\
	 	Abell\,2255                       & 0.081                   &                     2&.64       &                      12&.0               & \multicolumn{2}{c}{too faint}             & gischt can. ($+$ halo)        &     Fer12 \,&$\cdot$ Piz08     \\
	 	Abell\,2256                       & 0.058                   &                     3&.75       &                     462&.0               &                      183&.9               & gischt can. ($+$ halo)        &     Cla06 \,&$\cdot$ Cla06     \\
	 	Abell\,2345                       & 0.177                   &                     5&.90       &                      59&.0               &                       78&.7               & gischt double                 &     Bon09 \,&$\cdot$ Bon09     \\
	 	Abell\,2443                       & 0.108                   &                     1&.9        &                       6&.5               & \multicolumn{2}{c}{not gischt}            & phoenix                       &     Fer12 \,&$\cdot$ Coh11     \\
	 	Abell\,2744                       & 0.308                   &                    12&.86       &                      18&.2               &                        9&.4               & gischt ($+$ halo)             &     Fer12 \,&$\cdot$ Gov01     \\
		Abell\,3365                       & 0.093                   &                     0&.86       &                      47&.9               &                       38&.9               & gischt double can.            &    Wee11c \,&$\cdot$ Wee11c    \\
	 	Abell\,3376                       & 0.046                   &                     1&.08       &                     302&.0               & \multicolumn{2}{c}{too extended}          & gischt double                 &     Fer12 \,&$\cdot$ Bag06     \\ 
	 	Abell\,3411                       & 0.169                   &                    53&.9$^a$    &                      53&.0               &                       53&.6               & gischt can. ($+$ halo)        &     Ebe02 \,&$\cdot$ Wee13     \\
		Abell\,3527-bis                   & 0.200                   &                     1&.9         &                      35&.0               &                       22&.6               & gischt                        &     Gas17 \,&$\cdot$ Gas17     \\
	 	Abell\,3667                       & 0.056                   &                     4&.73       &                    2400&.0               & \multicolumn{2}{c}{low dec.}              & gischt                        &     Fer12 \,&$\cdot$ Roe97     \\
	 	Abell\,4038                       & 0.030                   &                     1&.97       &                      49&.0               & \multicolumn{2}{c}{not gischt}            & phoenix                       &     Fer12 \,&$\cdot$ Sle01     \\
	 	ACT-CL\,J0102$-$49: {\it El Gordo}& 0.870                   &                    64&.6$^e$    &                      8&.6$^D$           & \multicolumn{2}{c}{low dec.}              & gischt double                 &     Lin14 \,&$\cdot$ Lin14     \\
	 	CIZA\,J0107$+$54                  & 0.107                   &                     3&.9        &                      30&.0               &                        9&.2               & gischt double can.            &     Ran16 \,&$\cdot$ Ran16     \\
	 	CIZA\,J0649$+$18                  & 0.064                   &                     2&.38       &                      27&.5$^E$           &                       21&.3               & gischt can.                   &    Wee11c \,&$\cdot$ Raj17     \\
	 	CIZA\,J2242$+$53: {\it Sausage }  & 0.192                   &                     6&.80       &                     144&.0               &                      107&.2               & gischt                        &     Fer12 \,&$\cdot$ Str13     \\
	 	CL\,0217$+$70                     & 0.066                   &                     0&.63       &                        &-                & \multicolumn{2}{c}{{contam.}}             & gischt double can. ($+$ halo) &     Fer12 \,&$\cdot$ Bro11     \\
	 	CL\,1446$+$26                     & 0.370                   &                     3&.42       &                       5&.3               & \multicolumn{2}{c}{contam.}               & gischt can. ($+$ halo)        &     Fer12 \,&$\cdot$ Gov12     \\
		MACS\,J0025$-$1222                & 0.586                   &                     8&.8        &                       2&.3$^F$           & \multicolumn{2}{c}{too faint}             & gischt double                 &     Ebe07 \,&$\cdot$ Ris17     \\
	 	MACS\,J0717$+$37                  & 0.555                   &                     8&.6$^a$    &                      83&.0               &                      108&.5               & gischtlet can. ($+$ halo)      &     Fer12 \,&$\cdot$ Pan13     \\
	 	MACS\,J1149$+$22                  & 0.544                   &                    14&.0        &                      10&.3               &                       13&.3               & gischt double can.            &     Bon12 \,&$\cdot$ Bon12     \\
	 	MACS\,J1752$+$44                  & 0.366                   &                     8&.0$^e$    &                      101&.8$^G$           &                       91&.2               & gischt double ($+$ halo)      &     Bon12 \,&$\cdot$ Bon12     \\
	 	MACS\,J2243$-$09                  & 0.447                   &                    11&.56       &                       1&.9$^D$           & \multicolumn{2}{c}{too faint}             & gischt can.                   &     Can16 \,&$\cdot$ Can16     \\
	 	MaxBCG\,138$+$25                  & 0.324                   &                      &-         &                      24&.7               & \multicolumn{2}{c}{not gischt}            & phoenix can.                  &        -- \,&$\cdot$ Wee11a    \\
	 	MaxBCG\,217$+$13                  & 0.160                   &                     1&.0        &                        &-                & \multicolumn{2}{c}{not gischt}            & phoenix can.                  &     Wee09 \,&$\cdot$ Wee09    \\
	 	PLCK\,G004$-$19                   & 0.516                   &                    16&.0        &                      37&.0               &                       36&.2               & gischt can.                   &     Sif14 \,&$\cdot$ Sif14     \\
	 	PLCK\,G287$+$32                   & 0.390                   &                    17&.2        &                      58&.0               &                       59&.3               & gischt double                 &     Fer12 \,&$\cdot$ Bag11     \\
	 	PSZ1\,G096$+$24                   & 0.300                   &                     3&.7        &                      27&.2               &                       20&.1               & gischt double                 &     Gas14 \,&$\cdot$ Gas14     \\
	 	PSZ1\,G108$-$11                   & 0.335                   &                     7&.5        &                     113&.1               &                      104&.2               & gischt double ($+$ halo)      &    Gas15b \,&$\cdot$ Gas15b    \\
	 	RXC\,J0225$-$29                   & 0.060                   &                     0&.41       &                      37&.0               &                       31&.2               & gischt can.                   &     Sha16 \,&$\cdot$ Sha16     \\
	 	RXC\,J1053$+$54                   & 0.070                   &                     3&.69       &                      15&.0               &                        7&.0               & gischt can.                   &     Fer12 \,&$\cdot$ Wee11c    \\
	 	RXC\,J1234$+$09                   & 0.229                   &                     6&.32       &                       3&.1               & \multicolumn{2}{c}{too faint}             & gischt can. ($+$ halo)        &     Kal15 \,&$\cdot$ Kal15     \\ 
	 	RXC\,J1314$-$25                   & 0.247                   &                    10&.75       &                      30&.3               &                       40&.3               & gischt doube ($+$ halo)       &     Fer05 \,&$\cdot$ Fer05     \\
	 	S1081                             & 0.220                   &                      &-         &                       2&.4               & \multicolumn{2}{c}{too faint}             & gischt can.                   &        -- \,&$\cdot$ Mid08     \\
	 	ZwCl\,0008$+$52                   & 0.103                   &                     0&.5        &                      67&.0               &                       43&.6               & gischt double                 &     Fer12 \,&$\cdot$ Wee11b    \\
	 	ZwCl\,2341$+$00                   & 0.270                   &                     2&.4        &                      18&.5               &                       23&.0               & gischt double can.            &     Gas14 \,&$\cdot$ Gio10     \\
	 	1E\,0657$-$55: {\it Bullet}       & 0.296                   &                    22&.59       &                      82&.6               & \multicolumn{2}{c}{low dec.}              & gischt ($+$ halo)             &     Gio09 \,&$\cdot$ Shi15     \\
	 	1RXS\,J0603$+$42: {\it Toothbrush}& 0.225                   &                     7&.7        &                     364&.2$^H$           &                      301&.2               & gischt                        &     Ogr13 \,&$\cdot$ Wee16     \\
	 	24P73                             & 0.150                   &                      &-         &                      12&.0               & \multicolumn{2}{c}{not gischt}            & phoenix                       &        -- \,&$\cdot$ Wee11a    \\ 	
	 	\hline                      
	 \end{tabular}
	 	\end{scriptsize}
	 	\label{tab:clusters}
\end{table*}

\begin{table*}
	\contcaption{Radio relic collection.}
	 	\begin{minipage}{\textwidth}
	 	\begin{small}	
		    $^a$Corrected for cosmology; 
		    $^b$Estimated via $M_{\rm vir} = 2.05 \times 10^{14} M_\odot$ (Bos08) and Eq.~\ref{eq:LMscaling};
		    $^c$Estimated via $M_{\rm vir} = 1.5  \times 10^{14} M_\odot$ (Sub03) and Eq.~\ref{eq:LMscaling};
		    $^d$Converted to $0.1-2.4\,$keV band via {\tt mekal} model with $T = 6.39 \: \rm keV$ and a metallicity of 0.26 (Gas15a);
		    $^e$Converted to $0.1-2.4\,$keV band via {\tt mekal} model with $T = 14.5 \: \rm keV$ (Lin14) and a metallicity of 0.25;
		    $^A$Extrapolated from $323\,$MHz adopting a spectral index of $-1.2$;
		    $^B$Extrapolated from $240\,$MHz adopting a spectral index of $-1.62$;
		    $^C$Mean of the $1.360\,$GHz and $1.465\,$GHz flux values;
		    $^D$Extrapolated from $610\,$MHz adopting a spectral index of $-1.2$;
		    $^E$Extrapolated from $1.47\,$GHz adopting a spectral inded of $-1.15$;
		    $^F$Extrapolated from $325\,$MHz adopting a spectral index of $-1.3$;
		    $^G$Extrapolated from $1.714\,$GHz adopting a spectral index of $-1.18$;
		    $^H$Extrapolated from $1.5\,$GHz adopting a spectral index of $-1.09$.\\

		    References: 
		    Bag06  -- \citet{2006Sci...314..791B};
		    Bag11  -- \citet{2011ApJ...736L...8B};
		    B{\"o}h00 -- \citet{2000ApJS..129..435B};
		    Bon09  -- \citet{2009A&A...494..429B};
		    Bon12  -- \citet{2012MNRAS.426...40B};
		    Bon15  -- \citet{2015MNRAS.454.3391B};
		    Bos08  -- \citet{2008A&A...487...33B};
		    Bro11  -- \citet{2011ApJ...727L..25B};
		    Can16  -- \citet{2016MNRAS.458.1803C};
		    Cla06  -- \citet{2006AJ....131.2900C};
		    Coh11  -- \citet{2011AJ....141..149C};
		    Ebe96  -- \citet{1996MNRAS.281..799E};
		    Ebe98  -- \citet{1998MNRAS.301..881E};
		    Ebe00  -- \citet{2000MNRAS.318..333E};
		    Ebe02  -- \citet{2002ApJ...580..774E};
		    Ebe07  -- \citet{2007ApJ...661L..33E};
		    Far13  -- \citet{2013ApJ...779..189F};
		    Fer01  -- \citet{2001A&A...373..106F};
		    Fer05  -- \citet{2005A&A...444..157F}; 
		    Fer06  -- \citet{2006MNRAS.368..544F};
		    Fer12  -- \citet{2012A&ARv..20...54F};
		    Gas14  -- \citet{2014MNRAS.444.3130D};
		    Gas15a -- \citet{2015MNRAS.448.2197D};
		    Gas15b -- \citet{2015MNRAS.453.3483D};
		    Gas17  -- \citet{2017A&A...597A..15D};
		    Gia06  -- \citet{2006NewA...11..437G};
		    Gia09  -- \citet{2009ApJ...704L..54G};
		    Gio00  -- \citet{2000NewA....5..335G};
		    Gio09  -- \citet{2009A&A...507.1257G};
		    Gio10  -- \citet{2010A&A...511L...5G};
		    Gov01  -- \citet{2001A&A...376..803G};
		    Gov11  -- \citet{2011A&A...529A..69G};
		    Gov12  -- \citet{2012A&A...545A..74G};
		    Kal15  -- \citet{2015A&A...579A..92K};
		    Kem01  -- \citet{2001ApJ...548..639K};
		    Kom94  -- \citet{1994A&A...285...27K};
		    Lin14  -- \citet{2014ApJ...786...49L};
		    Mac11  -- \citet{2011ApJ...728...82M};
		    Mid08  -- \citet{2008AJ....135.1276M};
		    Ogr13  -- \citet{2013MNRAS.433..812O};
		    Pan13  -- \citet{2013A&A...557A.117P};
		    Piz08  -- \citet{2008A&A...481L..91P};
		    Raj17  -- Rajpurohit et al. (in preparation);
		    Ran10  -- \citet{2010ApJ...722..825R};
		    Ran16  -- \citet{2016ApJ...823...94R};
		    Ris17  -- \citet{2017A&A...597A..96R};
		    Roe97  -- \citet{1997MNRAS.290..577R};
		    Sif14  -- \citet{2014A&A...562A..43S};
		    Sha16  -- \citet{2016MNRAS.459.2525S};
		    Shi15  -- \citet{2015MNRAS.449.1486S};
		    Sle01  -- \citet{2001AJ....122.1172S};
		    Str13  -- \citet{2013A&A...555A.110S};
		    Sub03  -- \citet{2003AJ....125.1095S};
		    Ven13  -- \citet{2013A&A...551A..24V};
		    Wee09  -- \citet{2009A&A...508...75V};
		    Wee11a -- \citet{2011A&A...527A.114V};
		    Wee11b -- \citet{2011A&A...528A..38V};
		    Wee11c -- \citet{2011A&A...533A..35V};
		    Wee13  -- \citet{2013ApJ...769..101V};
		    Wee16  -- \citet{2016ApJ...818..204V}.
	\end{small}
  \end{minipage}
\end{table*}

	Interestingly, evidence for a spectral break has also been found for radio gischt textbook examples like 
	the `Toothbrush' and the `Sausage' relics \citep{2016MNRAS.455.2402S}. However, in these cases, the break occurs 
	at very high frequencies ($\nu>10\,$GHz) and its origin is still debated. In particular, \citet{2016A&A...591A.142B} 
	showed that the SZ effect may have a significant impact at these frequencies.

	Most radio relics in our compilation are unambiguously classified either as gischt or phoenix. However, there are 
	some cases where only little information is available and/or the emission properties present contradictory evidence. 
	In those cases, we add the attribute `candidate' to our classification. Specifically, the latter becomes particularly difficult 
	for diffuse emission that is only moderately extended, located rather close to the cluster centre, and that shows a complex 
	morphology. Three scenarios for its origin are conceivable: (i) it is produced by a radio galaxy that 
	recently went into passive phase and is significantly separated from the plasma ejecta due to a high velocity 
	w.r.t. the ambient medium; (ii) the plasma has been ejected long time ago and is now re-energised by shock compression, 
	i.e. it is a phoenix, or (iii) it might actually be a small merger shock at an early stage, e.g. see the radio 
	ridges in Abell\,1351 \citep{2009ApJ...704L..54G} and Abell\,1682 \citep{2013A&A...551A..24V}. 
	To highlight that some cluster mergers may cause small, complex shock fronts giving rise to radio emission according to 
	the gischt scenario, we introduce a new class: the `gischtlet'. For these objects, several aspects of the standard relic 
	classification do not apply, e.g. they are evidently not located at the cluster outskirts, hence all gischtlets are 
	only `candidates' in our nomenclature. We stress that, both our gischt and gischtlet classifications, are based on the 
	spectral and morphological properties of the diffuse radio emission only, regardless of the origin of the (re-)accelerated 
	electron population.

	Additionally, we note that, for some objects formerly listed as radio relics, evidence has been found that the emission 
	is actually produced by a radio galaxy. For the diffuse emission near the galaxy cluster Abell\,786, \citet{2012ApJ...744...46K} 
	reported a compact radio source near the centre of the radio structure. The diffuse emission near the galaxy cluster Abell\,2069 has been 
	also identified to be part of a radio galaxy (Drabent et al. in preparation). Therefore, we do not include these sources in our list.

	It has been found that the radio luminosity of relics correlates with X-ray luminosity of the host galaxy cluster. Because of this, 
	we also include cluster X-ray luminosities in the $0.1-2.4\,$keV energy band in our compilation. For most clusters, we adopted the 
	values given in \citet{2012A&ARv..20...54F}. If necessary, we corrected published X-ray luminosities to the cosmology and/or 
	energy band adopted here. For some clusters, where no X-ray luminosity is known but estimates for the virial mass have been published, 
	we adopt the luminosity-to-mass relation given in Section~\ref{sec:Xray_clusters} to estimate the X-ray emission.

	Our compilation contains a total of 69 clusters hosting radio gischt, gischtlet, or phoenix emission including candidates 
	(see Table~\ref{tab:clusters}). The majority of the relic-type diffuse emission is classified as gischt. In 13 of our clusters, 
	the emission is classified as phoenix, whereas in 7 as gischtlets.

\subsection{Analysis of NVSS images}
\label{sec:ImageAnalysis}

	%
	
\begin{table*}
\begin{center}
\caption{Radio relic emission `islands' identified in the NVSS images. {Column 1:} Cluster name and island identifier. 
	{Columns 2 and 3:} Right ascension and declination.
	{Columns 4 and 5:} Flux density and rest-frame luminosity at $1.4\,$GHz.
	{Columns 6 to 10:} Largest angular size, largest linear size, solid angle, shape and projected distance to cluster centre.
	{Column 11:} Island orientation (see Section~\ref{sec:shapes}).}
\scriptsize
\begin{tabular}{l r r r@{}l r r r@{}l r@{}l r r r}
	\hline
	Cluster                  & RA     & Dec     & \multicolumn{2}{c}{$S_{1.4}$}  & log$_{10}(P_{1.4})$     & LAS     & \multicolumn{2}{c}{LLS} & \multicolumn{2}{c}{$\Omega$}             & $\lambda_2/\lambda_1$ & $D_\mathrm{proj}$ & $\phi$ \\
	                          & (deg)    & (deg)    & \multicolumn{2}{c}{(mJy)}        & (W\,Hz$^{-1}$)                    & (arcmin)  & \multicolumn{2}{c}{(kpc)} & \multicolumn{2}{c}{($\Omega_\text{beam}$)} &    (\%)                 & (kpc)               & (deg)    \\
	(1)                       &  (2)   &  (3)   & \multicolumn{2}{c}{(4)}        &    (5)                  &  (6)    & \multicolumn{2}{c}{(7)} & \multicolumn{2}{c}{(8)}                  &   (9)                 &  (10)             &  (11)  \\
	\hline
	Abell 115                      & 14.00  & +26.46 &     68&.4$\pm$14.3           & $24.90^{+0.08}_{-0.10}$ & $10.12$ &  1982&$\pm$78        &      20&.7$\pm$1.4                       &    8                  & 1602              & 64.2   \\
	Abell 521                      & 73.59  & -10.28 &     13&.7$\pm$2.9            & $24.45^{+0.09}_{-0.11}$ & $ 3.69$ &   874&$\pm$95        &       5&.7$\pm$0.8                       &   12                  & 895               & 66.0   \\
	Abell 523                      & 74.79  & +8.81  &     29&.1$\pm$8.7            & $23.91^{+0.11}_{-0.15}$ & $ 7.63$ &   875&$\pm$46        &      16&.0$\pm$1.3                       &   39                  & 373               & 58.6   \\
	Abell 610                      & 119.87 & +27.12 &     15&.5$\pm$3.2            & $23.56^{+0.08}_{-0.10}$ & $ 2.94$ &   311&$\pm$42        &       6&.2$\pm$0.8                       &   39                  & 310               & 88.6   \\
	Abell 746                      & 137.23 & +51.58 &     12&.4$\pm$3.3            & $24.32^{+0.10}_{-0.14}$ & $ 4.39$ &   977&$\pm$89        &       6&.8$\pm$0.8                       &   11                  & 1488              & 67.0   \\
	Abell 781                      & 140.13 & +30.47 &     16&.2$\pm$2.7            & $24.70^{+0.07}_{-0.08}$ & $ 3.10$ &   830&$\pm$107       &       4&.8$\pm$0.7                       &   56                  & 628               & 61.0   \\
	Abell 1240 N                   & 170.86 & +43.17 &      6&.0$\pm$1.8            & $23.63^{+0.11}_{-0.15}$ & $ 3.22$ &   531&$\pm$66        &       3&.7$\pm$0.6                       &   35                  & 642               & 68.8   \\
	\phantom{Abell 1240} S         & 170.96 & +43.01 &      5&.4$\pm$1.5            & $23.59^{+0.11}_{-0.14}$ & $ 2.16$ &   356&$\pm$66        &       3&.1$\pm$0.6                       &   39                  & 1169              & 50.0   \\
	Abell 1300                     & 172.95 & -19.95 &     19&.8$\pm$3.5            & $24.81^{+0.07}_{-0.09}$ & $ 3.63$ &   987&$\pm$109       &       6&.4$\pm$0.8                       &   34                  & 1284              & 45.6   \\
	Abell 1351                     & 175.59 & +58.50 &     16&.8$\pm$2.8            & $24.79^{+0.07}_{-0.08}$ & $ 2.86$ &   805&$\pm$112       &       5&.0$\pm$0.7                       &   51                  & 713               & 68.8   \\
	Abell 1443                     & 180.27 & +23.10 &      8&.8$\pm$1.6            & $24.32^{+0.07}_{-0.09}$ & $ 2.10$ &   520&$\pm$99        &       2&.9$\pm$0.5                       &   45                  & 1268              & 78.0   \\
	Abell 1612                     & 191.96 & -2.88  &     61&.3$\pm$8.6            & $24.76^{+0.06}_{-0.07}$ & $ 5.19$ &   942&$\pm$73        &      13&.4$\pm$1.2                       &   19                  & 1071              & 77.1   \\
	Abell 1664                     & 195.88 & -24.37 &     96&.6$\pm$21.2           & $24.63^{+0.09}_{-0.11}$ & $ 7.90$ &  1084&$\pm$55        &      41&.9$\pm$2.0                       &   75                  & 1053              & 20.7   \\
	Abell 1682-NW                  & 196.68 & +46.57 &     20&.0$\pm$4.1            & $24.50^{+0.08}_{-0.10}$ & $ 2.25$ &   492&$\pm$87        &       3&.5$\pm$0.6                       &   62                  &   323             & 27.6   \\      
        Abell 1758N E                  & 203.23 & +50.54 &     14&.5$\pm$4.9            & $24.58^{+0.13}_{-0.18}$ & $ 3.31$ &   848&$\pm$102       &       5&.7$\pm$0.7                       &   23                  & 679               & 61.8   \\
        \phantom{Abell 1758N} W        & 203.16 & +50.56 &      7&.6$\pm$1.8            & $24.30^{+0.10}_{-0.12}$ & $ 3.21$ &   823&$\pm$102       &       3&.6$\pm$0.6                       &   20                  & 133               & 27.6   \\
        Abell 2034 W                   & 227.42 & +33.52 &     28&.0$\pm$3.6            & $23.98^{+0.05}_{-0.06}$ & $ 2.56$ &   315&$\pm$49        &       5&.0$\pm$0.7                       &   49                  & 807               & 54.1   \\
	Abell 2061                     & 230.00 & +30.53 &      5&.8$\pm$1.5            & $22.95^{+0.10}_{-0.13}$ & $ 2.88$ &   255&$\pm$35        &       3&.0$\pm$0.5                       &   19                  & 1753              & 82.7   \\
	Abell 2163                     & 244.06 & -6.09  &     12&.8$\pm$2.0            & $24.20^{+0.07}_{-0.08}$ & $ 2.25$ &   452&$\pm$80        &       3&.5$\pm$0.6                       &   43                  & 2008              & 26.3   \\
	Abell 2256                     & 255.85 & +78.73 &    183&.9$\pm$29.8           & $24.18^{+0.07}_{-0.08}$ & $12.07$ &   814&$\pm$27        &      52&.1$\pm$2.3                       &   19                  & 87                & 81.8   \\
	Abell 2345 E                   & 321.90 & -12.19 &     41&.6$\pm$8.1            & $24.58^{+0.08}_{-0.10}$ & $ 6.90$ &  1236&$\pm$72        &      15&.3$\pm$1.2                       &    9                  & 1173              & 70.9   \\
	\phantom{Abell 2345} W         & 321.68 & -12.13 &     32&.8$\pm$6.1            & $24.48^{+0.08}_{-0.09}$ & $ 5.39$ &   966&$\pm$72        &      11&.9$\pm$1.1                       &   19                  & 1227              & 69.6   \\
	                               & 321.63 & -12.12 &      4&.3$\pm$1.1            & $23.59^{+0.10}_{-0.13}$ & $ 2.33$ &   418&$\pm$72        &       2&.2$\pm$0.5                       &   14                  & 1759              & 62.4   \\
	Abell 2744                     & 3.67   & -30.35 &      9&.4$\pm$2.6            & $24.49^{+0.11}_{-0.14}$ & $ 5.60$ &  1526&$\pm$109       &       5&.3$\pm$0.7                       &    9                  & 1408              & 74.0   \\
	Abell 3365 E                   & 87.27  & -21.78 &     38&.9$\pm$8.3            & $23.94^{+0.08}_{-0.10}$ & $ 6.20$ &   644&$\pm$42        &      16&.2$\pm$1.3                       &   14                  & 1620              & 80.0   \\
	Abell 3411 a                   & 130.52 & -17.58 &     27&.5$\pm$5.5            & $24.35^{+0.08}_{-0.10}$ & $ 4.06$ &   703&$\pm$69        &      10&.6$\pm$1.0                       &   43                  & 1258              & 20.4   \\
	\phantom{Abell 3411} b         & 130.55 & -17.63 &     18&.3$\pm$3.0            & $24.18^{+0.07}_{-0.08}$ & $ 2.65$ &   459&$\pm$69        &       5&.1$\pm$0.7                       &   35                  & 1843              & 12.3   \\
	\phantom{Abell 3411} c         & 130.56 & -17.54 &      7&.8$\pm$1.5            & $23.80^{+0.08}_{-0.09}$ & $ 1.97$ &   341&$\pm$69        &       2&.8$\pm$0.5                       &   41                  & 1245              & 84.9   \\
        Abell 3527-bis                 & 192.77 & -37.03 &     22&.6$\pm$4.6            & $24.43^{+0.08}_{-0.10}$ & $ 5.04$ &   1001&$\pm$80       &       8&.8$\pm$0.9                       &    8                  & 1387              & 86.9   \\
	CIZA J0107+54 NE               & 16.96  & +54.15 &      4&.0$\pm$1.1            & $23.07^{+0.11}_{-0.14}$ & $ 2.40$ &   282&$\pm$47        &       2&.3$\pm$0.5                       &   71                  &   77              & 23.4   \\
	\phantom{CIZA J0107+54} SW     & 16.92  & +54.09 &      5&.2$\pm$1.7            & $23.19^{+0.12}_{-0.17}$ & $ 2.73$ &   321&$\pm$47        &       3&.7$\pm$0.6                       &   67                  &  381              & 68.8   \\
	CIZA J0649+18                & 102.16 & +18.02 &     16&.4$\pm$3.9            & $23.22^{+0.09}_{-0.12}$ & $ 5.72$ &   422&$\pm$30        &       7&.9$\pm$0.9                       &   14                  & 784               & 84.8   \\
	                          & 102.19 & +17.97 &      4&.9$\pm$1.0            & $22.69^{+0.08}_{-0.10}$ & $ 2.79$ &   206&$\pm$30        &       2&.0$\pm$0.4                       &   24                  & 702               & 60.7   \\
	CIZA J2242+53 N              & 340.74 & +53.15 &     91&.2$\pm$13.0           & $25.00^{+0.06}_{-0.07}$ & $10.79$ &  2072&$\pm$77        &      23&.8$\pm$1.5                       &   10                  & 1468              & 81.7   \\
	                          & 340.91 & +53.09 &     11&.9$\pm$2.2            & $24.11^{+0.07}_{-0.09}$ & $ 2.94$ &   564&$\pm$77        &       4&.5$\pm$0.7                       &   22                  & 1550              & 27.5   \\
	\phantom{CIZA J2242+53} S    & 340.64 & +52.94 &      4&.1$\pm$1.0            & $23.65^{+0.10}_{-0.13}$ & $ 2.36$ &   454&$\pm$77        &       2&.1$\pm$0.5                       &   16                  & 1102              & 86.2   \\
	MACS J0717+37                & 109.40 & +37.76 &    108&.5$\pm$12.0           & $26.17^{+0.06}_{-0.06}$ & $ 3.93$ &  1524&$\pm$155       &       8&.7$\pm$0.9                       &   40                  & 20                & 56.6   \\
	MACS J1149+22 NW             & 177.35 & +22.39 &      7&.4$\pm$1.4            & $24.99^{+0.08}_{-0.11}$ & $ 2.08$ &   797&$\pm$154       &       2&.6$\pm$0.5                       &   33                  & 1139              & 41.3   \\
	\phantom{MACS J1149+22} SE   & 177.44 & +22.36 &      5&.9$\pm$1.4            & $24.89^{+0.10}_{-0.13}$ & $ 2.65$ &  1019&$\pm$154       &       2&.8$\pm$0.5                       &   31                  & 1268              & 5.9    \\
	MACS J1752+44 NE             & 268.09 & +44.70 &     64&.2$\pm$7.6            & $25.50^{+0.05}_{-0.06}$ & $ 3.95$ &  1208&$\pm$122       &       9&.0$\pm$0.9                       &   21                  & 1093              & 89.6   \\
	\phantom{MACS J1752+44} SW   & 267.98 & +44.63 &     27&.0$\pm$3.5            & $25.13^{+0.06}_{-0.07}$ & $ 3.32$ &  1016&$\pm$122       &       5&.0$\pm$0.7                       &   45                  & 1058              & 70.2   \\
        PLCK G004-19                 & 289.26 & -33.51 &     36&.2$\pm$4.3            & $25.63^{+0.06}_{-0.07}$ & $ 2.52$ &   954&$\pm$151       &       5&.3$\pm$0.7                       &   62                  & 274               &  1.2   \\
	PLCK G287+32 N               & 177.71 & -28.05 &     38&.9$\pm$6.9            & $25.35^{+0.07}_{-0.09}$ & $ 6.16$ &  1961&$\pm$127       &      12&.6$\pm$1.1                       &   17                  & 517               & 49.6   \\
	\phantom{PLCK G287+32} S     & 177.81 & -28.19 &     20&.4$\pm$4.8            & $25.07^{+0.10}_{-0.12}$ & $ 5.24$ &  1669&$\pm$127       &       9&.8$\pm$1.0                       &   16                  & 2854              & 71.5   \\
	PSZ1 G096+24 N               & 284.15 & +66.44 &      7&.1$\pm$1.8            & $24.34^{+0.10}_{-0.13}$ & $ 2.73$ &   731&$\pm$107       &       3&.6$\pm$0.6                       &   14                  & 985               & 75.4   \\
	\phantom{PSZ1 G096+24} S     & 284.18 & +66.33 &     13&.0$\pm$3.0            & $24.60^{+0.09}_{-0.12}$ & $ 3.50$ &   938&$\pm$107       &       5&.9$\pm$0.8                       &   24                  & 850               & 75.3   \\
	PSZ1 G108-11 NE              & 350.72 & +48.82 &     60&.8$\pm$7.6            & $25.38^{+0.06}_{-0.07}$ & $ 5.13$ &  1471&$\pm$115       &      10&.2$\pm$1.0                       &   16                  & 1353              & 89.6   \\
	\phantom{PSZ1 G108-11} SW    & 350.56 & +48.66 &     43&.4$\pm$6.7            & $25.23^{+0.07}_{-0.08}$ & $ 3.80$ &  1091&$\pm$115       &      11&.4$\pm$1.1                       &   39                  & 2069              & 70.9   \\
	RXC J0225-29                 & 36.47  & -29.61 &     31&.2$\pm$4.6            & $23.45^{+0.06}_{-0.07}$ & $ 4.03$ &   284&$\pm$28        &       7&.5$\pm$0.9                       &   20                  & 890               & 49.1   \\
	RXC J1053+54                 & 163.21 & +54.95 &      7&.0$\pm$1.5            & $22.93^{+0.08}_{-0.10}$ & $ 2.54$ &   203&$\pm$32        &       2&.8$\pm$0.5                       &   23                  & 738               & 88.3   \\
	RXC J1314-25 E               & 198.70 & -25.26 &     13&.4$\pm$2.5            & $24.42^{+0.08}_{-0.09}$ & $ 4.36$ &  1016&$\pm$93        &       4&.8$\pm$0.7                       &   12                  & 975               & 69.3   \\
	\phantom{RXC J1314-25} W     & 198.58 & -25.26 &     26&.9$\pm$4.6            & $24.72^{+0.07}_{-0.08}$ & $ 4.50$ &  1049&$\pm$93        &       8&.2$\pm$0.9                       &   22                  & 483               & 69.1   \\
	ZwCl 0008+52 E               & 3.11   & +52.61 &     33&.8$\pm$7.9            & $23.97^{+0.09}_{-0.12}$ & $ 9.49$ &  1079&$\pm$45        &      16&.0$\pm$1.3                       &    4                  & 1200              & 87.0   \\
	\phantom{ZwCl 0008+52} W     & 2.79   & +52.51 &      9&.8$\pm$2.7            & $23.43^{+0.11}_{-0.14}$ & $ 3.57$ &   406&$\pm$45        &       5&.7$\pm$0.7                       &   23                  & 328               & 85.3   \\
	ZwCl 2341+00 a               & 355.95 & +0.23  &     14&.0$\pm$3.5            & $24.53^{+0.10}_{-0.13}$ & $ 3.57$ &   887&$\pm$99        &       7&.2$\pm$0.8                       &   44                  & 1293              & 79.8   \\
	\phantom{ZwCl 2341+00} c     & 355.91 & +0.35  &      9&.0$\pm$1.9            & $24.34^{+0.08}_{-0.11}$ & $ 2.72$ &   675&$\pm$99        &       3&.7$\pm$0.6                       &   22                  & 557               & 72.7   \\
	1RXS J0603+42 S                    & 90.90  & +42.17 &      6&.2$\pm$1.3            & $23.99^{+0.08}_{-0.10}$ & $ 1.66$ &   362&$\pm$87        &       2&.5$\pm$0.5                       &   74                  & 1005              & 6.0    \\
	\phantom{1RXS J0603+42} SW         & 90.94  & +42.22 &      5&.7$\pm$1.6            & $23.96^{+0.11}_{-0.14}$ & $ 3.01$ &   655&$\pm$87        &       3&.3$\pm$0.6                       &   13                  & 1247              & 89.4   \\
	\phantom{1RXS J0603+42} Toothbrush & 90.82  & +42.30 &    289&.3$\pm$30.7           & $25.66^{+0.04}_{-0.05}$ & $ 8.82$ &  1915&$\pm$87        &      22&.7$\pm$1.5                       &    6                  & 1132              & 58.8   \\
	\hline                                                                                                                                     
\end{tabular}    
\label{tab:islands}
\end{center}
\end{table*}

	%
	The relics listed in Table~\ref{tab:clusters} have been studied using different 
	telescopes, observing configurations and/or frequencies. Relics are extended 
	low-surface brightness objects. The deeper the observation is, the more extended 
	faint emission might be revealed. Therefore, properties such as flux density and 
	size reported in the literature crucially depend on instrumental parameters 
	like beam size and sensitivity of the observation.

	%
	Our aim is to measure relic properties for all known relics in a homogeneous way, 
	even if that implies recovering less flux than reported in the literature. 
	The NVSS is well suited for this purpose because it covers about 82\% of the sky 
	and the survey sensitivity is sufficient to recover most of the known relics. 
	The NVSS has been carried out at 1.4\,GHz, and images are made with a Gaussian restoring 
	beam of $\theta_\text{FWHM} = 45\,\text{arcsec}$. The root mean square (rms) of the 
	surface brightness fluctuations amounts, in average, to 
	$\sigma_{\rm NVSS} = 0.45\,{\rm mJy\,beam}^{-1}$. We use radio images obtained with 
	the NVSS Postage Stamp Server\footnote{cv.nrao.edu/nvss/postage.shtml} and measure 
	flux density, size and shape for the relics listed in Table~\ref{tab:clusters}.

	%
	The measurement of properties for the known relics in a homogeneous fashion is 
	optimally carried out by implementing a semi-automated analysis of the NVSS images. 
	We will later apply the same procedure to analyse our simulated relic sample as it 
	will be discussed in the next section. For every object, we first define a region 
	closely encircling the relics reported in the literature. This is done manually in order 
	to fully comprise the regions of interest, as well as to exclude compact sources 
	and false positives due to noise. Compact sources within the relic regions 
	have been subtracted from the images using the NVSS restoring beam and adopting 
	flux density values from the high-resolution radio survey FIRST \citep{Becker95} 
	or by estimating the flux density in better resolved observations 
	available in the literature. The gischt double relic in Abell\,2345 is an example 
	of an object where we have subtracted a bright compact source.

	%
	For the threshold encircling the radio emission we adopt 
	the $I_\text{th} = 2 \times \sigma_{\rm NVSS}$ contour as the surface brightness 
	limit to define the relic boundaries. We note that there is no danger of detecting a significant 
	amount of spurious emission since the manually defined regions closely encircle
	the true flux. We have chosen such a low threshold to recover as much emission as possible. 
	Interestingly, even after adopting this value, some relics may decompose into several pieces, 
	e.g. see the diffuse emission in Abell\,3411. In what follows, we will denote every 
	radio-emitting piece as an `island'. Fig.~\ref{fig:NVSSimage} shows three examples of relics in 
	the NVSS where it can be seen that recovered morphologies are quite diverse. 
	The most prominent is the `Toothbrush' (top panel) that
	not only excels through its extremely linear, yet non-symmetric appearance, 
	but also through its surface brightness. In fact, the `Toothbrush' has the 
	highest flux density in our sample. Fainter structures, also reported as relics, 
	are well visible in NVSS within this cluster too.

	%
	Obviously, not all relics in our compilation will be accesible to NVSS. 
	In this respect, several reasons can be given: (i) if the object is too far south, 
	resulting in a declination lower than $-40^{\circ}$, it will not be covered; 
	(ii) if the surface brightness is too low the relic may not be detected; 
	(iii) if the relic emission is confused with a bright source the flux and/or 
	the morphological properties will not be easily measured, 
	and (iv) if the relic is too extended it could not be fully recovered due 
	to missing short spacings in the interferometer. The NVSS was carried out with the 
	Very Large Array (VLA) in D and DnC configurations, hence, the largest possible angular scale accessible 
	to the survey is 970\,arcsec\footnote{science.nrao.edu/facilities/vla/docs/manuals/oss/performance/\\resolution}. 
	Therefore, for clusters at redshift $z \lesssim 0.05$, relics are too extended 
	for a reliable flux density measurement, i.e. they are ``resolved out''. 
	We note, however, that even at the largest angular scale configurations, 
	still less than 50\% of the object's flux is collected by the interferometer. 
	As a result, somewhat less extended sources may still not be fully recovered 
	\citep{2013ApJ...779..189F}. In fact, for the extended relics in Abell\,2256, 
	Abell\,115 and CIZA\,2242$+$53, we measure less flux in NVSS than reported in the literature 
	(see Fig.~\ref{fig:FluxesLAS}).

	\subsubsection{Flux density, size and area}
	
	%
	For each individual island, we measure the total solid angle, $\Omega$, from the sum 
	of its constituent pixels. We also determine the largest angular size (LAS), $\theta_\text{LAS}$,
	using the smallest enclosing circle. From the latter, we also derive 
	the LLS of the islands adopting our fiducial cosmology.  
	To determine the flux density, $S_{1.4}$, one has to sum the surface 
	brightness over all pixels across the island and then divide by 
	the beam solid angle. In radioastronomy, the latter is commonly defined as  
	\begin{equation}
		\Omega_\text{beam} 
		= 
		\frac{1}{ \ln 2 } \Omega_\text{FWHM}
		=
		\frac{1}{ \ln 2 } \, \frac{\pi}{4} \theta_\text{FWHM}^2\, ,
	\end{equation}

	\noindent where $\Omega_\text{FWHM}$ is the restoring beam solid angle and 
	$\theta_\text{FWHM}$ its associated angular scale. For simplicity, we adopt an angular 
	scale of $45\,\text{arcsec}$ for the restoring beam width of the survey even if there 
	are small variations across the large sky area covered by NVSS. Since the beam has a 
	Gaussian shape, the maximum flux density of an island can be related to its solid angle by
	\begin{equation}
		S_{1.4,\text{max}} (\Omega)
		=
		I_\text{th} \Omega_\text{beam}
		\left\{
		  \exp\left( \frac{\Omega}{\Omega_\text{beam}} \right)
		  -1
		\right\} .
		\label{eq:Smax}
	\end{equation}
	
	To avoid including very small islands that might be governed 
	by noise, we discard all objects with a flux density below 
	$S_{1.4,\text{min}} = 8 \times \sigma_\text{NVSS}$, i.e. $3.6\,$mJy. 
	This value has been chosen to obtain reliable flux measurements while keeping 
	the maximum number of islands possible.
	From the flux density of each island we also derive its radio luminosity, $P_{1.4}$, 
	assuming an average spectral index of $-1.2$ for the $k$-correction.

	\subsubsection{Shape, position and orientation}
	\label{sec:shapes}

	%
	We quantify the shape and orientation of islands via image moments, which are commonly 
	used to characterize images \citep{1980JBIS...33..323S,1996A&AS..117..393B}. 
	First, we determine the emission-weighted centre, $(x_c,y_c)$, of each island. 
	Then, the central moment, $\mu_{pq}$, of order $p+q$ can be computed as
	\begin{equation}	
		\mu_{pq} 
		= 
		\sum\limits_{(i,j) \in \text{island}} (x_i-x_c)^p (y_j-y_c)^q I_{ij} ,
	\end{equation}
	where $(x_i,y_j)$ and $I_{ij}$ are the coordinates and surface brightness of the 
	pixels, respectively. 
	We describe the shape $s$ of the relic through the ratio
	\begin{equation}
		s 
		\equiv 
		 \lambda_2/\lambda_1 {\rm ,}
	\end{equation}
	where $\lambda_1$ and $\lambda_2$ are the first and second eigenvalues 
	of the covariance matrix 
	\begin{equation}
		\frac{1}{\mu_{00}}
		\left( 
		\begin{array}{cc}
		\mu_{20}  & \mu_{11}  \\ 
		\mu_{11}  & \mu_{02}  
		\end{array}
		\right)
		.
	\end{equation}
	As an example, we note that idealized linear and circular objects would have shape 
	parameters of $s = 0,1$, respectively. It is evident that the shape of the manually 
	defined relic regions might have some effect on the relic shape. 
	Nevertheless, we checked that the impact of the latter is negligible.

	To characterize the location of islands within their host galaxy clusters, we measure 
	their projected distance, $D_\text{proj} \equiv |\bf{r_{cc}}|$, where $\bf{r_{cc}}$ denotes 
	the position vector joining the cluster and island centres. For the cluster centres, 
	we adopt positions from the SIMBAD catalogue \citep{2000A&AS..143....9W}. 
	We note, however, that these positions are not homogeneously determined, some being 
	derived from the X-ray surface brightness distribution and others via the cluster 
	SZ effect.

	Additionally, we determine the orientation of the 
	relic with respect to the cluster. The first eigenvector of the covariance 
	matrix, $\bf{v_1}$, is parallel to the major axis of the radio island. 
	Hence, the angle between $\bf{v_1}$  and the position vector $\bf{r_{cc}}$, i.e.
	\begin{equation}
		\phi 
		= 
		\angle( \bf{v_1}, \bf{r_{cc}}){\rm ,} 
	\end{equation}
	indicates how any given island is oriented within its host galaxy cluster, where 
	radial and tangential elongations are given by the angles 
	$\phi=0^{\circ}$ and $90^{\circ}$, respectively (see purple solid lines in Fig.~\ref{fig:NVSSimage}).

	\subsubsection{Error estimates}
	\label{sec:error}

	%
	We estimate the uncertainties of island flux densities, luminosities, 
	sizes, projected distances and solid angles. In the case of the flux, we take 
	into account the three dominant error sources contributing to the island 
	flux measurement, namely: 
	(i) the image noise; (ii) the uncertainties caused by point source subtraction, 
	and (iii) the errors in the absolute flux calibration. We simply assume that the flux 
	of the subtracted point sources, $S_\text{ps}$, is uncertain by 5\% 
	and that the absolute flux calibration is uncertain by 10\%.  
	Since the errors are uncorrelated we estimate the flux density uncertainty as

	\begin{equation}
		\Delta S_{1.4}
		 = 
		 \sqrt{  (0.1 S_{1.4})^2  
		 	      +   (0.05 S_\text{ps})^2  
		 	      + \sigma^2_\mathrm{NVSS} \left(\frac{\Omega}{\Omega_\text{beam}}\right) }
		 	   .
	\nonumber
	\end{equation}

	The uncertainty of the luminosity is obtained by error propagation of the flux 
	density. We assume that measurement of the LAS has an error of about 
	half the beam width, i.e. we take $\Delta\theta_{\text{LAS}} = 0.4\,$arcmin. From this 
	uncertainty we compute the corresponding quantity for the LLS. For the projected 		
	spatial distance, we simply assume a fiducial error of 100\,kpc for all islands. 
	This value is large enough to encompass typical errors resulting from 
	the measurement of the images within the redshift range of the sample.  
	
		\begin{figure}
		\hspace{-0.8cm}
	                \includegraphics[width=0.51\textwidth]{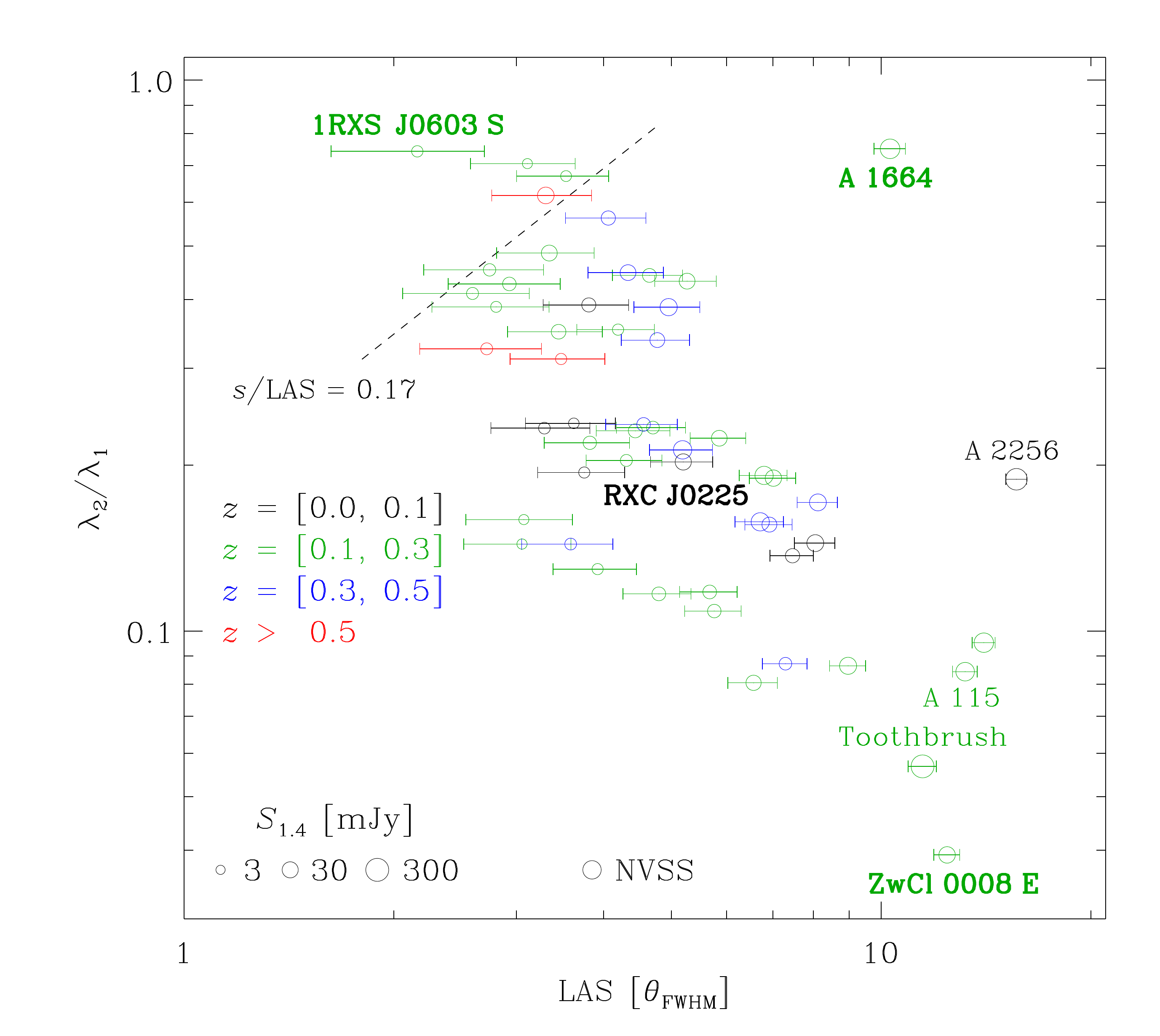}
			\caption{   
				Shape parameter, $s\equiv\lambda_2/\lambda_1$, versus the 
				largest angular size, LAS, for NVSS relics. Symbol sizes are 
				scaled according to radio flux. Redshift bins are indicated using 
				different colours. The dashed line indicates the empirical relation 
				$s/\text{LAS}=0.17$.
			}
			\label{fig:Shape-vs-LAS-NVSS}
	\end{figure}

	\subsection{Characterizing the relic samples}
	\label{sec:CharRelics}
	
	%
	The measured parameters of all NVSS islands fulfilling our selection criteria are 
	listed in Table~\ref{tab:islands}. To better characterize the relic shape, 
	in Fig.~\ref{fig:Shape-vs-LAS-NVSS} we plot the shape parameter versus the LAS 
	for all NVSS islands (open circles).
	
\begin{figure}
		\hspace{-0.85cm}
                        \includegraphics[width=0.57\textwidth]{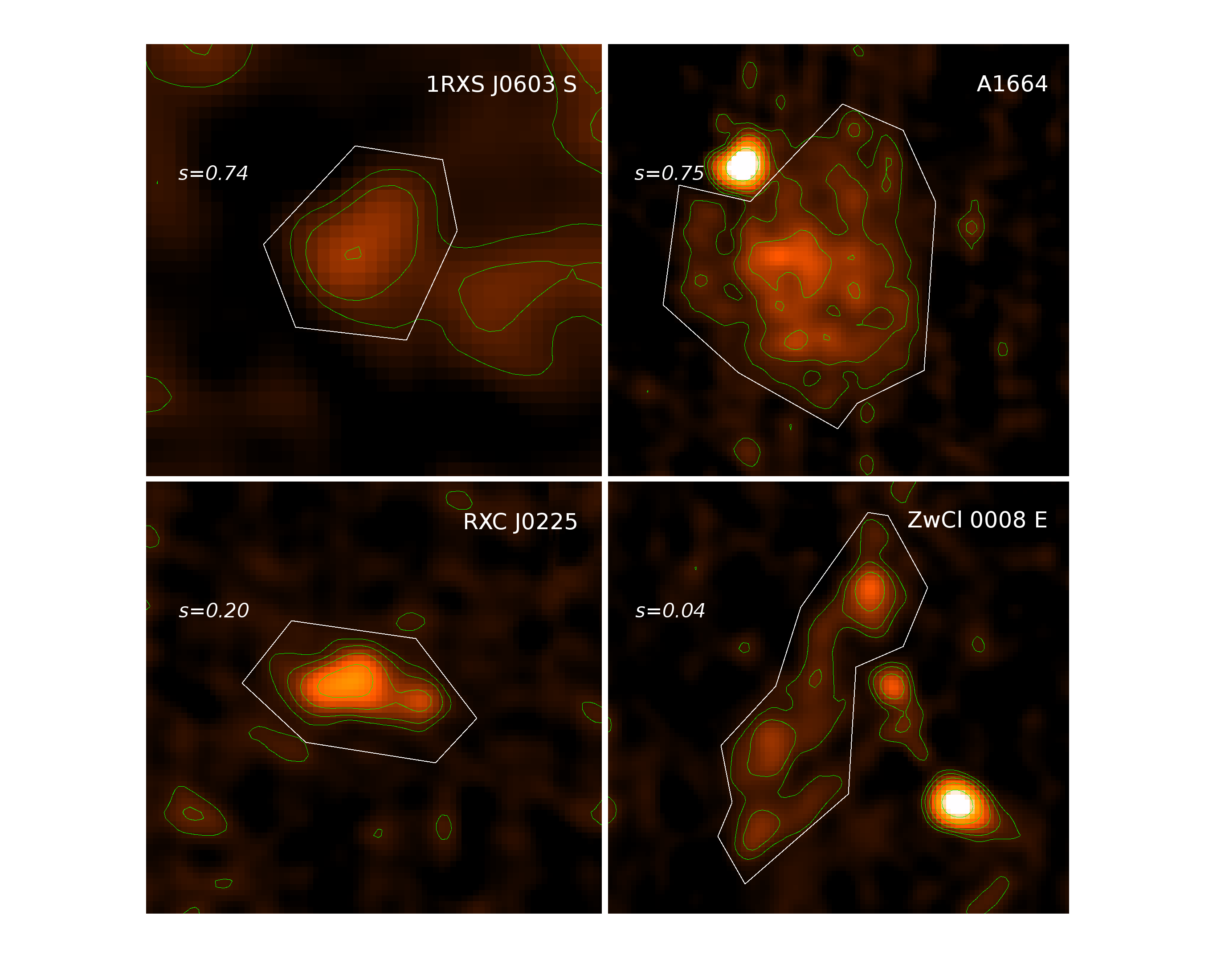}
			\caption{   
				Relic examples drawn from the $s-$LAS correlation (shown in boldface in Fig.~\ref{fig:Shape-vs-LAS-NVSS}): 
				{\bf 1RXS J0603 S:} most roundish and compact object; {\bf A1664:} largest outlier; 
				{\bf RXC J0225:} object with median $s/{\rm LAS}$, and {\bf ZwCl 0008 E:} most elongated object. 
				Relic regions are indicated by white solid lines. 
			}
			\label{fig:Shapes-NVSS}
	\end{figure}

	As seen in the plot, the relic shape is anti-correlated with the LAS, 
	meaning that, the larger (smaller) the relic is, 
	the more elongated (roundish) results. Interestingly, there are a few outliers to this relation
	that are clearly not fragments of a larger relic. One quite roundish for its LAS is 
	the gischt candidate in Abell\,1664, whose classification is somewhat uncertain. 
	This object shows an extended emission coming from the cluster periphery and no evidence of a 
	spectral break suggesting a gischt interpretation. However, the cluster's cool core might be 
	considered as an argument against an ongoing merger. Additionally, proximity to a cluster member 
	galaxy suggests that the relic could be a lurking radio lobe; although 
	deeper observations are needed to confirm or discard the association \citep{2012ApJ...744...46K}. 
	Similarly, we also found a somewhat roundish island in Abell\,1351.  
	In this cluster the ridge is difficult to disentangle from the halo emission \citep{2009ApJ...704L..54G}. 
	Another example is the gischt double relic in MACS J1149$+$22, which appears roundish in the NVSS images 
	because the cluster is rather distant and the brightness of the relics is low. 
	The most roundish object in the sample is the southern emission feature in the `Toothbrush' 
	cluster due to its unusual compactness. For compact emission, the minimum LAS 
	introduced by the surface brightness limit given by Eq.~\ref{eq:Smax} 
	is $\theta_\text{LAS}=1.52\times\theta_\text{FWHM}$, which lies below the corresponding 
	angular scale of NVSS islands. A visual impression of the islands can be seen in Fig.~\ref{fig:Shapes-NVSS} where four 
	extreme cases drawn from the shape-LAS correlation are presented. 
	
	In general, we find that there are only a few islands with 
	$s/\theta_\text{LAS} > 0.17$ (see dashed line in Fig.~\ref{fig:Shape-vs-LAS-NVSS}). 
	We consider this ratio as an empirical detection limit in NVSS, possibly caused by the 
	fact that point-like sources are usually not classified as radio relics. 
	For instance, relics in distant clusters may have escaped attention in systematic 
	searches based on NVSS, as they might appear as faint point sources in the 
	radio maps. Therefore, in what follows, we use the ratio $s/\theta_\text{LAS}=0.17$ 
	to distinguish between `small-roundish' and `elongated' objects. We stress that this 
	distinction will be particularly important when discussing the simulated mock relic 
	sample (see Section~\ref{sec:Compare}).
	
		\begin{figure}
		\hspace{-0.8cm}
		\includegraphics[width=0.52\textwidth]{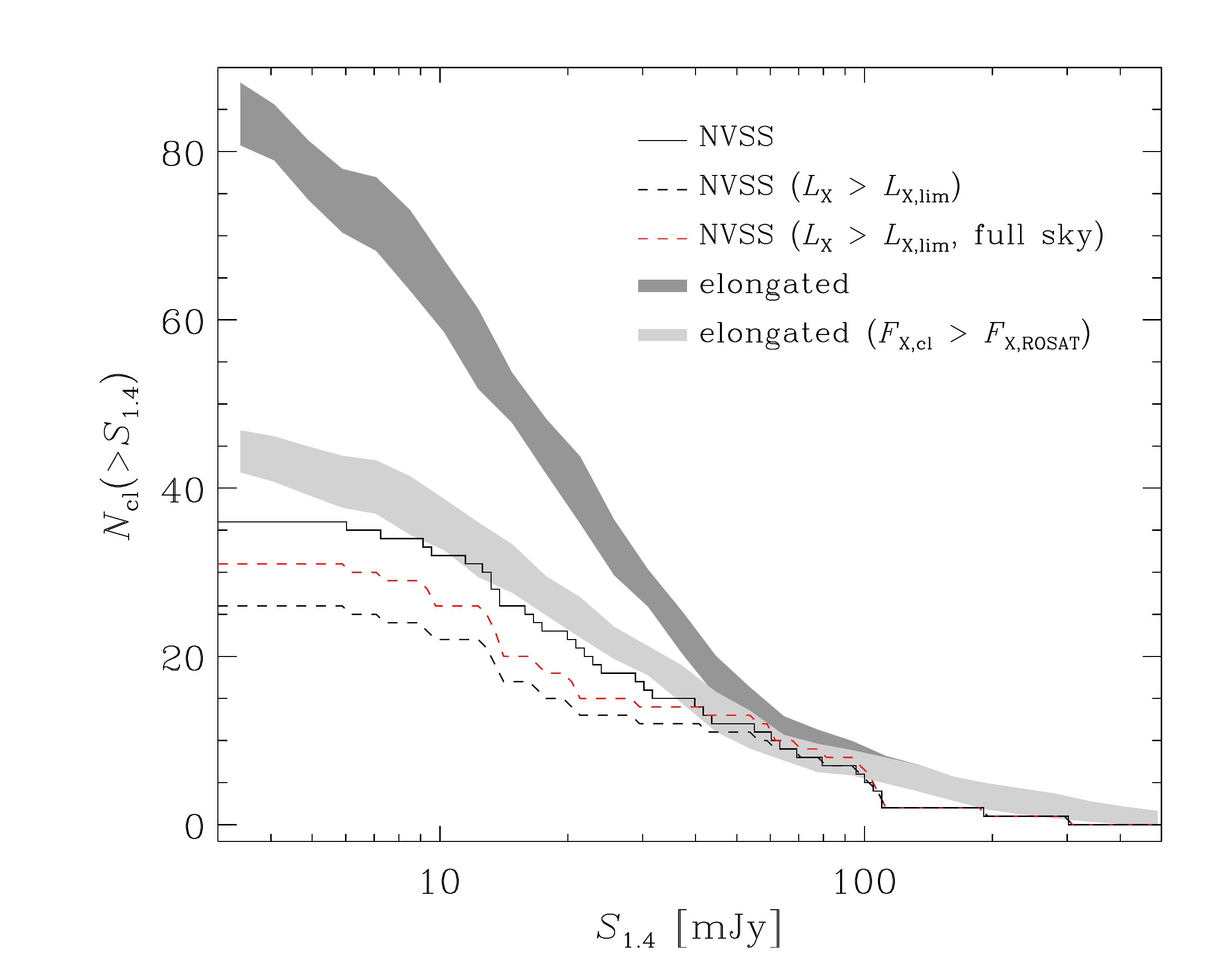}
		\caption{
			Cumulative number of clusters versus total radio flux density. 
			The cluster flux density is computed adding the contribution 
			of all individual islands obtained from the image analysis described 
			in Section~\ref{sec:ImageAnalysis}. 
			Shown are the NVSS (solid line), idem but only for 
			clusters above the X-ray luminosity completeness limit of the 
			MUSIC-2 simulation (dashed line; see Section~\ref{sec:MUSIC_clusters}), 
			and the latter including a correction for sky incompleteness (red line). 
			Shaded regions indicate the mean and standard deviation for a set 
			of MUSIC-2 trials (see Section~\ref{sec:zsample}). 
		}
		\label{fig:FluxClusterCumulative}
	\end{figure}

	\subsection{NVSS completeness}
	\label{sec:NVSScomp}
	%
	%
	Fig.~\ref{fig:FluxClusterCumulative} shows the cumulative number of clusters hosting 
	relics versus their total radio flux for the NVSS sample (solid line). 
	For future reference, the cumulative counts of the `elongated' simulated cluster sample are 
	also shown (shaded regions). These curves represent full-sky extrapolations computed 
	from a set of MUSIC-2 mock cluster samples. A detailed discussion concerning the generation of 
	mock catalogues and relic abundance is postponed until Sections~\ref{sec:zsample} and~\ref{sec:relic_number}, 
	respectively. We note that the total radio flux of all galaxy clusters, 
	both in observations and mocks, is computed adding all diffuse emission coming from their associated islands.

	As seen in Fig.~\ref{fig:FluxClusterCumulative}, the NVSS curve rapidly flattens below a flux value of
	about 10\,mJy. Several effects may contribute to this behaviour. 
	The most obvious is related to detection limit: to be observed, an object's surface 
	brightness has to be larger than the threshold, $I_\text{th}$. For instance, a relic covered 
	by 10 beams has to have, at least, a flux density larger than $9\,$mJy to be listed in 
	Table~\ref{tab:islands}. It is also possible that faint relics could reside in low-mass clusters 
	that have not been identified yet. In \citet{2012MNRAS.420.2006N} we estimated that about 
	50\% of relics with a flux density below 10\,mJy may reside in clusters below 
	the detection limit of the REFLEX cluster sample. This supports the idea that a significant 
	fraction of them still remains unobserved.

	\begin{figure}
		\hspace{-0.8cm}
			\includegraphics[width=0.52\textwidth]{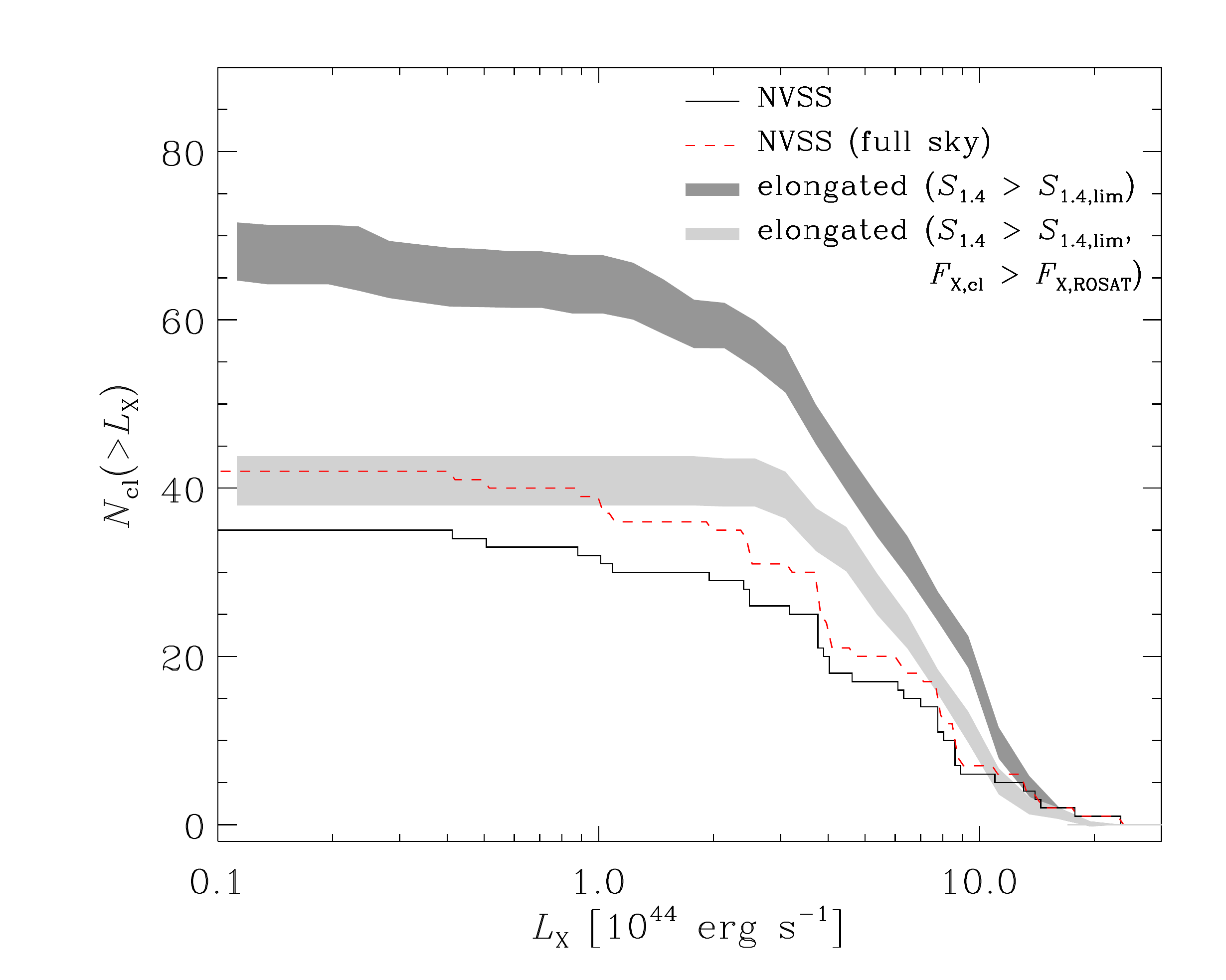}
			\caption{
				Cumulative number of clusters hosting radio relics versus cluster X-ray luminosity. 
				The original (solid line) and expected (red solid line) NVSS samples are also shown. 
				The latter has been obtained after correcting the original NVSS counts by sky incompleteness 
				as explained in the text. 
				Shaded regions indicate the mean and standard deviation of a set of MUSIC-2 trials 
				(see Section~\ref{sec:zsample}) for a total radio flux above the effective lower 
				limit of the NVSS.
			 } 
		\label{fig:LxClusterCumulative}
	\end{figure}

	In addition to the latter, some relics may escape discovery for other reasons, 
	including: confusion due to bright objects; the presence of a radio halo close to the relic position; 
	low surface brightness as a result of the very extended emission of close-by objects; 
	roundish morphology when located in distant clusters, or simply misclassification, e.g. 
	the relic is too small and is located close to the cluster centre. 
	Despite all these biases, it is evident that the NVSS completeness will mainly depend on relic flux density. 
	It is reasonable to think that objects having flux of about 100\,mJy are almost all detected. 
	However, one has to be cautious, yet a few relics of this class do not have been 
	discovered until recently (e.g., PSZ1\,G108$-$11).

\section{Radio relics in a cosmological simulation}
\label{sec:MockRelics}

\subsection{Simulated galaxy clusters} 
\label{sec:MUSIC_clusters}

	%
	In this work, we use the non-radiative MUSIC-2 galaxy cluster sample\footnote{music.ft.uam.es} of \citet{2013MNRAS.429..323S} to 
	generate a set of synthetic radio relic observations that can be compared with the NVSS images. The galaxy cluster sample consists of 
	282 regions hosting a massive galaxy cluster extracted from the {\sc multidark} $N$-body cosmological simulation \citep{Prada12}. 
	This simulation is performed within the context of the concordance cosmology and comprises a comoving volume of $1\,\hGpcc$. 
	Specifically, the adopted cosmological parameters correspond to a flat universe with a matter density of $\Omega_{\rm M}=0.27$, 
	a baryon density of $\Omega_{\rm b}=0.0469$, an amplitude of mass fluctuations of $\sigma_8=0.82$, a scalar spectral index of $n=0.95$ and 
	$h=0.7$, i.e. the Hubble constant amounts to $70\:\rm km\,s^{-1}\,Mpc^{-1}$.

	%
	The zooming technique \citep{2001ApJ...554..903K} was used to produce the initial conditions (ICs) necessary for the 
	resimulation of Lagrangian regions within a distance of $6\hMpc$ around the cluster mass centre at $z=0$. 
	The ICs were evolved until the present time using the parallel TreePM+SPH {\sc Gadget} hydrodynamical 
	code \citep{Springel05}. Additionally, haloes and subhaloes have been identified and characterized using the 
	hybrid MPI+OpenMP parallel halo finder {\sc ahf} \citep{Knollmann09}. The mass resolution of these high-resolution 
	regions corresponds to a dark-matter and gas particle mass of $m_{\rm DM}=9\times10^8\hMsun$ and $m_{\rm g}=1.9\times10^8\hMsun$, 
	respectively. This implies that most of our resimulated massive clusters are typically described with more than a million 
	gas particles. The gravitational softening length in the high-resolution zones was set to $6\hkpc$ both for the gas and 
	dark matter particles.

	%
	Several low mass clusters have been found close to the most massive systems in our sample. Therefore, the total 
	number of resimulated clusters considerable exceeds the number of selected cluster volumes. In total, the MUSIC-2 sample 
	contains 535 clusters with $M_{\rm vir}>10^{14}\hMsun$ and more than 2000 group-like objects with virial masses in the 
	range $10^{13}-10^{14}\hMsun$ at $z=0$. Although most of these systems are located outside the virial radius ($R_{\rm vir}$) 
	of our target galaxy clusters any possible interaction between them is consistently described by our simulations.

	%
	The cumulative mass function of the cluster sample between redshifts $0 \leq z \leq 1$ can be seen in 
	Fig. 2 of \citet{2013MNRAS.429..323S}. The mass completeness limit of the MUSIC-2 data set goes from 
	$4.5\times10^{14}\hMsun$ at $z=1$ to $8.5\times10^{14}\hMsun$ at $z=0$ (see their Table 1). Since the progenitors of the 
	clusters are also included in the analysis, the final sample is effectively complete at somewhat lower masses. Here, we 
	adopt a low-mass threshold of $5.5\times10^{14}\hMsun$.

	%
	In what follows, we consider the whole set of 282 resimulated galaxy cluster volumes at eight different 
	times corresponding to redshifts of $z=0,0.11,0.25,0.33,0.43,0.67,1,1.50$, thus producing a catalogue of 2256 outputs.

\subsection{Cluster X-ray luminosity} 
\label{sec:Xray_clusters}

	%
	One of the most evident observational features of galaxy clusters is their extended X-ray emission.  
	To compare the simulated clusters with those hosting an NVSS relic it is necessary to assign X-ray luminosities 
	to the simulations. We avoid computing it directly from the simulated galaxy clusters 
	since the non-radiative nature of MUSIC-2 would bias the total luminosity towards artificial higher values. 
	Instead, we use an empirical scaling relation linking the cluster mass to the total X-ray output. 
	We need to compute the rest-frame emission in the $0.1-2.4\,$keV energy band, as in the case of the 
	observed galaxy clusters listed in Table~\ref{tab:clusters}. 
	To this end, we adopt the scaling relation given by \citet{2014A&A...570A..31B}
	\begin{equation}
		\frac{ L_{500,0.1-2.4} }
		{ 10^{44} \: h_{70}^{-2} \: \rm erg \, s^{-1} }
		=
		\frac{ 0.1175 \: 	E(z)^\alpha }{ 	h_{100}^{2-\alpha} }
		\left( 
			\frac{M_{\rm 200}}{10^{14}\,h_{70}^{-1}\,\rm{M_\odot}} 
		\right)^\alpha
		,
		\label{eq:LMscaling}
	\end{equation}
	where $\alpha=1.51$, $E(z) = \sqrt{\Omega_{\rm M}(1+z)^3 + \Omega_\Lambda}$ 
	and $M_{200}$ is the cumulative cluster mass within $R_{200}$, i.e. the radius 
	for which the overdensity is 200 times the critical density of the universe. 
	For simplicity, throughout this paper, we will use $L_{\rm X}$ instead of $L_{500,0.1-2.4}$ to indicate cluster 
	X-ray luminosities within $R_{500}$ in the $0.1-2.4\,$keV band. 

	%
	We note that MUSIC-2 clusters are characterized by their virial masses, although masses at 
	different radii are also available. Therefore, to evaluate this scaling relation, we use the $M_{200}$ value 
	corresponding to each cluster. If $M_{200}$ is not available, which only happens a few times per snapshot, 
	we estimate $M_{200}$ from the virial mass. For the cosmological parameters adopted in our simulations, 
	the cumulative density contrast with respect to the critical density of the universe at the radius of 
	virialization is approximately $\Delta=100$ at $z=0$. To convert between $M_{\Delta}$ and $M_{200}$ we assume 
	an NFW profile adopting the halo concentrations of \citet{2008MNRAS.390L..64D}.

	%
	Using this scaling for the effective mass completeness given in the previous section, we estimate 
	that the simulated cluster sample is complete for X-ray luminosities above 
	$2.6 \times 10^{44} \: \rm erg \, s^{-1}$, adopting an average redshift 
	of $z=0.3$. This result evidently agrees with the comparison of the number of 
	clusters in the NVSS and MUSIC-2 samples. This is shown in Fig.~\ref{fig:LxClusterCumulative}, 
	where we plot the cumulative number of clusters hosting radio relics as a function of 
	X-ray luminosity. The curves corresponding to the NVSS and MUSIC-2 samples start to flatten 
	at an X-ray luminosity of about $3 \times 10^{44} \: \rm erg \, s^{-1}$ in agreement with the X-ray limit 
	estimated above. More details concerning this figure will be presented in Section~\ref{sec:relic_number} 
	where we will discuss on the relic abundance.

\subsection{Shock finder algorithm}
\label{sec:shock_finder}

	%
	We identify shock fronts in the simulated galaxy clusters at all available 
	redshifts following \citet{2008MNRAS.391.1511H} and \citet{2012MNRAS.420.2006N}. In this section, 
	we present a brief description of the shock detection scheme. 
 
	For every gas particle we evaluate the pressure gradient and define its 
	{\it shock normal} as ${\bf n}\equiv-\nabla{P}/|\nabla{P}|$.
	We then search for true shocks by imposing the following conditions: (i) $\nabla\cdot{\bf v} < 0$, 
	(ii) $\rho_{\rm u}<\rho_{\rm d}$ and (iii) $S_{\rm u}<S_{\rm d}$, where $\bf{v}$ is the velocity field, 
	$\rho$ is the gas density and $S$ is its entropy for the upstream and downstream regions respectively.
	To determine the Mach number we use the Rankine-Hugoniot 
	equations for hydrodynamical shocks \citep[see e.g.,][]{1959flme.book.....L} for each one 
	of the above conditions and then take the minimum resulting value as a conservative estimate. 
	This is done in order to avoid an overestimation of the Mach number that could lead to strong 
	spurious radio emission.

\subsection{Magnetic field model}
\label{sec:Bmodel}

	Very little is known about the strength and structure of magnetic fields in radio
	relics. One of the most stringent lower limits for the field strength has been derived
	by \citet{2009PASJ...61..339N} for the relic in A3667. Based on upper limits of the IC 
	emission from the relic region they concluded that the field strength must 
	exceed $1.6\:\rm \mu G$. In radio relics, magnetic fields could be dominated by those 
	generally present in the ICM and then compressed by the shock front. In addition,
	the field might be amplified, for instance, by upstream instabilities \citep{2014ApJ...794..153G}
	or downstream turbulence \citep{2016MNRAS.462.2014D}.

	%
	It is beyond the scope of this work to model relic magnetic fields in detail. 
	Therefore, we adopt a simple parametrization linking the  magnetic field strength $B$ to the local 
	electron density $n_{\rm e}$. Following \cite{2012MNRAS.420.2006N}, we assume a scaling of the form
	\begin{equation} 
		B
		=
		B_0 
		\left(\frac{n_{\rm e}}{10^{-4}\,{\rm cm}^{-3}}\right)^{\eta}{\rm ,}
		\label{eq:B_scaling}
	\end{equation}
	in agreement with previous works (see e.g., \citealt{2001A&A...378..777D,2010MNRAS.408..684S}). 
	The parameters adopted correspond to the best-fitting model of \cite{Bonafede10} for the Coma cluster, 
	i.e. $B_0=0.8\,\mu$G and $\eta=1/2$. These values lead to magnetic field 
	strengths of the order of $1 \: \rm \mu G$ in cluster outskirts. It is worth noting
	that the field strength adopted here is generally below the lower limits derived by 
	\citet{2009PASJ...61..339N} for A3667 and \citet{2010Sci...330..347V} for the `Sausage' relic, 
	namely 1.6 and $5\:\rm \mu G$ respectively. However, the field strength parameter 
	$B_0$ and the electron acceleration efficiency in our model (see below) are basically 
	degenerated in radio luminosity. This implies that, if magnetic field strengths in relics 
	are higher than adopted here, the true acceleration efficiency will be lower.

\subsection{Lighting up the shocks}
\label{sec:radio_shocks}

	%
	As for the magnetic field, our knowledge about the origin and acceleration mechanism 
	of the relativistic electrons that give rise to the observable synchrotron emission is incomplete. 
	It is generally assumed that diffusive shock acceleration (DSA) at the merger shock fronts leads 
	to the relativistic energies of the `observable' electrons, although many details of the acceleration 
	mechanism are under discussion. For instance, pre-existing populations of cosmic ray electrons 
	may significantly enhance the acceleration efficiency \citep{2012ApJ...756...97K}. Alternatively, shock drift 
	acceleration may result in an efficient mechanism to accelerate electrons at low Mach number shocks \citep{2014ApJ...794..153G}.

	%
	Since up to now the details of the acceleration mechanism are unknown, we follow the 
	general framework of \citet{2007MNRAS.375...77H}. This model is based on two main assumptions: (i) the slope 
	of the energy distribution of the accelerated electrons follows the predictions of test particle DSA and 
	(ii) only a fixed fraction, $\xi_{\rm e}$, of the energy dissipated at the shock front is used to accelerate 
	the electrons. The resulting spectrum of relativistic electrons may be considered as the average 
	outcome of an acceleration process that is, in detail, much more complicated than test particle DSA. 
	\citet{2007MNRAS.375...77H} compute the total radio emission coming from a shock assuming that electrons accelerated 
	at the front are advected with the downstream plasma. In this region, electrons loose energy via synchrotron 
	and IC emission, the latter being caused by collisions with cosmic microwave background (CMB) 
	photons.

	\begin{figure*}
		\begin{center} 
		  \includegraphics[width=0.23\textwidth]{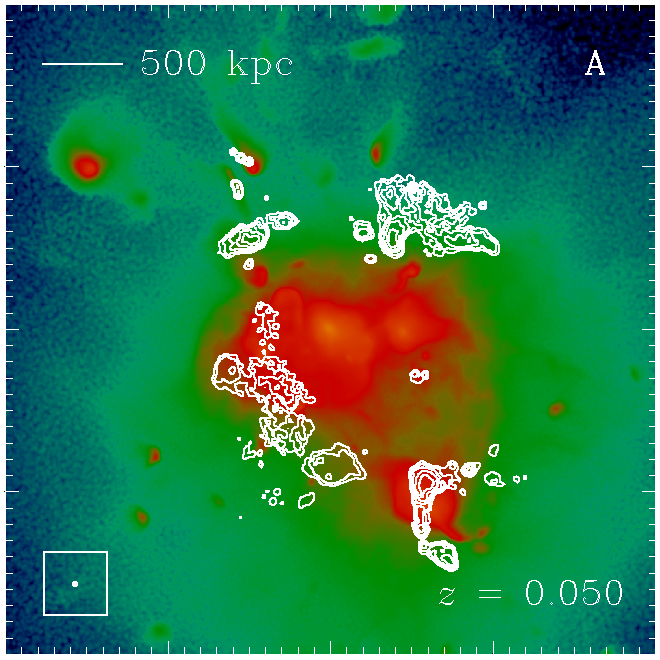}
		  \hspace{-0.15cm}
		  \includegraphics[width=0.23\textwidth]{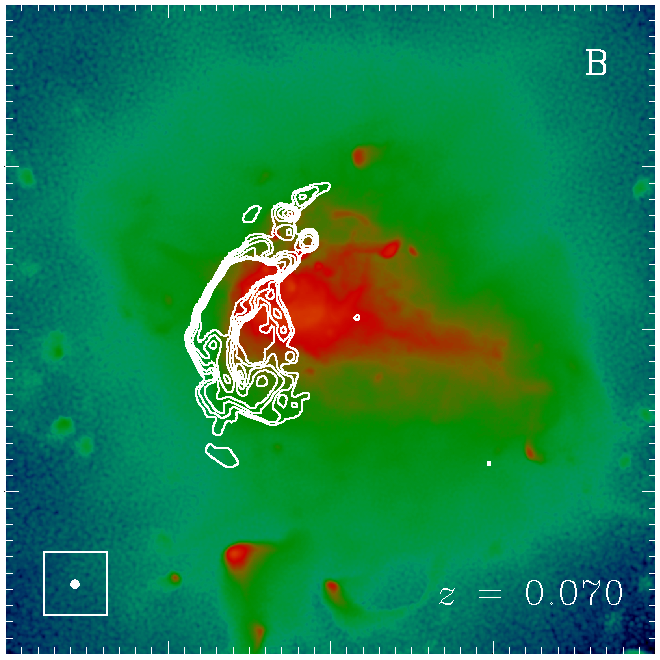}
		  \hspace{-0.15cm}
		  \includegraphics[width=0.23\textwidth]{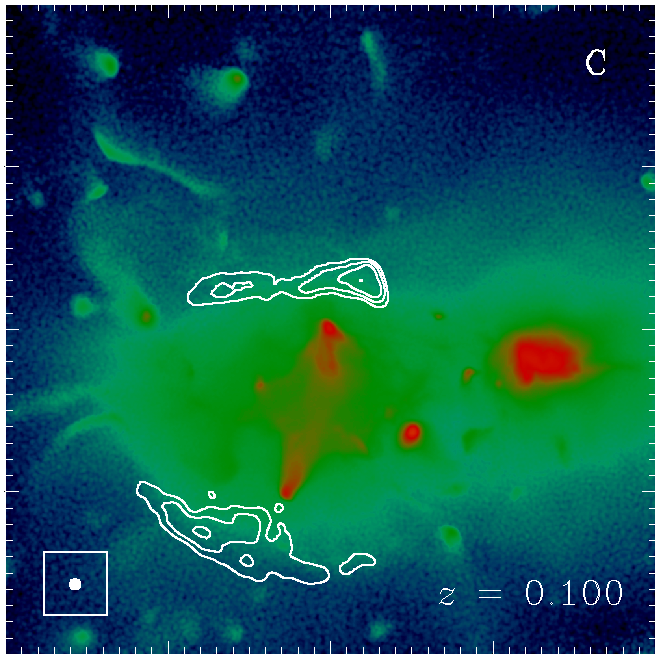}
		  \hspace{-0.15cm}
		  \includegraphics[width=0.23\textwidth]{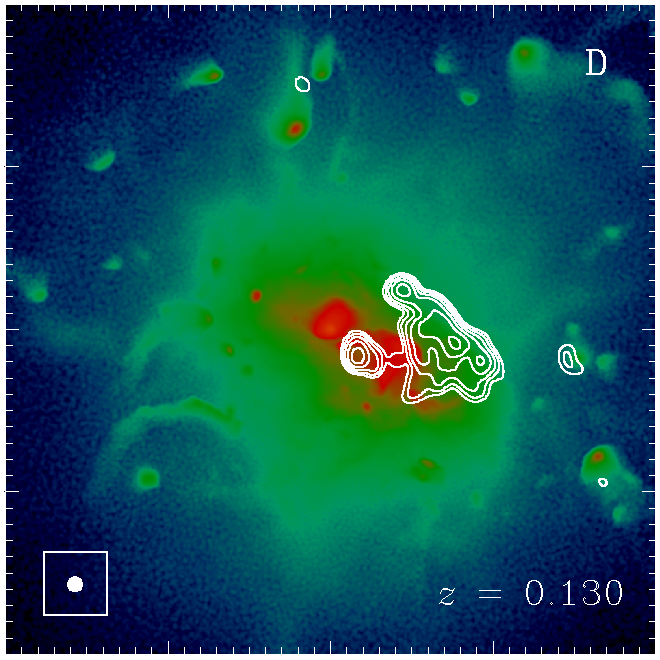}
		  
		  \includegraphics[width=0.23\textwidth]{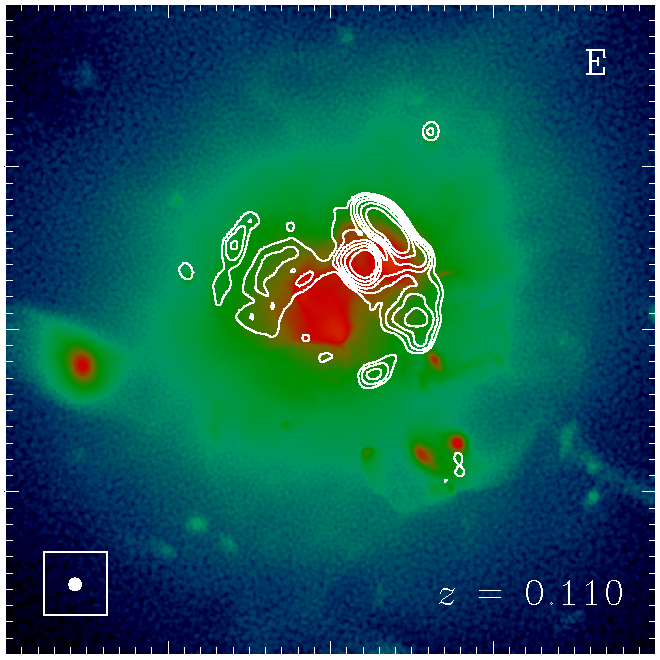}
		  \hspace{-0.15cm}
		  \includegraphics[width=0.23\textwidth]{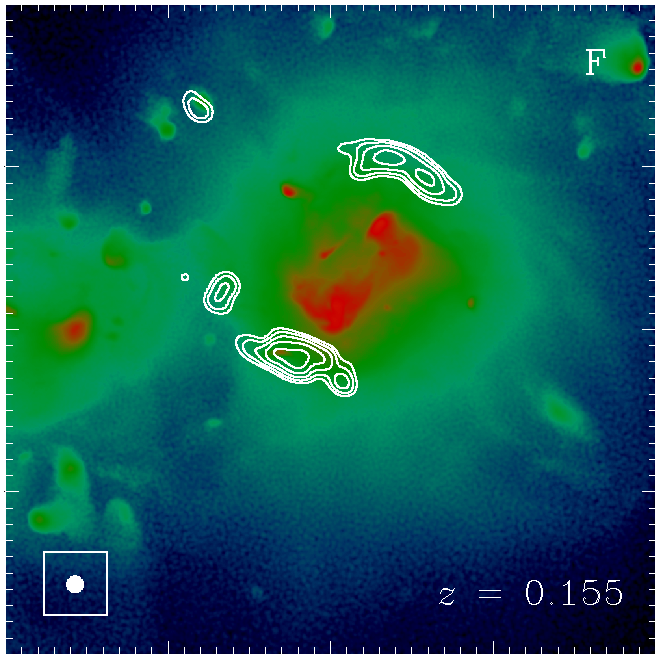}
		  \hspace{-0.15cm}
		  \includegraphics[width=0.23\textwidth]{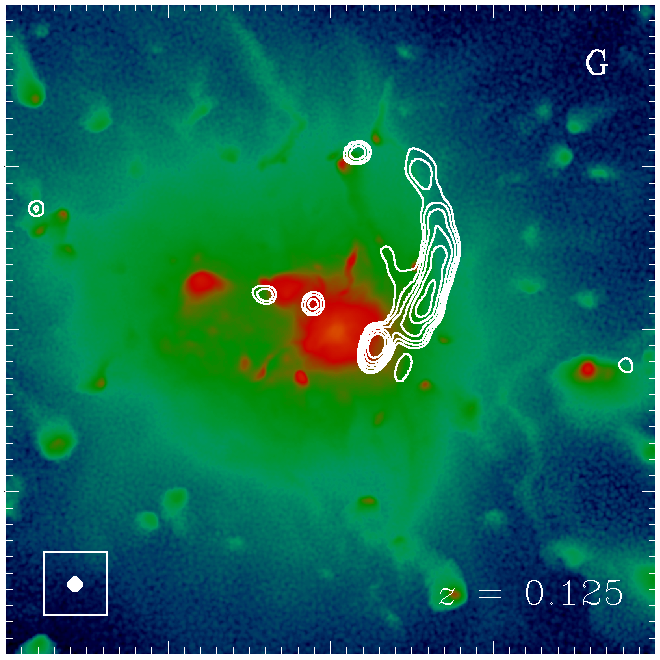}
		  \hspace{-0.15cm}
		  \includegraphics[width=0.23\textwidth]{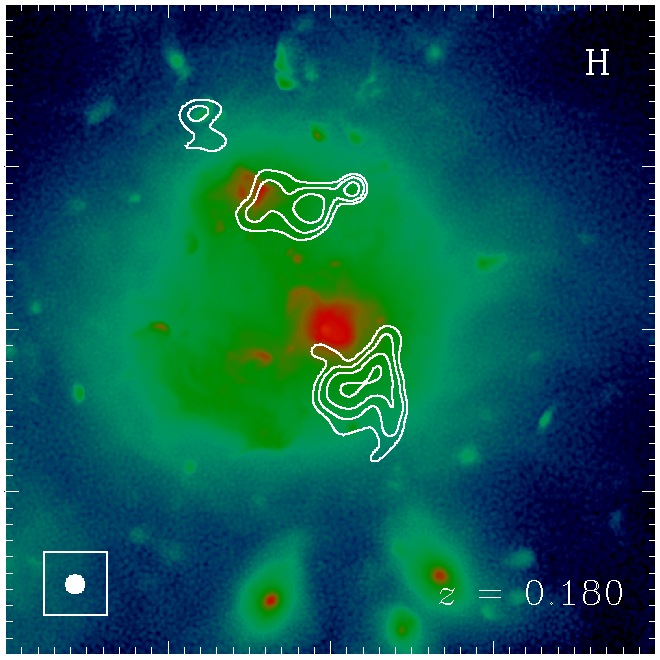}
		  
		  \includegraphics[width=0.23\textwidth]{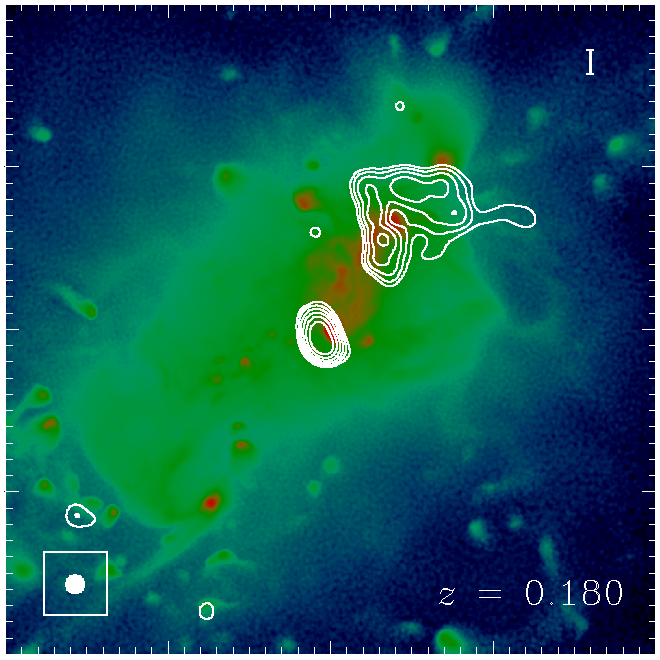}
		  \hspace{-0.15cm}
		  \includegraphics[width=0.23\textwidth]{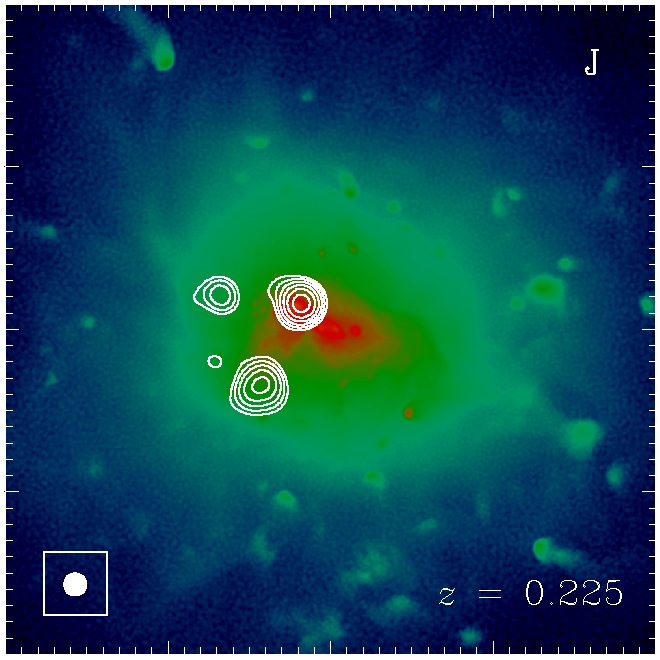}
		  \hspace{-0.15cm}
		  \includegraphics[width=0.23\textwidth]{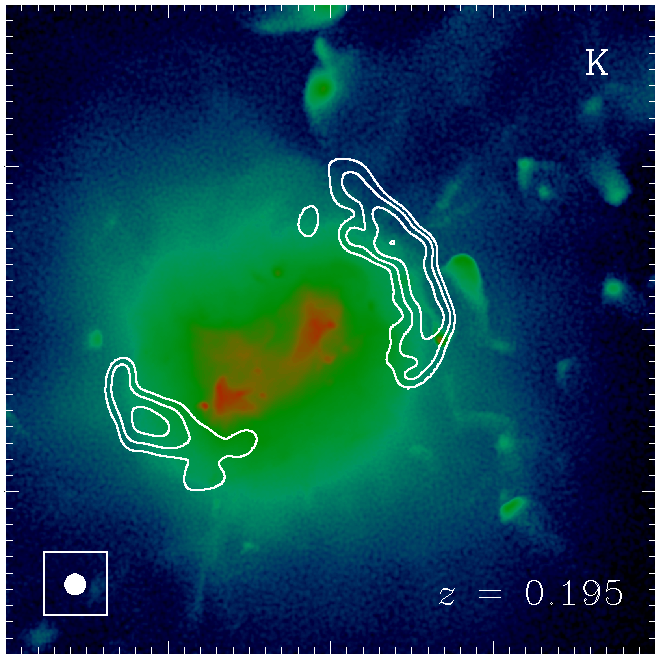}
		  \hspace{-0.15cm}
		  \includegraphics[width=0.23\textwidth]{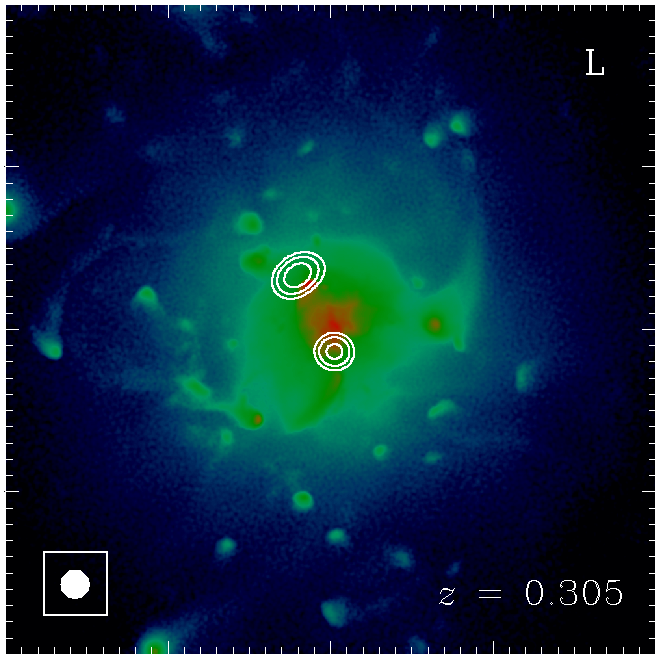}
		  
		  \includegraphics[width=0.23\textwidth]{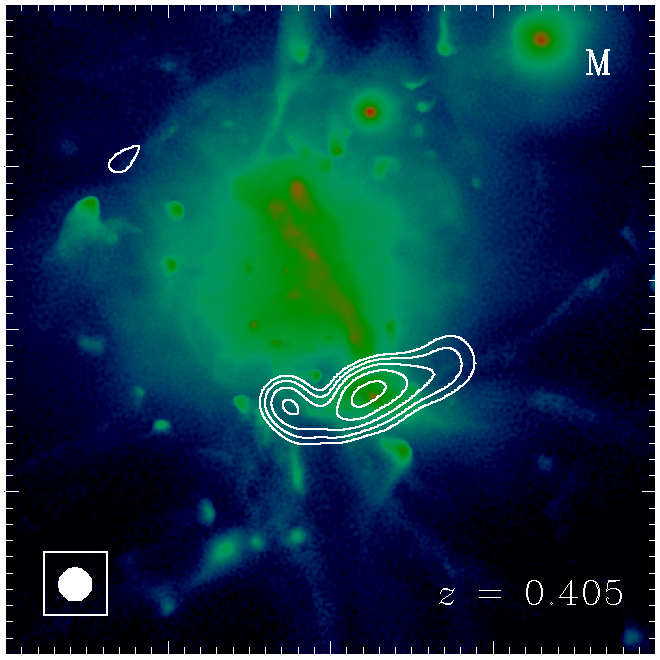}
		  \hspace{-0.15cm}
		  \includegraphics[width=0.23\textwidth]{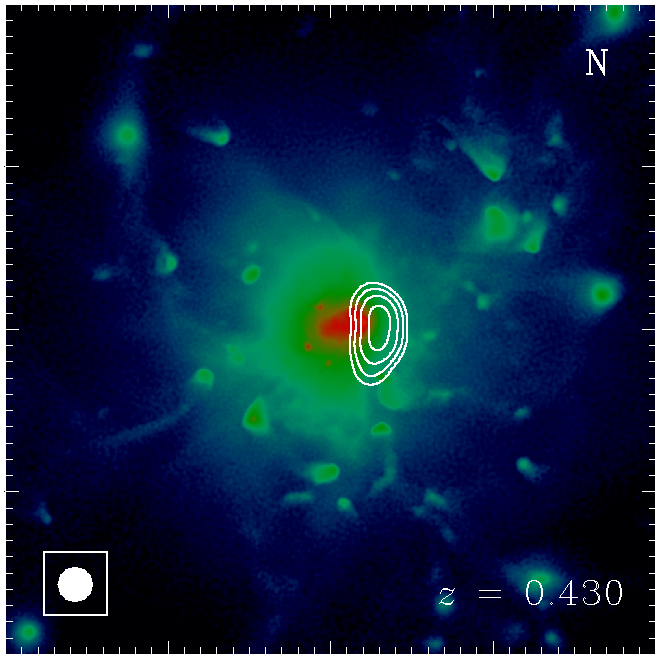}
		  \hspace{-0.15cm}
		  \includegraphics[width=0.23\textwidth]{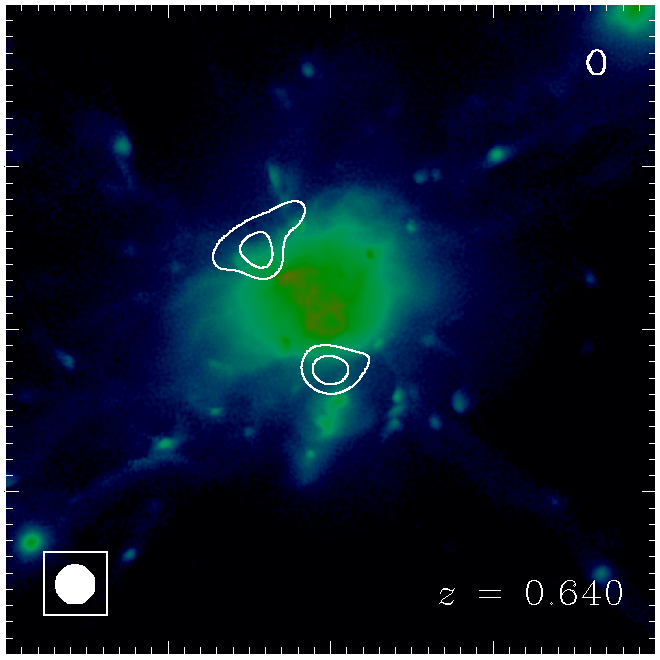}
		  \hspace{-0.15cm}
		  \includegraphics[width=0.23\textwidth]{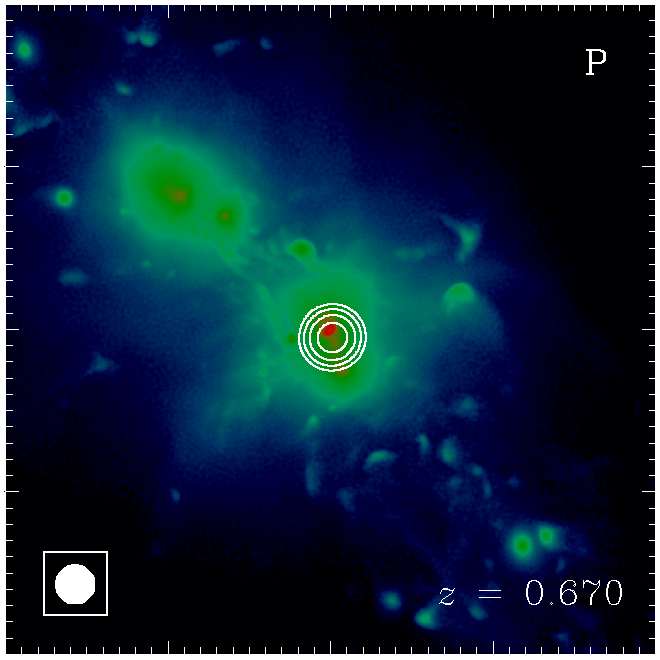}
		  \begin{center}
		   \includegraphics[width=0.3\textwidth]{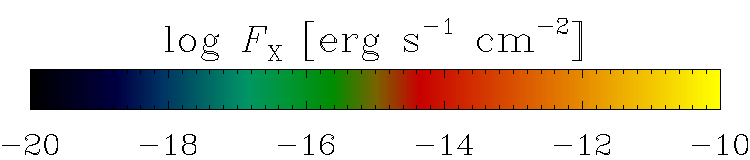}
		  \end{center}

		\end{center}
			\caption{
				Examples of mock radio relic observations for clusters at different redshifts. The colour scale indicates 
				the X-ray surface brightness of simulated clusters. Synthetic relic surface brightness 
				at $1.4\,$GHz observing frequency is shown using contours drawn at 
				$[2,4,8,16,...]\times \sigma_\text{NVSS}$ with $\sigma_\text{NVSS}=0.45\,$mJy\,beam$^{-1}$. 
				Every panel spans a proper area of $4\times4\,$Mpc$^2$ whereas the filled white circles 
				indicate the FWHM beam area. A large morphological diversity of relics can be observed.
			}
		\label{fig:relicCollection}
	\end{figure*}

	\begin{figure*}
			\includegraphics[width=0.44\textwidth]{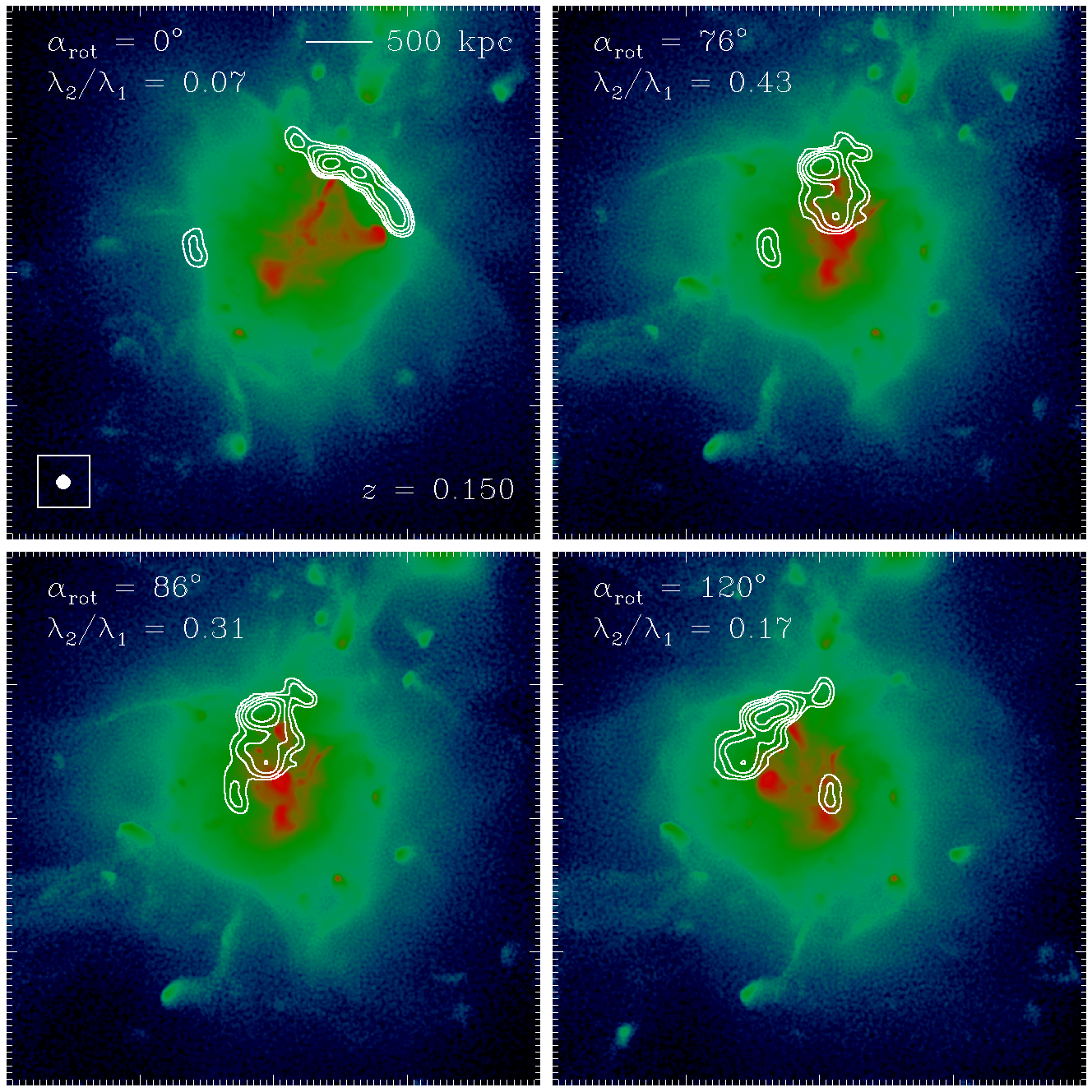}\hspace{1cm}
			\vspace{0.5cm}\includegraphics[width=0.31\textwidth]{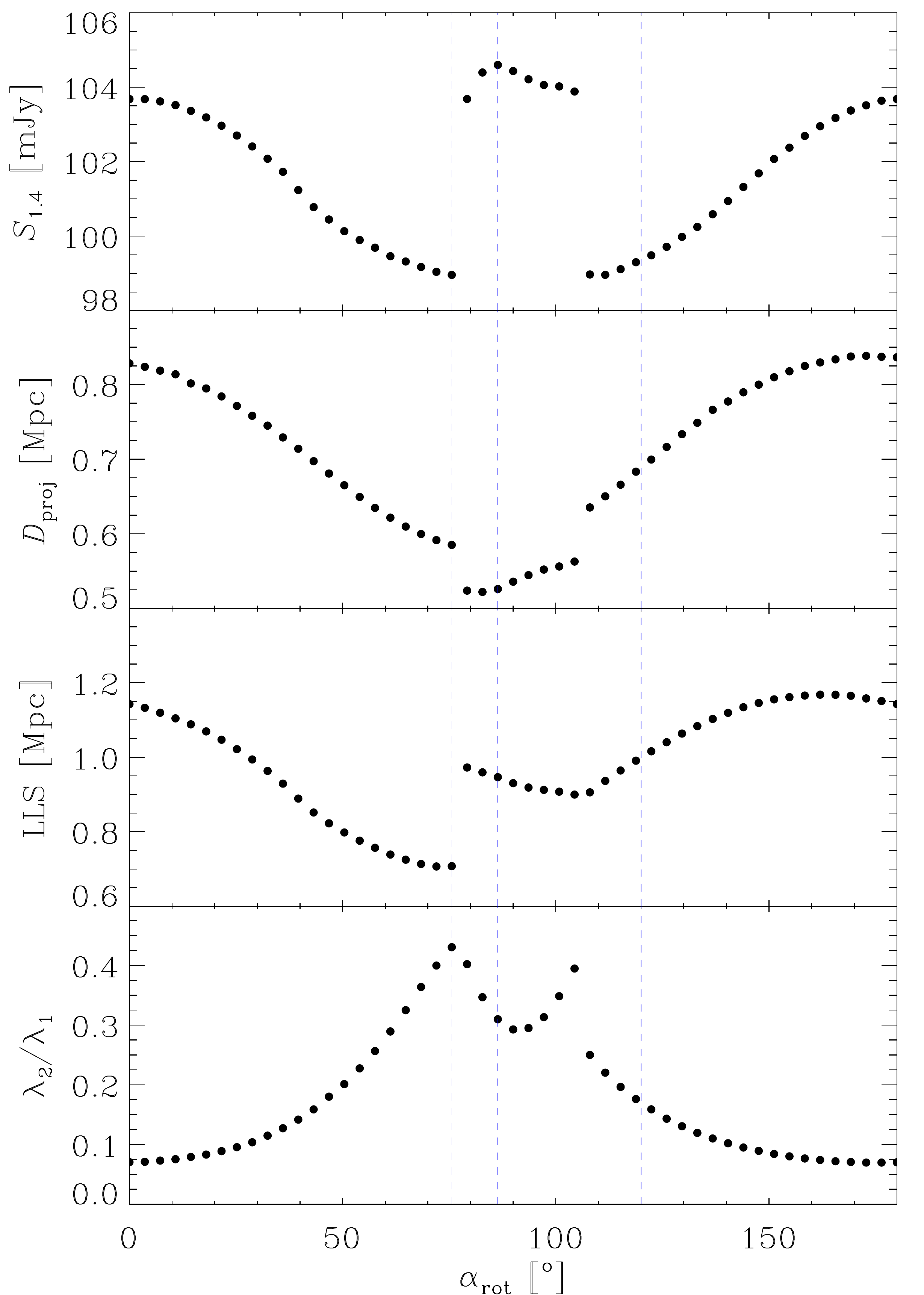}
			\caption{{\it Left-hand panel:} Idem as  Fig.~\ref{fig:relicCollection} for mock observations of one of the simulated 
				clusters at $z=0.15$ and four different rotation angles, $\alpha_{\rm rot}$.  
				Also shown is the shape parameter of the largest relic in each panel. 
				{\it Right-hand panel:} From top to bottom: flux density, projected distance, largest linear size, 
				and shape parameter of the largest relic island as a function of rotation angle. The vertical dashed 
				lines indicate the rotation angles shown in the left-hand panel.
			}
		\label{fig:clusterrotation}
	\end{figure*}

	%
	As a result, the total radio luminosity is given by a few parameters, namely, the Mach number of the shock 
	and the thermodynamical properties and magnetic field strength of the downstream plasma. For the smoothed-particle 
	hydrodynamics (SPH) particles we identify those that are located at a shock front using the shock finder described above. 
	Since the cooling of electrons takes place on much smaller length scales than is possible to resolve, we can 
	attribute the entire radio luminosity of simulated relics to SPH particles at the shock front. Therefore, the output radio luminosity per unit 
	frequency contributed by every shocked particle will also be related to its associated shock area element. 
	The radio power per unit frequency contributed by an SPH gas particle $i$ can be written as                                              
	\begin{eqnarray}
		P_{\nu,i}
		& = &
		6.4 \times 10^{34}\, {\rm erg\,s^{-1}\,Hz^{-1}} \;\;
		\frac{A_i}{{\rm Mpc^2}} \,
		\frac{ n_{{\rm e},i}}{\rm 10^{-4} cm^{-3}}\;
		\nonumber                 
		\\
		&& \quad
		\times
		\frac{\xi_{\rm e}}{0.05} \:
		\left(\frac{\nu}{\rm 1.4\,GHz} \right)^{-\frac{s_i}{2}}
		\left(\frac{{T_{{\rm d},i}}}{\rm 7\,keV} \right)^{\frac{3}{2}} \:
		\label{eq:radio_power}     
		\\
		&& \quad
		\times
		\frac{(B_{{\rm d},i}/{\rm \mu G})^{1+\frac{s_i}{2}} }
		{(B_{\rm CMB}/{\rm \mu G})^2 + (B_{{\rm d},i}/{\rm \mu G})^2}
		\;
		\Psi({\cal M}_i)
		{\rm ,}
		\nonumber
	\end{eqnarray}
	where $A_i$ represents the surface area associated with the particle, $n_{{\rm e},i}$ is the electron density, 
	$\xi_{\rm e}$ is the electron acceleration efficiency, $s_i$ is the slope of electron energy distribution as 
	given by DSA, $T_{{\rm d},i}$ is the post-shock temperature, $B_{{\rm d},i}$ is the post-shock magnetic field, 
	$B_{\rm CMB}$ is the magnetic measure of the CMB energy density and $\Psi({\cal M}_i)$ is a function that depends 
	on the shock strength. The area corresponding to each SPH particle is proportional to the square of the 
	smoothing length divided by the number of particles within the kernel.
	The electron acceleration efficiency denotes the fraction of energy  dissipated at the shock 
	front that is transferred to suprathermal particles. We note that, in practice, this parameter acts as a normalization 
	factor that does not depend on the strength of the shock. The latter is taken into account by the $\Psi({\cal M}_i)$ 
	function that, for weak shocks in the range ${\cal M}_i\gtrsim2-$4, gives an additional factor of about 0.01$-$0.5 
	(see Fig. 4 of \citealt{2007MNRAS.375...77H}). Shock fronts with lower Mach numbers have, in general, very 
	low radio luminosities due to the steep decline of $\Psi({\cal M}_i)$. Within this context, 
	an {\it effective} Mach-dependent acceleration efficiency can be 
	constructed as the product of these two quantities, i.e., $\xi'_{\rm e}({\cal M})\equiv\xi_{\rm e}\Psi({\cal M})$. 
	The interested reader is referred to \cite{2007MNRAS.375...77H} and \cite{2008MNRAS.391.1511H} 
	for more details concerning our radio emission scenario.

	%
	The combination of shock finder, magnetic field prescription and radio luminosity model allows us to `illuminate' 
	merger shocks in the simulation.  Although this model clearly simplifies the 
	complex processes taking place in relics, it allows us to estimate the position and morphology of merger shocks 
	in our cosmological simulation in a way that suits our needs. 
	Throughout this paper, we assume $\xi_{\rm e}=5\times10^{-5}$ as a working efficiency value that is enough to 
	reproduce the number of NVSS clusters with $S_{1.4,{\rm tot}}=100\,$mJy and is also consistent with the findings 
	of \citet{2012MNRAS.420.2006N}. A more detailed discussion concerning relic abundance will be presented in Section~\ref{sec:relic_number}.

\subsection{Mock cluster and radio relic observations}
\label{sec:mock}

	%
	Our aim is to produce a set of mock relics that can be used to study the properties predicted by our model, as well as to 
	compare results with available observations. Here, the mock instrumental parameters will be chosen to resemble those of 
	an NVSS-like survey, as described in Section~\ref{sec:ImageAnalysis}. Therefore, in what follows, we will consider an 
	observational frequency of $\nu_{\rm obs}=1.4\,$GHz, a telescope beam size of $45\,$arcsec and a survey sensitivity of 
	$\sigma_{\rm NVSS}=0.45\,$mJy\,beam$^{-1}$.

	%
	We generate a set of synthetic observations for the galaxy clusters in our sample from surface brightness maps obtained by 
	projecting all cluster radio emission inside $2\times R_{\rm vir}$ on to a plane. In all cases, the rest frame radio frequency 
	has been $k$-corrected to obtain our mock observations at frequency $\nu_{\rm obs}$. To produce radio images resembling those 
	obtained by actual radio telescopes all projected emission has been convolved with a Gaussian filter at a scale 
	$\theta_{\rm FWHM}$ that corresponds to the beam size resolution at the given redshift of the cluster. Then, the emission 
	has been converted into flux per beam. We identify island boundaries using $2\times\sigma_{\rm NVSS}$ contour lines and 
	discard all islands with a flux below $8\times\sigma_{\rm NVSS} = 3.6\,\text{mJy}$, as done for the NVSS sample.

	%
	In our images, the X-ray emission of the clusters is shown as colour-coded maps. The X-ray luminosity of each particle has 
	been computed using the software {\sc xspec} adopting a {\sc Mekal} emission model. We note that X-ray maps are only used to 
	illustrate the dynamical state of the cluster and no attempt is made of computing the emission directly from the simulations.  
	This is due to the fact that non-radiative simulations tend to overestimate the bremsstrahlung luminosity output since they lack 
	the necessary energy feedback to lower the ICM gas density \citep{2007ApJ...668....1N}.

	\begin{figure*}
		\hspace{-0.7cm}
			\includegraphics[width=0.85\textwidth]{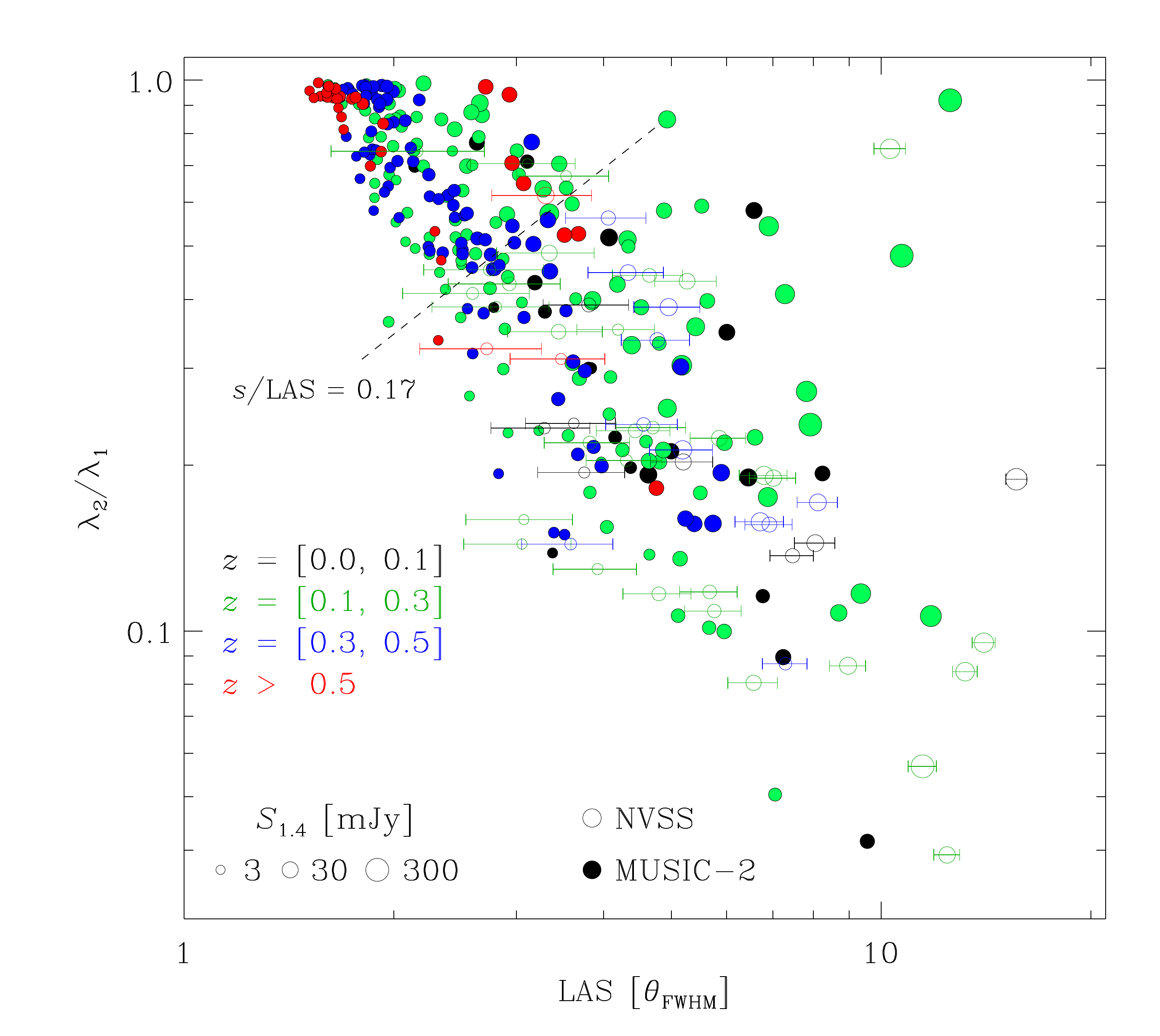} 
			\caption{
				Shape parameter, $s\equiv\lambda_2/\lambda_1$, versus the 
				largest angular size, LAS, for the NVSS (open circles) and MUSIC-2 (filled circles) 
				relic samples. Symbol sizes are scaled according to radio flux. Redshift bins are indicated 
				using different colours. We distinguish the `elongated' from the `small-roundish' islands 
				using the empirical criterion $s/\text{LAS} < 0.17$ (see dashed line). 
			}
		\label{fig:Shape-vs-LAS}
	\end{figure*}

	%
	Fig. \ref{fig:relicCollection} shows some examples of relics found in the MUSIC-2 cluster sample that are `observed' assuming NVSS 
	specifications. All clusters show clear evidence of recent merger activity in the X-ray surface brightness distribution. Some of 
	these clusters host single, double or more irregular relic structures. For instance, clusters `C', `F', and `K', show nice 
	double gischt morphologies as those found in observations. Very often the two cores of the progenitor clusters are well 
	visible in the X-ray maps. An spectacular example of a single relic is shown in cluster `G'. Large extended objects are also seen 
	at high redshift, an example of which is cluster `M' that hosts a quasi-linear radio structure of about 1 Mpc long at $z=0.405$. 
	Other clusters show that radio relic morphologies may significantly deviate from classical textbook examples, i.e. elongated structures 
	tangentially oriented in the cluster outskirts. For instance, the relic in cluster `B' is located very close to the cluster centre with 
	a morphology resembling the enigmatic object found in Abell\,523. Another example is the complex relic in cluster `D', which is not at 
	all positioned along the apparent merger axis. Complex merger scenarios can give rise to peculiar emission features such as those 
	seen in cluster `A'. Moreover, a line-of-sight along the merger axis, or close, can reveal how relic surfaces can break into several 
	pieces, as shown, e.g., in cluster `E'. Finally, small and/or poorly resolved emission features typically appear as more roundish objects 
	as in clusters `J', `L' and `P'. 
	
	%
	We speculate that a significant fraction of the roundish-like islands are caused by spurious effects in the simulation, with possibly several 
	origins. The simulation has been carried out without any heating of the ICM except for the dissipation of structure formation 
	shocks. It is known that this kind of treatment underestimates the temperature of the ICM, most significantly in the cluster 
	centre. Therefore, owing to the lower sound speed, fast moving dense clumps may more easily generate a small shock. Moreover, since 
	SPH is intrinsically noisy and the Mach number depends in a very non-linear way on the particle properties, one may also find 
	some outliers. 
	Fig.~\ref{fig:clusterrotation} shows how the morphology and other island parameters change with viewing angle. To study this, 
	we have rotated one of the clusters in our MUSIC-2 set along the vertical axis. We plot four panels, one for each rotation angle. 
	This particular example consists of a large linear gischt structure of about 1.2 Mpc long located at about 0.9 Mpc from the 
	cluster centre when the relic is seen edge-on, i.e. $\alpha_\text{rot}=0^{\circ}$. A hint of a second relic structure in the 
	opposite side of the cluster can also be seen. This scenario changes dramatically when rotating the cluster. 
	The right-hand panel of Fig.~\ref{fig:clusterrotation} plots the evolution of several island parameters for the main 
	relic as a function of rotation angle. As expected, radio flux, projected distance and LLS decrease as 
	the rotation angle tends to $90^{\circ}$. The opposite is true for the shape parameter, meaning that, 
	when the relic is seen face-on, its shape appears rounder to the observer. Moreover, when the projected contours of the two radio 
	features merge (separate) at $\alpha_\text{rot}\approx80^{\circ}\,(105^{\circ})$ there are clear discontinuities in the island 
	parameters. These results show that special care must be taken when interpreting observations as projection effects can significantly 
	affect the location and morphological properties of relics.  
	
	%
	We find that the MUSIC-2 cluster sample leads to a large variety of radio relics, covering textbook examples of single and 
	double relics to systems with a very peculiar morphology. In the next section, we will use these relics to statistically 
	compare their location, morphology and emission properties with those found in NVSS.

\section{Properties of the samples} 
\label{sec:Compare}

	%
	Many spectacular radio relics have been observed and their properties studied in considerable detail. 
	However, we can still only speculate if all relics are caused by merger shocks in a homogeneous way. Here, 
	we would like to know if the radio luminosities of relics found in NVSS, their relation to the X-ray cluster 
	luminosity, their size, morphology and location within galaxy clusters can be reproduced by the 
	relics identified in the MUSIC-2 cluster sample. The location, size and morphology of the simulated 
	relics are determined by the shock fronts. Hence, both the merger history of galaxy clusters and the evolution 
	of the ICM determine their properties within this scenario.

\subsection{Constructing a representative cluster sample}
\label{sec:zsample}

	%
	First of all, we need to remember that the sample of relics identified in NVSS is flux limited, besides other limitations discussed 
	in Section~\ref{sec:NVSScomp}. Since the number of known relics is still small we cannot easily derive a volume limited sample 
	comprising a sufficient large number of relics by considering only a small redshift range. Therefore, in our modelling of relic 
	observations we reproduce the NVSS flux limited sample.

	%
	Radio relics in the redshift interval $z \in [z_i,z_{i+1}]$ are located within a comoving volume of
	\begin{equation}
		V_{\rm c}(z_i,z_{i+1}) 
		=  
		\frac{4 \pi}{3}\left( R_{\rm c}^3 (z_{i+1}) - R_{\rm c}^3 (z_i) \right)  
		\rm ,
		\label{eq:volShell}
	\end{equation}
	where $R_{\rm c}$ denotes the comoving radial distance and a flat universe is assumed. In order to generate the mock 
	relic sample we proceed as follows. We define a sequence of shells around the observer with comoving volumes according to 
	Eq.~\ref{eq:volShell}. For each redshift interval, we take the snapshot of the simulation with the closest redshift to the 
	shell mean value. 
	The ratio of the shell volume to the simulation one determines how many clusters from the MUSIC-2 sample should be chosen 
	to `populate' the shell. Then, we randomly pick these clusters from the sample. 

	%
	Since the redshift distribution within the shell is not uniform, the redshift of the selected cluster is randomly chosen 
	from the $[z_i,z_{i+1}]$ interval assuming a probability proportional to the comoving volume. Therefore, the normalized 
	probability density of finding a cluster at redshift $z$ within the shell can be written as
	\begin{equation}
		\mathcal{P}(z) 
		=  
		\frac{1}{V_{\rm c}(z_i,z_{i+1})}
		\frac{{\rm d}V_{\rm c}(0,z)}{ {\rm d} z}  
		\rm .
		\label{eq:redshShell}
	\end{equation}

	\begin{figure}
		\hspace{-0.8cm}
			\includegraphics[width=0.53\textwidth]{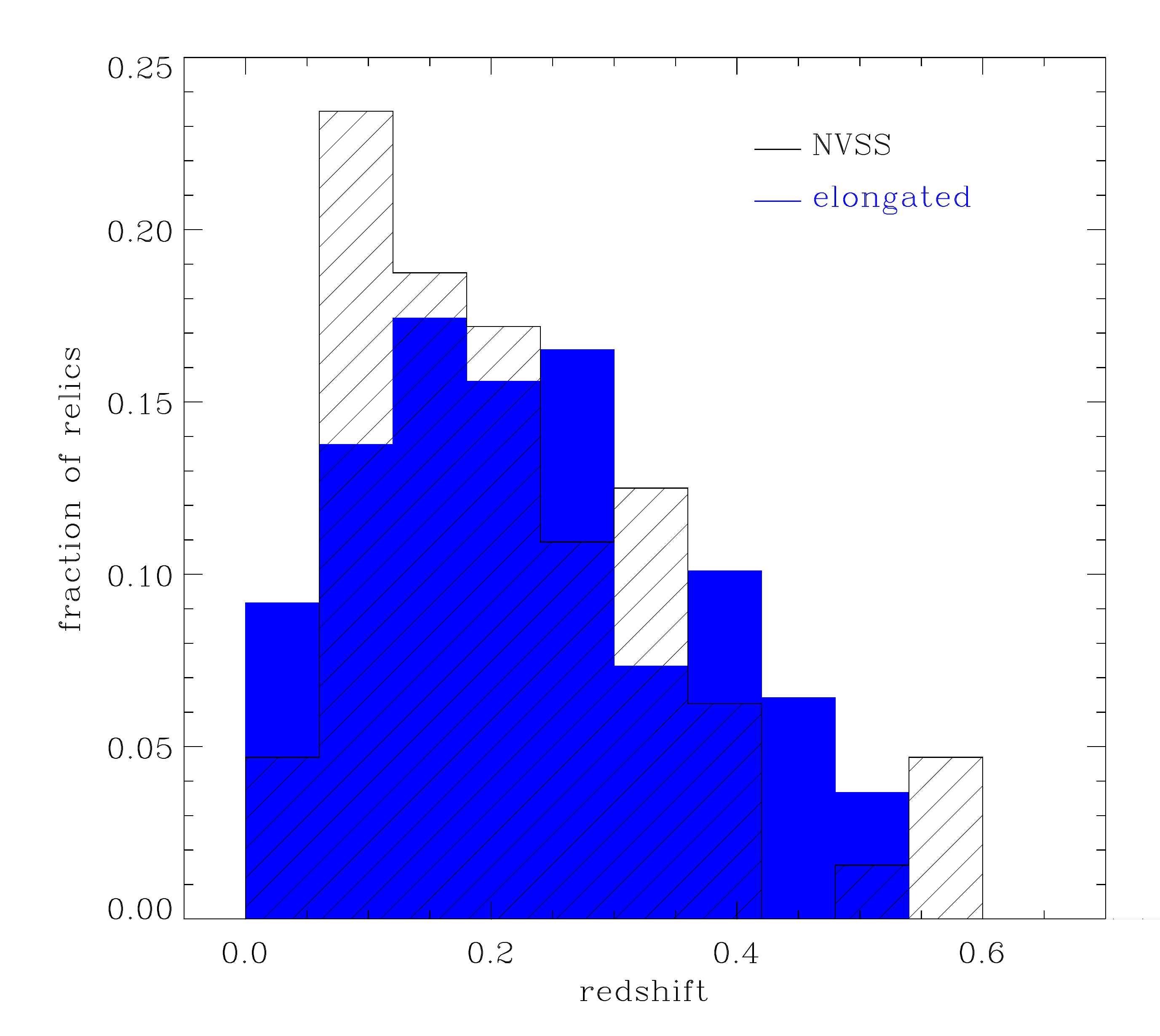} 
			\caption{Redshift distribution of the NVSS and `elongated' relic samples 
			(hashed and blue histograms, respectively).
			}
			\label{fig:zdistr}
	\end{figure}

	%
	The cluster selection procedure implies that only a few clusters are finally drawn from simulation snapshots at low redshifts. 
	At higher redshifts the situation is different. For example, the simulation snapshot at $z=0.43$ serves as a proxy for the 
	redshift interval $z \in [0.38,0.55]$. In this case, the comoving volume of the shell amounts to 
	23.4\,Gpc$^3$, indicating that every cluster in this snapshot needs to enter many times in the sample. 
	To avoid duplicating the same relic parameters, we randomly rotate the clusters in every iteration (see Fig.~\ref{fig:clusterrotation} for 
	the effect of rotation on the relic properties such as morphology and flux). 
	The fact that clusters at $z \gtrsim 0.5$ enter the sample a few tens of times shows the need of a large simulation 
	volume for our purpose. In this respect, the cosmological box and simulation technique of the MUSIC-2 set turn out to be 
	a good compromise: at the mean redshift of our radio relic samples, i.e. $z \sim 0.2-0.3$, clusters enter in the mock catalogues 
	only a few times at maximum, while the zoomed method permits us to resolve the shock fronts in a reasonable way.

	%
	As a result of this approach, the resulting mock relic sample will represent a unique realization of the selection process and, 
	therefore, relic properties will depend on which clusters are finally included in the sample. Then, if required, we can average 
	over several realisations, e.g. to get mean number counts that can be compared to observations. 
	%
	In what follows, we start by contrasting the properties of islands before considering the abundance and flux distribution of 
	the relic sample. 

	\begin{figure}
			\hspace{-0.7cm}
				\includegraphics[width=0.53\textwidth]{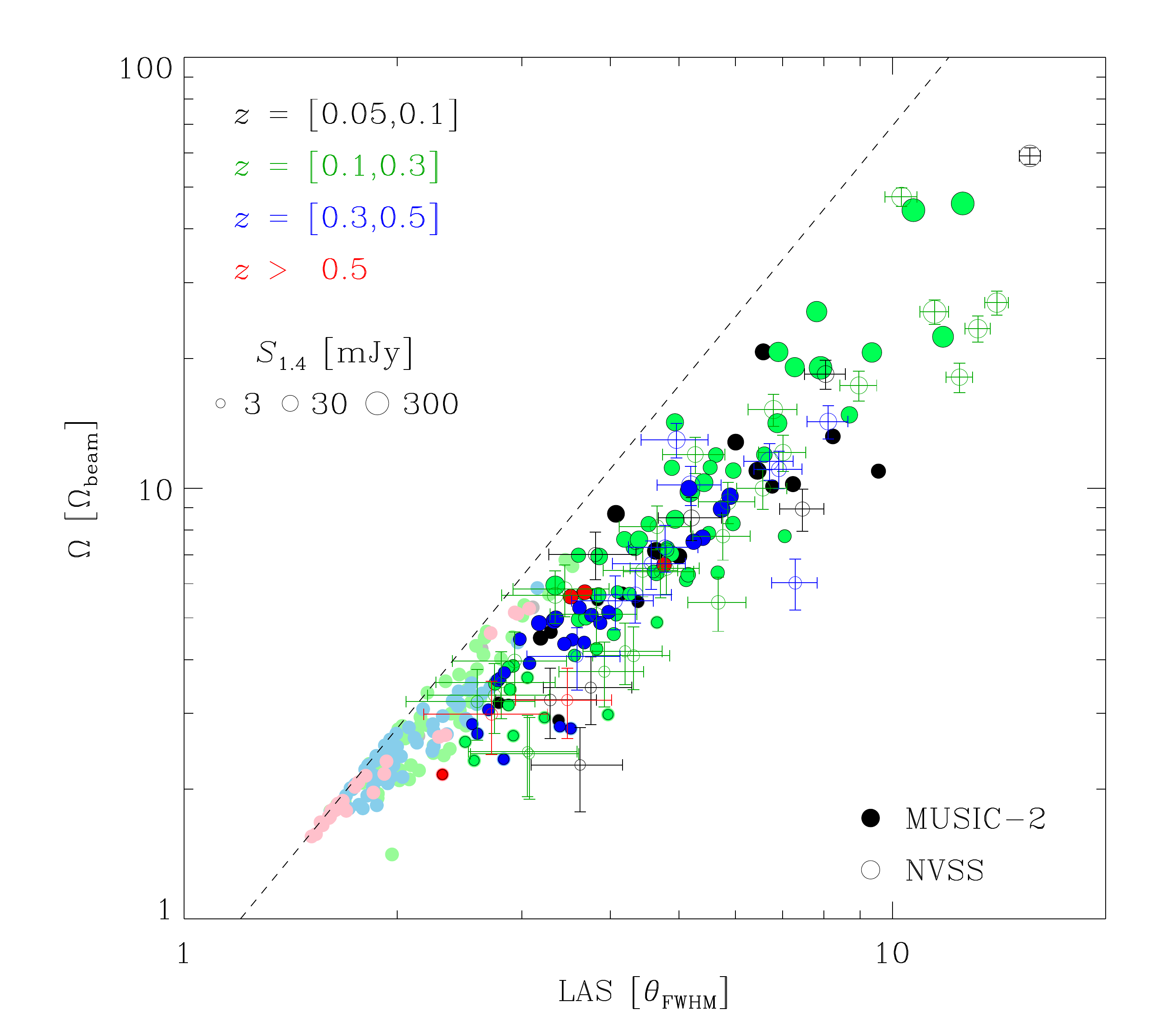} 
				\caption{
				Solid angle, $\Omega$, versus the largest angular size, LAS, for 
				the NVSS and mock samples. Shown are the NVSS islands (open circles) and the `elongated' 
				(filled circles) and `small-roundish' (light filled circles) mock relic samples. 
				Symbol sizes for the NVSS and `elongated' samples are scaled according to radio flux.  
				In all cases, redshift bins are indicated using different colours. 
				For reference, the relation for a circular object is shown as a dashed line. 
				}
				\label{fig:solidangle-vs-LAS}
	\end{figure}

\subsection{The radio islands}
\label{sec:islandsamples}

	%
	In Section~\ref{sec:ImageAnalysis} we have shown that relics, more precisely the radio emission islands, found in the NVSS images 
	show a strong correlation between shape and the largest angular size (LAS). As a first step of our comparison, we investigate if the NVSS 
	island sample shows a similar correlation. As a reminder, we summarize the relic selection criteria considered in our NVSS sample: (i) all 
	islands with a flux below $8 \times \sigma_\text{NVSS} = 3.6 \, \text{mJy}$ are discarded since image noise may have a 
	significant impact on their morphology, and (ii) clusters at $z<0.05$ are also discarded since, at these redshifts, the recovery of relics 
	with a typical extent of $1\,$Mpc already starts to be affected by the inner $uv$-gap of the VLA D configuration. 
	
	%
	We can now discuss the results of Fig.~\ref{fig:Shape-vs-LAS}. This plot shows the distribution of shape 
	parameter versus LAS for both samples. Evidently, the mock sample 
	displays a similar (anti-)correlation as observations. In contrast with the NVSS relics, the simulated sample shows many objects above 
	the empirical threshold $s/\text{LAS}=0.17$. In the following analysis, we subdivide the simulated relics into `small-roundish' and 
	`elongated' samples that comprise the islands above and below the threshold, respectively. When comparing to NVSS data we will 
	consider only the `elongated' sample since the small-roundish objects are mostly found in 
	the mock observations (see the discussion in Section~\ref{sec:Discussion}).

	%
	It is interesting to note, however, that the mock relic sample shows only very few {\it large} roundish objects that may 
	originate from relics seen face-on. It has been speculated that some giant radio haloes, which are centred on the X-ray cluster 
	emission and follow its morphology, might result from radio relics seen perfectly face-on. Our simulation confirms that such a 
	situation is very rare and cannot significantly contaminate the giant halo sample.

	%
	The redshift distribution of NVSS and `elongated' relics is shown in Fig.~\ref{fig:zdistr}. To compare with NVSS data 
	in an unbiased way, we need to exclude the `small-roundish' islands. Both distributions are quite similar with mean redshift 
	values of 0.23 and 0.21 for the simulated and NVSS samples, respectively. The main difference between the two is that we find a 
	smaller relic fraction in the mock sample at low redshift. This can be explained by the fact that many low-redshift 
	relics in NVSS reside in low-mass clusters, which are not included in MUSIC-2. However, at $z\sim 0.4-0.5$ we find 
	a higher relic fraction in comparison with observations. 
	We speculate that this is caused by the fact that some high-redshift relics in NVSS may escape discovery since their host 
	galaxy clusters are too faint in X-ray to be identified: most of the observed clusters are above the {\it ROSAT} X-ray flux 
	detection limit. In Sections~\ref{sec:distclusters} and~\ref{sec:relic_number}, we will give more details concerning 
	this point.

	%
	Fig.~\ref{fig:solidangle-vs-LAS} shows the correlation between solid angle and LAS for the mock (filled circles) and NVSS 
	(open circles) samples. Unless otherwise stated, besides the `elongated' simulated sample, in what follows we will also show 
	the `small-roundish' islands using lighter colours. For comparison, the relation for a circle, $\Omega_\text{circ}$, is 
	shown as a dashed line. As seen in the figure, the NVSS and `elongated' samples show a similar distribution. In particular, 
	the mean solid angles of the distributions are essentially the same but the LAS	are slightly higher in the NVSS case. 
	For the NVSS sample the filling factor, $\Omega / \Omega_\text{circ}$, amounts to about 35\%.  
	
	\begin{figure}
		\hspace{-1cm}
			\includegraphics[width=0.53\textwidth]{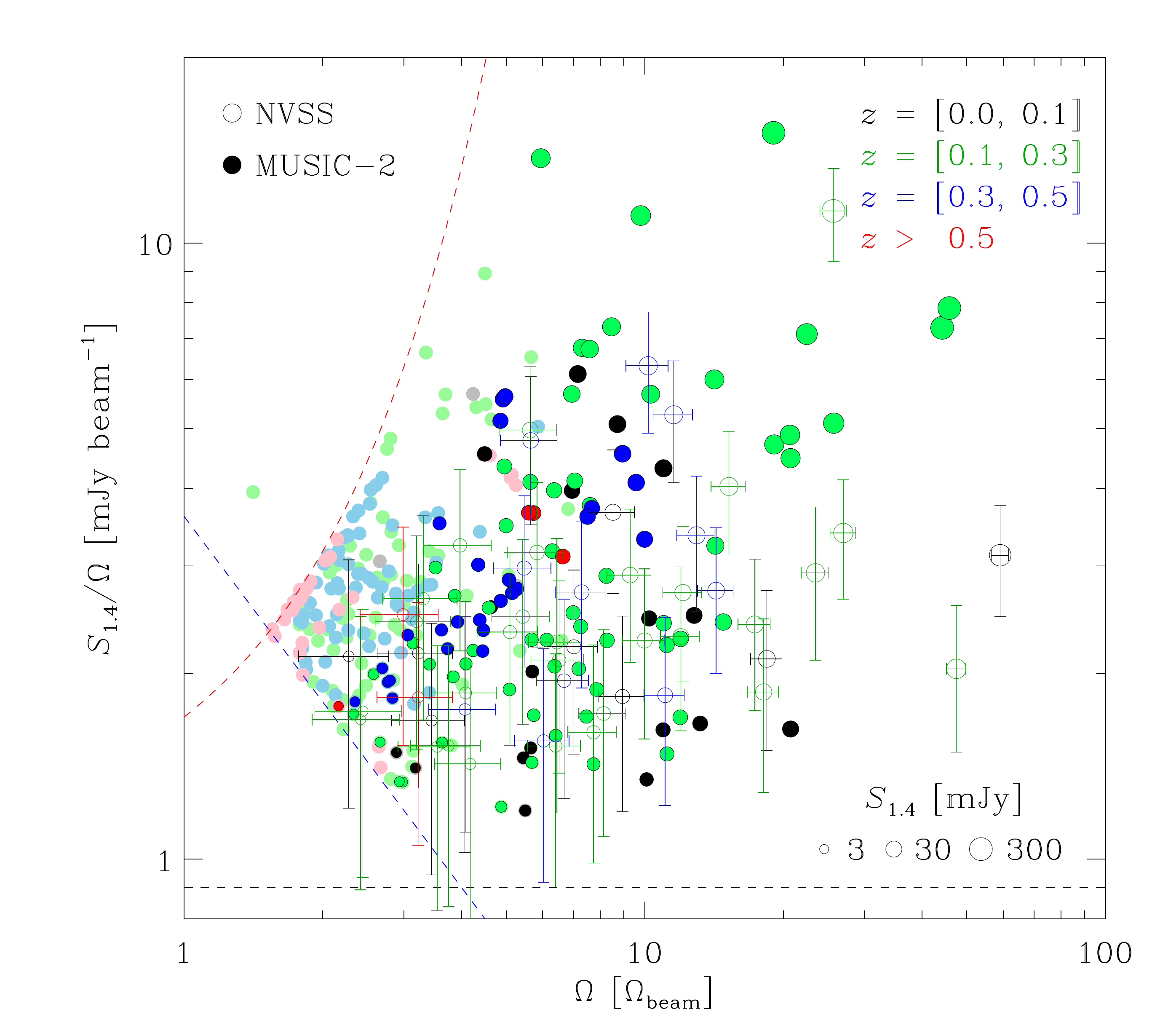} 
			\caption{Average surface brightness versus solid angle, $\Omega$, for 
				the relic samples. Shown are the NVSS islands (open circles) and the `elongated' 
				(filled circles) and `small-roundish' (light filled circles) mock relic samples. 
				Symbol sizes for the NVSS and `elongated' samples are scaled according to radio flux.  
				In all cases, redshift bins are indicated using different colours. 
				For reference, limits associated with the observation process are also drawn, namely, 
				the maximum flux given by Eq.~\ref{eq:Smax} (red dashed line); the cut associated 
				with the minimum flux density condition of an island, 
				i.e. $S_{1.4,{\rm min}}=8\times\sigma_{\rm NVSS}$ (blue dashed line), and the minimum possible 
				surface brightness of a beam, i.e $2\times\sigma_{\rm NVSS}$ (black dashed line), where 
				$\sigma_{\rm NVSS}$ is the NVSS sensitivity.   
			}
			\label{fig:SB-vs-solidangle}
	\end{figure}

	%
	Fig.~\ref{fig:SB-vs-solidangle} shows the average surface brightness versus the solid angle of the islands. The `elongated' 
	and NVSS scatter plots nicely match, being their median surface brightnesses 2.46 and 2.40 mJy$\,$beam$^{-1}$, respectively. 
	The lowest possible surface brightness is set by the adopted island boundary, 
	i.e. $2 \times \sigma_\text{NVSS}$ (black dashed line). The Gaussian smoothing in both the NVSS images and the mock observations imposes a maximum 
	surface brightness as a function of solid angle given by Eq.~\ref{eq:Smax} (red dashed line). Finally, an additional limit comes from our restriction 
	to consider only islands above $8 \times \sigma_\text{NVSS}$ (blue dashed line). These conditions evidently limit the observed surface brightness. 
	Still, it is remarkable that both the NVSS and mock island samples reside in a narrow range of average  
	brightness. This fact may reflect that relics are not compact: even if they comprise bright areas, there are also regions 
	with low surface brightness (see e.g., the `Toothbrush' relic in Fig.~\ref{fig:NVSSimage}). As a result, the deeper the observations are, the more 
	extended the relics may become. 
	
		\begin{figure*}
			\hspace{-0.7cm}
				\includegraphics[width=0.85\textwidth]{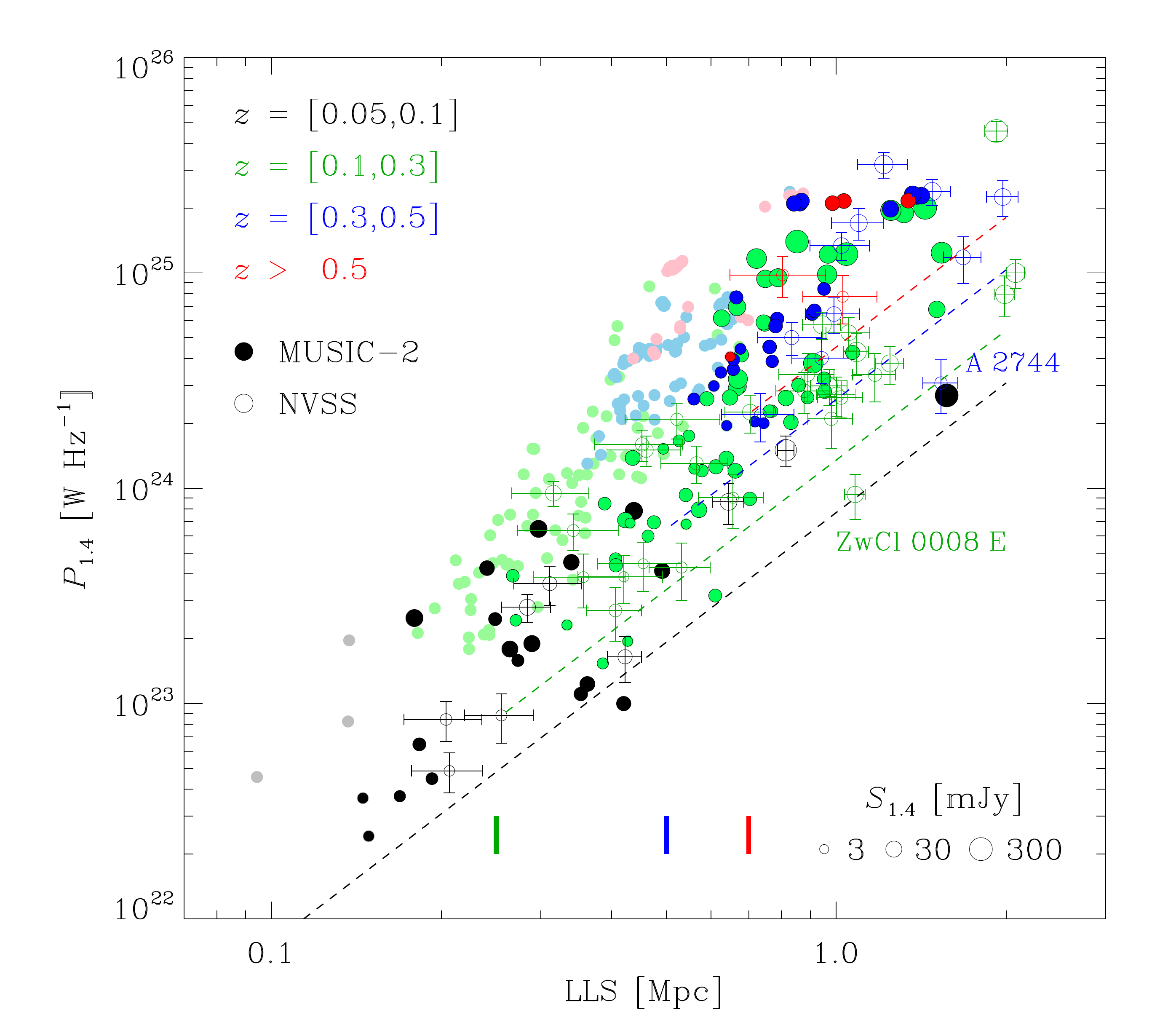} 
				\caption{Radio power, $P_{1.4}$, versus largest linear size, LLS, for 
				the relic samples. Shown are the NVSS islands (open circles) and the `elongated' 
				(filled circles) and `small-roundish' (light filled circles) mock relic samples. 
				Symbol sizes for the NVSS and `elongated' samples are scaled according to radio flux.  
				In all cases, redshift bins are indicated using different colours. 
				For reference, redshift-dependent lower limits for the flux density in 
				the $P_{1.4}-\text{LLS}$ plane are shown as colour-coded dashed lines (see text). 
				Additionally, we also indicate the corresponding spatial scales for 
				$1.5\times\theta_{\text{FWHM}}$ at the same redshifts, where 
				$\theta_{\text{FWHM}}$ is the NVSS beam angular size (vertical solid lines). 
				}
				\label{fig:P14-vs-LLS}
		\end{figure*}

	%
	Many authors have reported a correlation between radio power and LLS of relics 
	(e.g., \citealt{2012A&ARv..20...54F,2012MNRAS.426...40B}). Fig.~\ref{fig:P14-vs-LLS} shows this correlation for the NVSS and 
	mock relic samples. The `elongated' and NVSS samples agree reasonably well. As seen before for the LAS, the `elongated' mock sample 
	has slightly smaller sizes than observations, with a mean LLS value of $0.7\,$Mpc in comparison to $0.82\,$Mpc for the NVSS. 
	If one considers the whole mock relic sample, many objects have radio powers above the median luminosity of the NVSS as a result 
	of the more roundish shape of some of the islands (see Fig.~\ref{fig:solidangle-vs-LAS}). 
	To understand the origin of this correlation, we estimate a lower flux density limit for each redshift bin. 
	To convert LLS into LAS, we adopt the mean redshift of each bin. By using the solid angle versus LAS correlation, 
	we are able to obtain the average area related to LLS. Since the flux density has to be 
	larger than $2\times \sigma_\text{NVSS}$, we obtain a minimum flux limit that we then convert into rest-frame radio luminosity 
	(dashed-lines colour-coded by redshift). 
	We further indicate a diameter of $ 1.5 \times \theta_\text{FWHM}$ adopting the corresponding mean redshift value in each case 
	(vertical solid lines). 
	Basically, for each redshift bin, all islands are above the derived lower-limit radio power estimates. This indicates 
	that the evident correlation between relic luminosity and LLS actually originates from the fact that all observed relics 
	display a similar average surface brightness.

\subsection{Relics in distant clusters}
\label{sec:distclusters}

\begin{figure}
		\hspace{-0.7cm}  
			\includegraphics[width=0.53\textwidth]{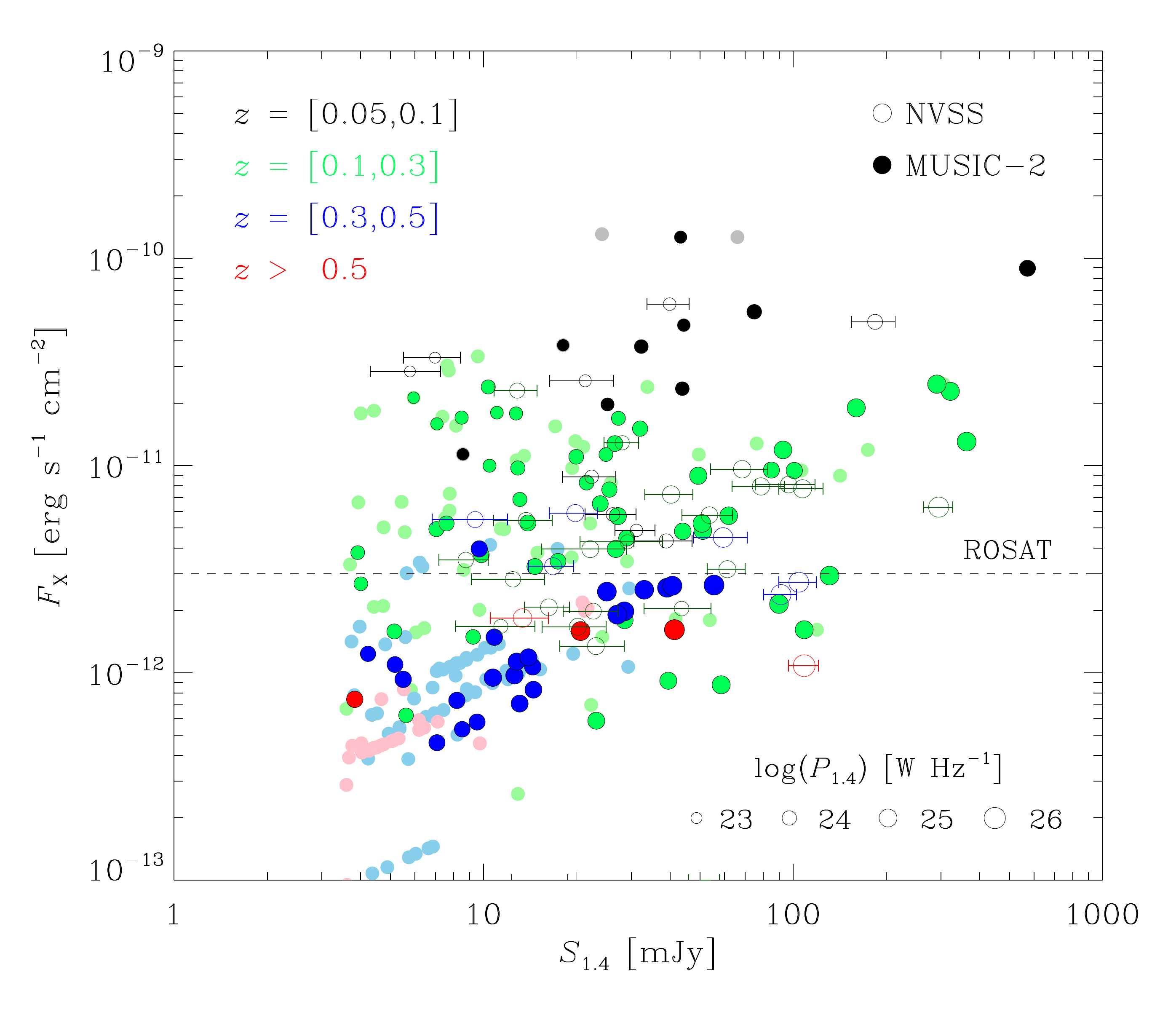} 
			\caption{X-ray flux, $F_X$, versus total radio flux, $S_{1.4}$, for clusters hosting relics. 
				Shown are the NVSS clusters (open circles) and those belonging to the `elongated' 
				(filled circles) and `small-roundish' (light filled circles) mock relic samples. 
				Symbol sizes for the NVSS and `elongated' samples are scaled according to radio power.  
				In all cases, redshift bins are indicated using different colours. 
				The {\it ROSAT} flux limit is included here for reference (dashed line). 
			}
			\label{fig:xflux-vs-radioflux}
	\end{figure}

	%
	For each cluster, we determine its radio flux as the sum of all island fluxes. 
	Hence, we can investigate the X-ray flux versus total radio flux correlation, which is shown in 
	Fig.~\ref{fig:xflux-vs-radioflux}. The X-ray flux is estimated from the luminosity via
	\begin{equation}
			F_\text{X} 
			=
			\frac{L_{\rm X} (1+z)}{4 \pi d_\text{L}^2},
	\end{equation}
	where $d_\text{L}$ denotes the luminosity distance. For simplicity, we do not consider here any $k$-correction. 
		
	%
	It is very well visible in the plot, that some clusters are `aligned'. This is caused by the fact that we considered the same simulation 
	snapshot of a cluster multiple times, using an arbitrary cluster rotation and a slightly different redshift in every iteration. 
	We note that this effect mainly occurs for clusters in the `small-roundish' sample as they enter more times 
	in the relic list as a result of their higher mean redshift. Interestingly, this shows that the $1\,h^{-1}\,$Gpc simulation box 
	we are considering here is only marginally sufficient to simulate radio relics in a representative cosmological volume 
	for the sensitivity limits applied.
	
	\begin{figure}
		\hspace{-0.75cm}
			\includegraphics[width=0.53\textwidth]{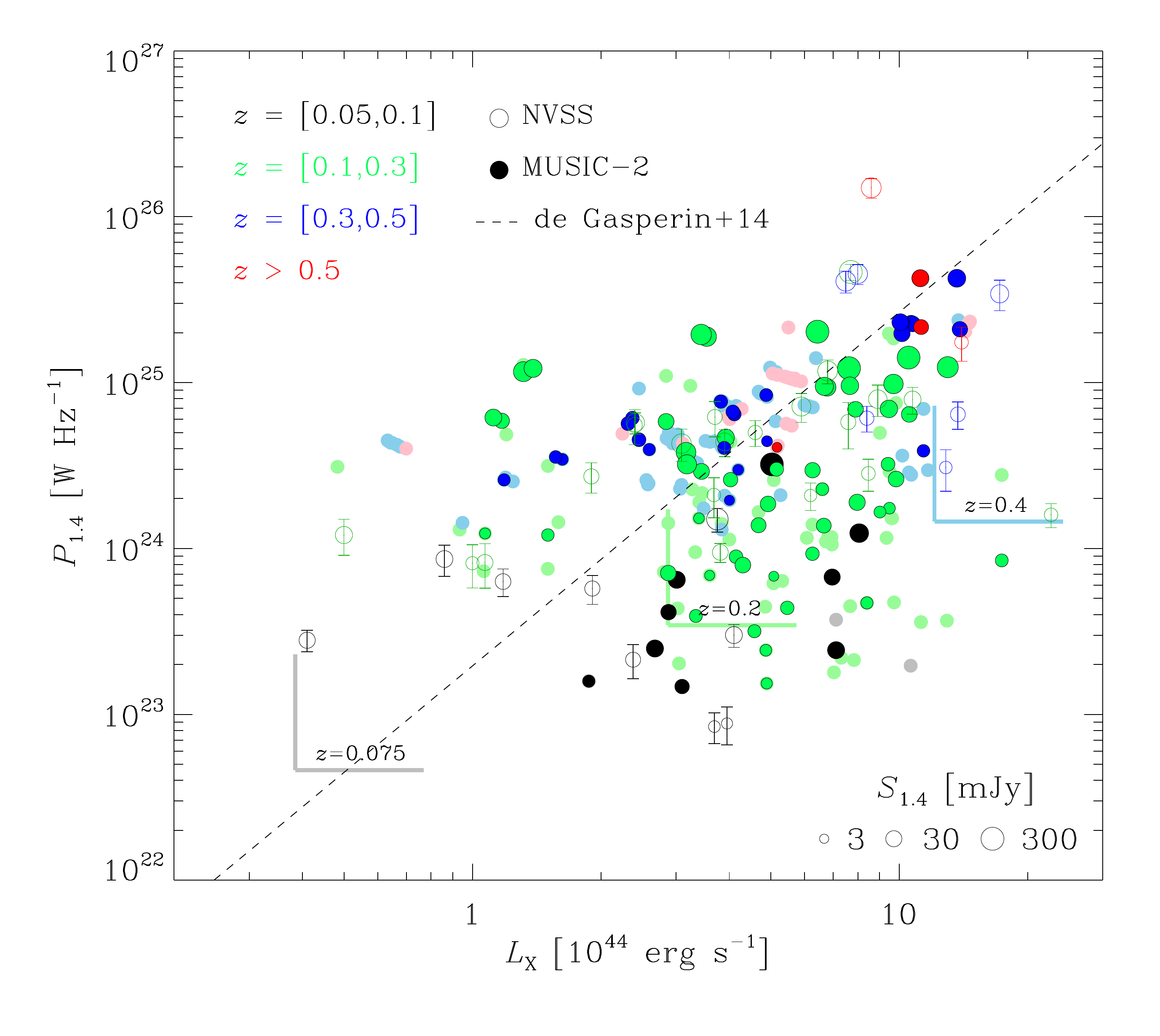} 
			\caption{Radio power, $P_{1.4}$, versus X-ray luminosity, $L_{\rm X}$, for clusters hosting relics. 
				Shown are the NVSS clusters (open circles) and those belonging to the `elongated' 
				(filled circles) and `small-roundish' (light filled circles) mock relic samples. 
				Symbol sizes for the NVSS and `elongated' samples are scaled according to radio flux.  
				In all cases, redshift bins are indicated using different colours. 
				The radio and X-ray luminosity thresholds associated with the minimum radio flux imposed 
				to our samples, $S_{1.4,{\rm min}}=8\times\sigma_{\rm NVSS}$ (horizontal solid lines), 
				and those corresponding to the {\it ROSAT} flux limit (vertical solid lines) for different redshifts, 
				are also shown. 
				For reference, we also include the \citet{2014MNRAS.444.3130D} relation after rescaling 
				cluster masses to the virial radius and transform to X-ray luminosity using 
				Eq.~\ref{eq:LMscaling} (dashed line). 
			}
			\label{fig:P14-vs-Lx}
	\end{figure}

	%
	Evidently, most of the clusters in the NVSS sample are above or in the vicinity of the {\it ROSAT} detection limit. 
	In contrast, many of the clusters of the mock sample are significantly below, which correspond to 
	systems at redshift $z \gtrsim 0.3$. We note that the MUSIC-2 cluster sample comprises rather massive clusters, 
	hence, a low X-ray flux can only originate from distant clusters. The fact that the `small-roundish' sample has no 
	NVSS counterpart and mainly populates distant clusters with low X-ray flux suggests that those relics are notoriously 
	difficult to discover: they are small, faint and primarily reside in clusters not yet found. As noted earlier,  
	some of these objects might be spurious. We will discuss this further below.

	%
	The correlation between radio power and cluster X-ray luminosity is well known. We show the resulting scatter plot for the NVSS and 
	the mock relic samples in Fig.~\ref{fig:P14-vs-Lx}. For comparison, we also include the best-fitting relation of \cite{2014MNRAS.444.3130D} 
	after scaling cluster masses to the virial radius (see Section~\ref{sec:Xray_clusters}) and transform to X-ray luminosities using 
	the mass-luminosity relation of Eq.~\ref{eq:LMscaling}. Also shown are the limits arising from the flux density threshold 
	of $8 \times \sigma_\text{NVSS}$ and the {\it ROSAT} X-ray flux limit by adopting the mean redshift of each bin. It is worth noting 
	that the {\it ROSAT} limit is not strict as several clusters below this threshold are known to host relics. In any case, the latter 
	suggests that this correlation is to a large extent determined by the detection thresholds.

\subsection{Relic orientation and projected distance to cluster centre}
\label{sec:DprojOrient}

\begin{figure*}
		\begin{center}
			\includegraphics[width=0.85\textwidth]{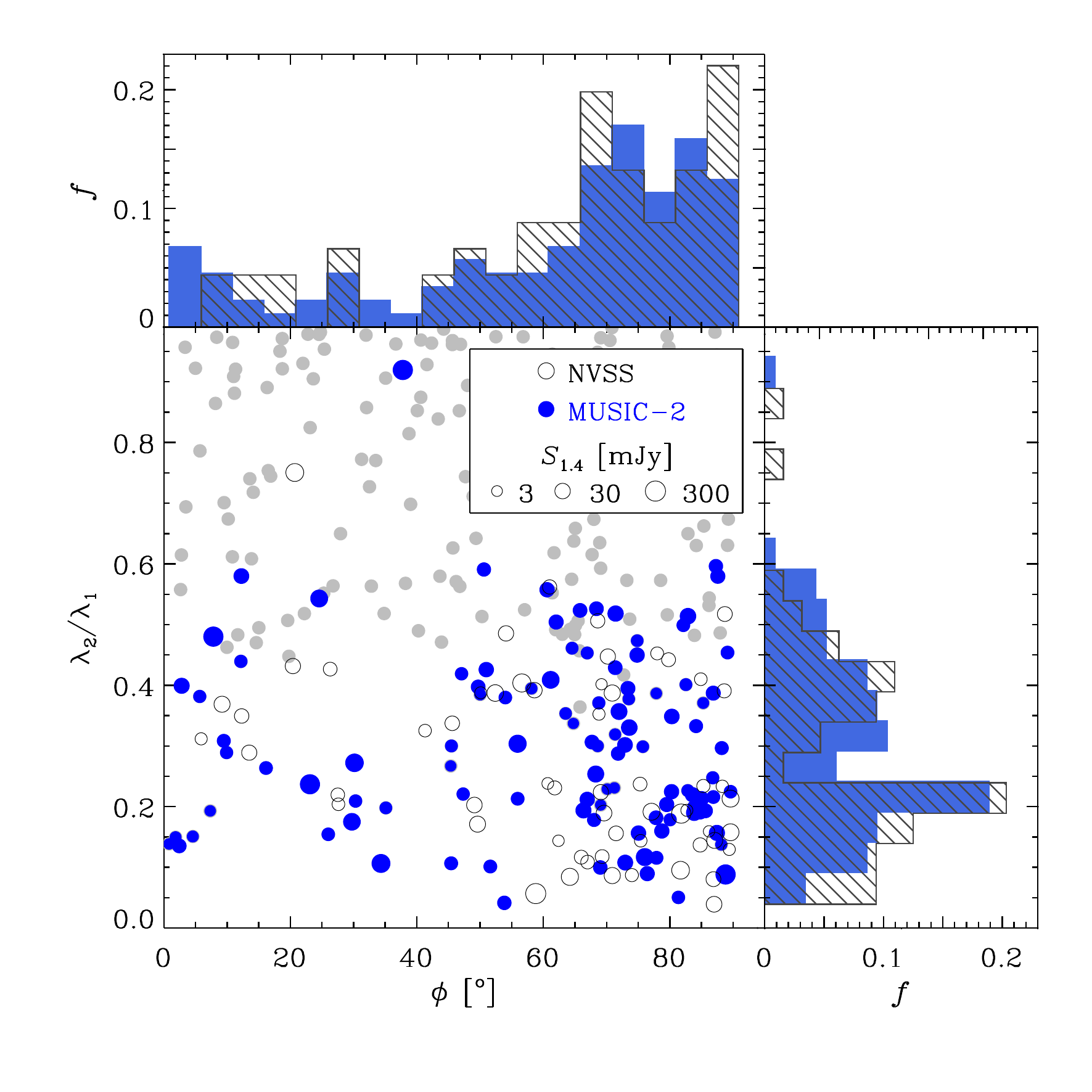}
			\vspace{-0.8cm}
			\caption{Shape, $s\equiv\lambda_2/\lambda_1$, versus orientation angle within 
				the cluster, $\phi$, for the relic samples. 
				Shown are the NVSS islands (open circles) and the `elongated' 
				(blue filled circles) and `small-roundish' (light filled circles) mock 
                                relic samples. 
				Symbol sizes for the NVSS and `elongated' samples are scaled according 
                                to radio flux. The upper and right-hand histograms show the comparison 
                                between the NVSS (hashed histogram) and `elongated' (blue histogram) 
                                samples for the orientation angle and shape parameter, respectively.   
			}
			\label{fig:shape-vs-angle}
		\end{center}
	\end{figure*}

	%
	In Fig.~\ref{fig:shape-vs-angle} we study the relation between relic shape and orientation. The latter is 
	characterized by the angle $\phi$ between the position and elongation axis of relics as explained 
	in Section~\ref{sec:shapes} (see also Fig.~\ref{fig:NVSSimage}). 
	The `elongated' MUSIC-2 and observed relic samples display a very close agreement. In general, symbols tend to cluster at larger 
	orientation angles, more significantly at $\phi \gtrsim 70^{\circ}$; evidently, this effect is stronger at the lowest 
	shape parameters.  
	As shape values increase (the `small-roundish' MUSIC-2 sample is shown using grey filled circles), 
	the clustering of data smooths out, eventually turning into a uniform distribution that 
	spans all possible orientations. This is most evident at shape values of $\lambda_2/\lambda_1 \gtrsim 0.6$, as the islands 
	become roundish and devoid of any preferred orientation.

	%
	The angle and shape distributions for the `elongated' and observed relic samples (upper and right-hand histograms 
	in Fig.~\ref{fig:shape-vs-angle}) are also shown. In both cases, a similarity between the simulated and observed 
	distributions can be seen. In particular, for the orientation angle distribution, the simulated and observed histograms 
	clearly peak towards angles close to $90^{\circ}$ confirming previous findings \citep{2011A&A...533A..35V}. 
	The `elongated' sample is in remarkable agreement with NVSS observations supporting the idea that relic orientation 
	is indeed linked to the direction of the merging structures. In particular, the median orientation angle for the `elongated' 
	and NVSS samples is $\langle \phi\rangle\simeq69^{\circ}$. In the same line, mean shape values of both distributions are 
	virtually identical with $\langle s\rangle\simeq0.31$.

	%
	These results can be interpreted in a simple way: the vast majority of the elongated structures are most likely 
	produced by mergers with low impact parameters (i.e., in head-on collisions or close). 
	In this case, pseudo-spherical surfaces of shocked material sweeping the 
	ICM are generated. When seeing edge-on, these surfaces are observed as arc-like radio features perpendicular to the line 
	joining the relic and cluster centres. This is what one would expect for shock fronts travelling outwards the merger's 
	axis direction.

\begin{figure*}
		\begin{center}
			\includegraphics[width=0.85\textwidth]{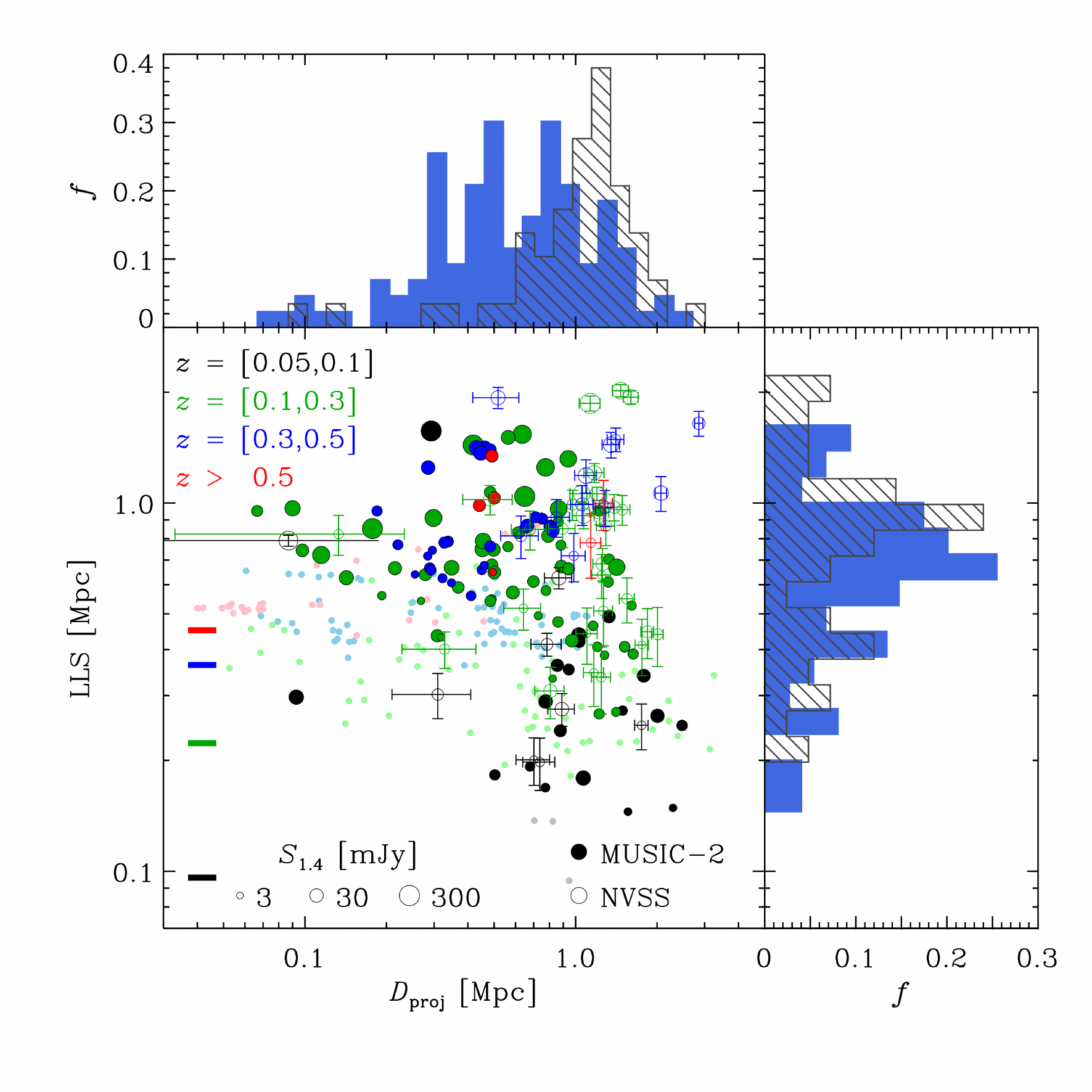} 
			\vspace{-0.8cm}
			\caption{Largest linear size, LLS, versus projected distance on to the 
                          cluster, $D_{\rm proj}$, for the relic samples. 
			  Shown are the NVSS clusters (open circles) and those belonging to the `elongated' 
			  (filled circles) and `small-roundish' (light filled circles) mock relic samples. 
			  Symbol sizes for the NVSS and `elongated' samples are scaled according to radio flux.  
			  In all cases, redshift bins are indicated using different colours. 
                          The upper and right-hand histograms show the comparison between the NVSS (hashed histogram) 
                          and `elongated' (blue histogram) samples for $D_{\rm proj}$ and LLS, respectively.   
			}
			\label{fig:LLS-vs-Dproj}
		\end{center}
	\end{figure*}

	%
	Fig.~\ref{fig:LLS-vs-Dproj} shows the LLS versus the projected distance, $D_\text{proj}$. As before, upper and right-hand histograms 
	of the corresponding quantities are also presented. In MUSIC-2 we obtain relics 
	spanning similar sizes and projected locations within galaxy clusters as those found in the NVSS sample. 
        However, the mock relics tend to be smaller and closer to the cluster centre, which is more evident in the 
	case of the `small-roundish' sample. For the `elongated' and NVSS samples the mean LLS values are $0.7$ and $0.82\,$Mpc. 
	In the case of projected distances differences are higher with mean values of $0.71$ and $1.07\,$Mpc, respectively. 
	These discrepancies become particularly striking when considering distant clusters at $z\gtrsim0.3$, which could originate 
	from the existence of too compact clusters at higher redshifts in our simulations. From the histograms of Fig.~\ref{fig:LLS-vs-Dproj}, 
	it can be seen that the two simulated distributions clearly peak towards smaller values in comparison to the NVSS. 
	A possible explanation of the latter may be related to the lack of ICM heating in our simulations that could artificially 
	increase the Mach number of the shocks. However, further work is needed to decide on this matter (see Section~\ref{sec:num_phys_ICM}).

	Our findings show that radio relics statistics are well suited to investigate the evolution of the ICM, because 
	the Mach number is sensitive to the ICM temperature. We want to stress that, observationally, relics are typically 
	searched as extended diffuse emission in the periphery of galaxy clusters. Therefore, there might also be a 
	classification bias in the NVSS sample, causing that more centrally located objects with the same physical origin 
	as large peripheral relics, are either not detected or not classified as such.

\subsection{Relic abundance}
\label{sec:relic_number}

\begin{figure}
		\hspace{-0.9cm}
		\includegraphics[width=0.53\textwidth]{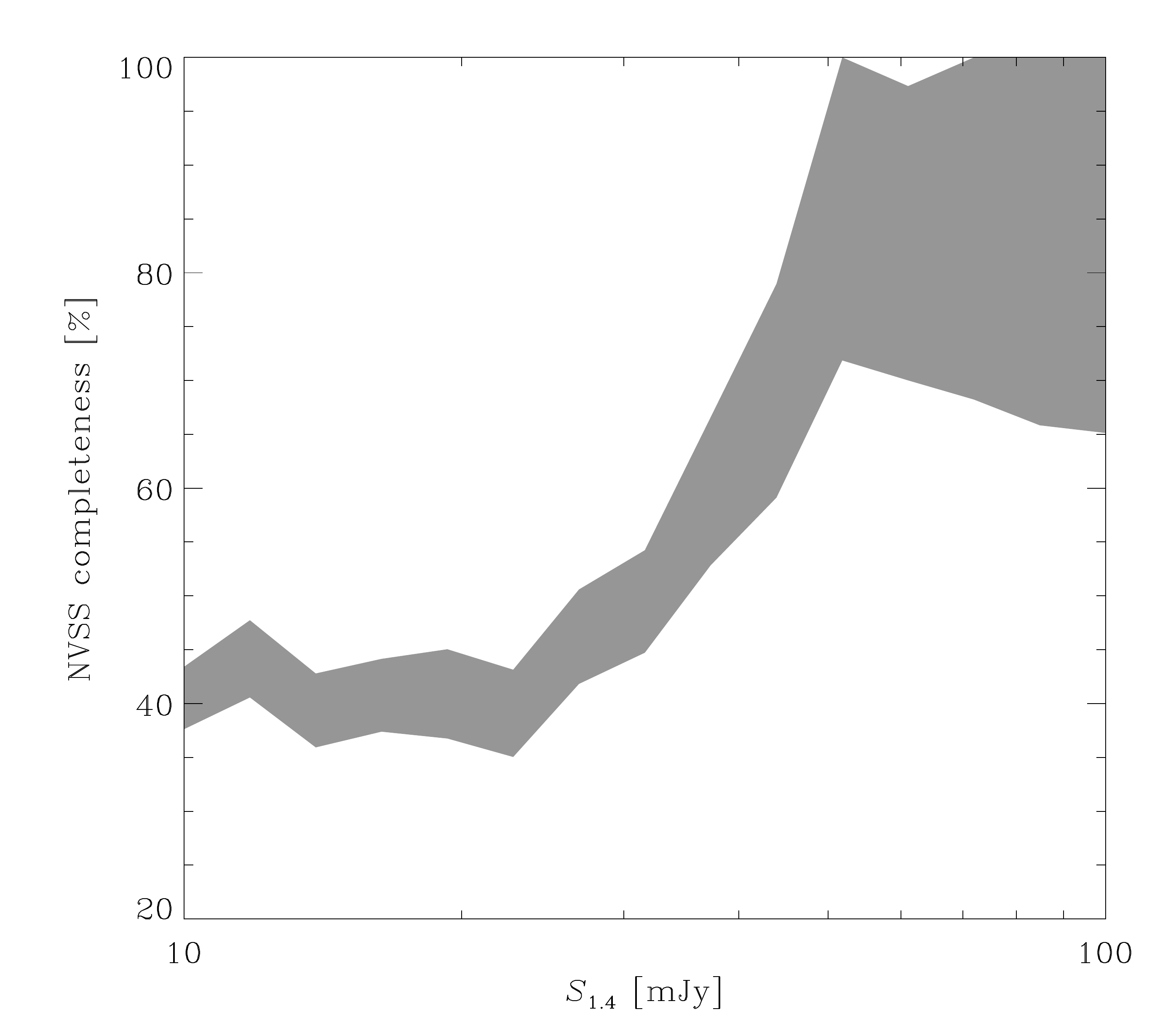}  
		\caption{Estimated completeness of the NVSS relic sample derived by the ratio between the cumulative numbers 
			of NVSS and `elongated' cluster samples (shaded region). The NVSS cluster number is scaled to full sky adopting 
			a sky coverage of 82\%.
		}
		\label{fig:efficiency}
	\end{figure}

	%
	So far, we have not addressed a crucial question, namely, if the MUSIC-2 simulation reproduces the number of relics 
	found in NVSS. The reason to postpone this discussion is that we needed to identify a region of the parameter space 
	where the simulated and observed cluster samples can be compared. The NVSS relic sample is essentially flux limited 
	with clusters spanning almost two decades in X-ray luminosities (see Fig.~\ref{fig:P14-vs-Lx}).
	On the other hand, the MUSIC-2 cluster sample is rather massive, hosting unambiguously detected radio islands well below 
	$10\,$mJy. 
	The comparison between the NVSS and `elongated' mock cluster samples extracted from the MUSIC-2 simulations 
	is shown in Fig.~\ref{fig:FluxClusterCumulative}, 
	where we plot the cumulative number counts with flux density larger than $S_{1.4}$. 
	As discussed in Section~\ref{sec:Xray_clusters}, the MUSIC-2 cluster sample is only complete 
	for $L_{\rm X} \gtrsim 3 \times 10^{44} \: \rm erg \, s^{-1}$, therefore, 
	we exclude all NVSS clusters below this limit from the plot (black dashed curve). We then take into account that NVSS 
	covers only 82\% of the sky and correct for incompleteness (red dashed curve). Evidently, in MUSIC-2 we find significantly 
	more relics than in observations. We emphasize that, in order to properly compare to NVSS, we have already excluded the `small-roundish' 
	islands found in some of the MUSIC-2 clusters. We speculate that the lower abundance of relics in NVSS reflects a flux-dependent 
	completeness. In fact, towards lower flux densities the discrepancy increases. At $S_{1.4}\lesssim10\,$mJy, the 
	MUSIC-2 predicts about two times more relics than found in NVSS (see Fig.~\ref{fig:efficiency}). Interestingly, if we 
	apply the {\it ROSAT} limit to the `elongated' sample we found that the number of clusters hosting relics in the 
	two samples roughly match. This suggests a natural explanation for the excess of objects in the `elongated' sample: 
	a significant fraction of the observed relics have simply not been detected or classified as their host clusters remain unidentified. 
	We note, however, that we are dealing with small number statistics 
	and only future X-ray and radio surveys can confirm or reject this hypothesis.

	%
	We are now in position to further discuss the results of Fig.~\ref{fig:LxClusterCumulative}, where we plotted the 
	cumulative number of clusters hosting relics versus $L_{\rm X}$. As before, results corresponding to the 
	`elongated' mock cluster samples extracted from the MUSIC-2 simulations are indicated by shaded regions.
	To properly compare with observations, we have excluded all mock clusters hosting relics with a flux below the NVSS threshold, 
	which, we set equal to $7\,$mJy (see solid line in Fig.~\ref{fig:FluxClusterCumulative}). 
	The NVSS and its all-sky corrected distributions are also shown (solid and red dashed lines, respectively). 
	As before, the mock cluster counts are larger than for observations by roughly 
	a factor of 2 (filled circles) and they nicely match if we exclude clusters below the {\it ROSAT} detection limit (open circles). 
	Interestingly, a few NVSS clusters have higher X-ray luminosities ($L_{\rm X}\gtrsim 10^{45}\,$erg\,s$^{-1}$) than for the `elongated' 
	sample. Two effects can contribute to this. We have adopted 
	a general mass-luminosity relation to assign X-ray luminosities to the simulated clusters. During a merger event, however, the cluster 
	X-ray luminosity can significantly deviate from the scaling relation and the system may boost its X-ray emission.  
	Thus, we possibly slightly underestimate the X-ray luminosities. Moreover, due to the finite size of the simulation volume, 
	we miss very massive clusters. This is particularly important at higher redshifts, where the observable volume is much 
	larger than the simulated one. 
	
	%
	As a final remark, we note that the abundance of relics in our model is essentially controlled by the magnetic field and the efficiency 
	parameter. For the adopted magnetic field model, we have deliberately chosen an efficiency parameter 
	of $\xi_{\rm e}=5\times10^{-5}$ to match the NVSS cluster counts at $S_{1.4}=100\,$mJy, a flux for which it is assumed 
	that almost all relics are known. As shown above, the resulting sample of mock relics reproduce the NVSS observed correlations 
	reasonably well. The most significant mismatch in the relic properties has been found for the sizes and projected distances of 
	relics in distant clusters at $z \gtrsim 0.3$. Clearly, a different magnetic field model --assuming another efficiency 
	parameter-- may have an impact on the size and location of relics. It is beyond the scope of the present work 
	to explore this in detail.

\subsection{Radio luminosity versus cluster mass}
\label{sec:radiopower-mass}

\begin{figure}
		\hspace{-0.8cm}
			\includegraphics[width=0.53\textwidth]{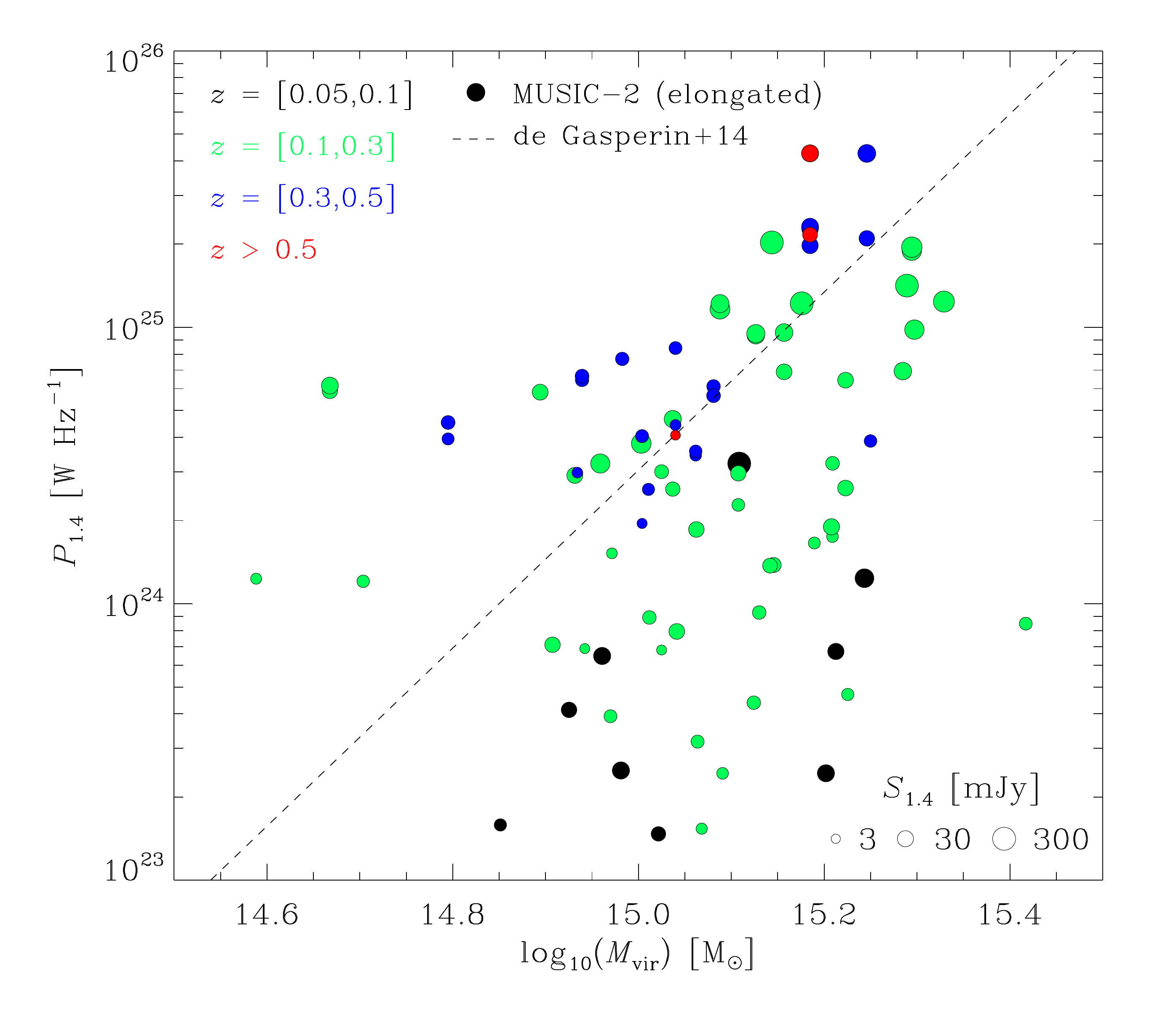} 
			\caption{Radio power, $P_{1.4}$, versus cluster virial mass, $M_{\rm vir}$, for 
				the `elongated' mock sample (filled circles). 
				Symbol sizes are scaled according to radio flux.  
				Redshift bins are indicated using different colours. 
				For reference, we also include the \citet{2014MNRAS.444.3130D} relation after 
				rescaling cluster masses to the virial radius (dashed line).   
			}
			\label{fig:P14-vs-Mvir}
	\end{figure}

	%
	The strong correlation between radio power and X-ray luminosity may also indicate that more massive clusters 
	could generate more luminous radio relics \citep{2014MNRAS.444.3130D}. We explore this scaling in Fig.~\ref{fig:P14-vs-Mvir}, 
	which shows total radio power versus cluster mass in the MUSIC-2 `elongated' sample. As before, we also 
	include the scaled best-fitting relation of \citet{2014MNRAS.444.3130D} for comparison.  
	The detection threshold (i.e., the radio flux limit) increases with redshift, 
	as reflected by the colour-coding of symbols. In particular, the interval $z\in[0.1,0.3]$ covers a fairly small redshift range 
	while having enough statistics to allow us investigate the existence of a luminosity--mass 
	correlation free of any Malmquist bias effect. It is evident that the most massive clusters in the sample host the most 
	luminous relics. Still, the scatter in radio luminosity is large. Therefore, less massive clusters 
	can eventually produce similarly bright relics as those with larger mass, 
	see e.g. luminosities corresponding to $M_\text{vir}\sim 10^{14.9}$ and $10^{15.2}\,$M$_{\odot}$ in Fig.~\ref{fig:P14-vs-Mvir}.

	%
	As mentioned above, it is evident that the radio power normalization of relics in more 
	distant clusters is higher. According to this, a correlation between radio luminosity and cluster mass would imply that only 
	the bright relics in massive clusters at high redshift are detected. As a result, the detection threshold could amplify the 
	observed correlation. 
	%
	The mass range covered in the MUSIC-2 simulations is too small to derive an average luminosity--mass correlation for 
	small redshift intervals, i.e. free of Malmquist bias. We speculate, however, that the {\it maximum} radio luminosity 
	at a given mass actually {\it increases} with cluster mass, but probably less steeply than found for samples affected 
	by the latter.

\section{Discussion}
\label{sec:Discussion}

\subsection{The `small-roundish' sample}

	%
	The small-roundish sample is, basically, exclusively found in MUSIC-2 with islands having the following properties: 
	(i) they are small and roundish with $s/\text{LAS} > 0.17$ (see Fig.~\ref{fig:Shape-vs-LAS}); (ii) they tend to be found 
	at higher redshifts, hence, predominantly residing in clusters with low X-ray flux; (iii) they show median values of LLS and 
	projected distance of about $500$ and $400\,$kpc respectively, and (iv) the median radio luminosities of these 
	objects are higher than in observations.

	%
	In the NVSS sample there are only a few objects in the `small-roundish' regime, namely, the recently discovered 
	gischt double candidate in CIZA\,J0107$+$54 \citep{2016ApJ...823...94R}; the north-western edge emission in Abell\,1682 
	\citep{2013A&A...551A..24V}; the southern steep-spectrum feature in the `Toothbrush' cluster \citep{2016ApJ...818..204V}, 
	and the recently discovered gischt candidate in PLCK\,G004$-$19 \citep{2014A&A...562A..43S}. All of these objects are also rather small in 
	size and the islands in Abell\,1682, the `Toothbrush' cluster, and PLCK\,G004$-$19 do not show the typical relic morphology. 
	This may indicate that the identification of these kind of objects in many NVSS clusters is not free of difficulties. Therefore, it is 
	possible that many of them are still not yet discovered. In particular, we found about 100 of these `small-roundish' islands in the MUSIC-2 
	simulations within the context of our model.

\subsection{Numerical scheme and ICM physics}
\label{sec:num_phys_ICM}

	%
	The abundance and location of shocks fronts is affected by the particular numerical scheme used. 
	For instance, \citet{2011MNRAS.418..960V} have shown that Eulerian codes tend to have a smaller fraction of shock fronts 
	with Mach numbers of $2-4$ in low-entropy gas compared to standard SPH simulations. This suggest that an Eulerian 
	numerical scheme would result in less shock fronts close to the cluster centre. However, recent SPH implementations have 
	been shown to better treat fluid mixing thus alleviating these problems \citep[e.g.,][]{2013MNRAS.428.2840H}.

	%
	The physical state of the ICM is also affected by galactic and AGN feedback, which is not included in the simulation we 
	use here. It has been shown that, in average, the feedback increases the ICM temperature \citep[see e.g.,][]{2016MNRAS.459.2973S}.
	This is particularly important for the inner parts of the cluster. A higher ICM temperature will result in lower Mach numbers 
	and, hence, make central radio relics less likely.

	%
	Therefore, the two major discrepancies found between the MUSIC-2 and the NVSS sample, i.e. too many small-roundish objects 
	and the presence of high-redshift elongated relics with somewhat smaller projected distances and sizes, might be partly 
	spurious due to the numerical scheme and missing physics of the simulation. However, there are also several additional ingredients 
	affecting the matching between the two samples: the uncertain completeness of the NVSS, the magnetic field distribution in the 
	ICM and the dependence of the radio luminosity on shock parameters. We have to leave to future studies the assessment of the impact 
	of all these factors in the comparison between observations and simulations.

\subsection{On the NVSS completeness}

	%
	Several fairly bright radio relics have not been discovered until recently, e.g. the relics in PLCK\,G287$+$32 
	\citep{2011ApJ...736L...8B} and PSZ1\,G108$-$11 \citep{2015MNRAS.453.3483D} with a flux density of $58\,$mJy and $113\,$mJy, respectively. 
	These objects are well visible in NVSS as extended sources. Evidently, since the clusters were not known prior to $Planck$ 
	observations, the emission was not classified as radio gischt. This suggests that there might be unclassified single or 
	double gischt objects already visible in NVSS. 
	
	The comparison between the NVSS and MUSIC-2 samples indicates that only about 
	40\% of the extended relics with a total flux smaller than 10\,mJy have been discovered in X-ray luminous clusters with 
	$L_X \gtrsim 3 \times 10^{44} \: \rm erg \, s^{-1}$. For example, the double gischt in Abell\,1240 is rather 
	difficult to identify in NVSS because each relic has a flux of about $5\,$mJy (see Fig.~\ref{fig:NVSSimage}). 
	The flux of the entire gischt emission is distributed over at least 10 beams. Only part of the 
	relic is recovered by the $2\times \sigma_\text{NVSS}$ contours. This illustrates the fact that systematic searches 
	for radio relics at $10\,$mJy and below in NVSS were incomplete.

\subsection{On the gischtlet class}

	%
	In Section~\ref{sec:relicCompile}, we introduced a class of radio relics with the same physical origin of gischts 
	but with a different morphology: the {\it gischtlets}. In contrast to classical gischts, these objects are rather small, 
	are located close to the cluster centre, and may show complex morphologies. Reference cases might be the ridges found 
	in Abell\,1682 \citep{2013A&A...551A..24V} and the large but central `chair-shaped' filament in MACS\,0717$+$37 
	\citep{2009A&A...503..707B,2009A&A...505..991V,2017ApJ...835..197V}. In both cases, the morphology does not obviously 
	agree with that of large merger shock fronts travelling outwards.

	%
	Fig.~\ref{fig:relicCollection} shows a few objects with rather unusual morphology that could lie within this classification, 
	e.g. the relic in cluster `B' that is rather close to the cluster centre and that of cluster `D' that is strangely not 
	aligned with the apparent merger axis. Hence, a non-negligible fraction of the objects found in the MUSIC-2 sample does not 
	resemble a classical large gischt caused by a binary merger. This again illustrates that the classification based on morphology 
	may cause that some relics are not recognized as a result of their unusual appearance.

\subsection{Upcoming surveys}

	Our approach is applied to the NVSS sample of relics, yet it is not tied to any specific survey. There are several upcoming 
	radio wide-field surveys, each yielding a large increase in both sensitivity and resolution compared to NVSS.

	For instance, Apertif-Wodan and ASKAP-EMU \citep{2011PASA...28..215N} have similar survey specifications and are highly 
	complementary because of their low-intersecting sky coverage. These surveys are going to probe the continuum radio sky at 
	frequencies of $1100-1400\,$MHz. With an expected survey depth of $10\,$mJy and a $10\,$arcsec resolution, the sensitivity for 
	extended structures will exceed NVSS by more than a factor of $10$.

	The LOFAR radio telescope is going to probe the northern radio sky at $120-168\,$MHz in three different 
	stages or `tiers'. Tier I will be the most shallow but largest survey with $5\,$arcsec resolution and $0.1\,$mJy 
	noise per beam. Sensitivity for both point sources and extended structures will exceed that of the 
	GMRT $150\,$MHz all-sky radio survey 
	\citep{2017A&A...598A..78I} by, at least, an order of magnitude, translating into 10 times better sensitivity 
	for extended objects with a spectral index of $\alpha=-1$.

	These surveys are expected to significantly increase the number of known radio relics. Ultimately, the relic sample will 
	be expanded towards lower surface brightnesses. The combination of all these new observations, together with more sophisticated 
	large-volume cosmological simulations, will allow for a more thorough comparison between models and reality. 
	Additionally, synergies with panchromatic surveys (e.g., X-ray; SZ) 
	and optical follow-up observations, will be necessary to identify galaxy clusters and determine their redshift.

	In this work, we explored one scenario to model the non-thermal radio emission of radio relics. However, our approach is 
	well suited to discriminate between different models, not only DSA, but also more complex acceleration mechanisms. 
	We leave this for future research.

\section{Summary}
\label{sec:Summary}

	The origin of radio relics is quite enigmatic. There is ample evidence indicating that relics are located at shock fronts 
	originated in merging galaxy clusters. However, it is not clear if all merger shocks are, in turn, able to produce radio relics. 
	To shed light on this issue, we compared the flux, size, morphology, orientation and location of mock relic structures 
	linked to simulated merger shocks with an homogeneous sample of known radio relics. 
	Moreover, the simulated host galaxy clusters were used to further study the well-known correlation 
	between radio power and X-ray luminosity and its most recent version involving cluster mass \citep{2014MNRAS.444.3130D}.

	%
	To perform the relic comparison, we compiled a comprehensive list of radio relics and candidates based on previous collections 
	\citep{2012A&ARv..20...54F,2012MNRAS.420.2006N,2015ApJ...813...77Y} and recently published relics. 
	We distinguished radio phoenixes and gischts according to the classification given in the literature. 
	In this work, a radio gischt is assumed to originate from cosmological merger shock fronts.

	%
	For all possible relics in our list, we determined their properties in NVSS images. To define the boundary of radio islands, we 
	used $2 \times \sigma_\text{NVSS}$ contours, where $\sigma_\text{NVSS}$ is the sensitivity limit of the NVSS. For each island, 
	we measured flux density, LAS, LLS, solid angle, shape based on image moments and orientation with 
	respect to the cluster. In total, we identified 59 islands in 39 galaxy clusters. A crucial aspect of our approach is to 
	determine relic parameters in the most homogeneous possible way.

	%
	To model relics, we used 282 resimulated galaxy cluster regions extracted from a cosmological comoving volume 
	of $1\,h^{-3}\,$Gpc$^3$ comprising the so-called MUSIC-2 simulations. The hydrodynamical evolution of the clusters has been 
	simulated using an SPH code including non-radiative physics only. For our purpose, it is mandatory to have as many massive clusters 
	as possible from a large cosmological volume. Still, every cluster has to be simulated with sufficient resolution to properly 
	describe the shock fronts. We applied the shock finder presented by \citet{2008MNRAS.391.1511H} to each snapshot in post-processing 
	to find the shock fronts and to estimate their associated radio emission.

	%
	For lighting up the shocks, we used the radio-emission model and parameters given by \citet{2012MNRAS.420.2006N}. 
	In particular, we adopted the \cite{Bonafede10} scaling (see \citealt{2012MNRAS.420.2006N}, their model `b') for estimating 
	the magnetic field strength as a function of local gas density and $\xi_{\rm e}=5\times10^{-5}$ for the efficiency parameter. 
	The latter has been chosen to restore the number of bright relics found in the NVSS sample.

	%
	For each snapshot, we carried out mock cluster observations adopting NVSS specifications and determined regions 
	of radio emission in the same way as for NVSS images. To account for the fact that the observable volume of the universe 
	increases with redshift, we computed a probability quantifying how often any given simulated cluster should be considered 
	part of the sample. To avoid duplicating relic parameters, we randomly rotated each cluster every time it enters more than 
	once into the mock sample. 
	
	Our main findings can be summarized as follows: 
	
	\begin{itemize}
	
	%
	\item For both the NVSS and the mock relic samples, we found that the shape, $s$, and LAS (anti-)correlate in a similar way. 
	In particular, mean shape values for these two samples are virtually identical with $\langle s\rangle\simeq0.31$.
	This shows that our simulations produce relics with a similar appearance to observations. Interestingly, our 
	mock relic sample does not show the presence of a significant population of large {\it and} roundish relics. This suggests 
	that current radio halo samples, which comprise diffuse radio emssion following the cluster X-ray distribution, 
	are not significantly contaminated by radio relics seen face-on.
	\\

	%
	\item In the mock relic sample, we found many `small-roundish' islands fulfilling the condition $s/\text{LAS} > 0.17$. 
	Only very few radio islands with similar properties are found in the NVSS sample. This `small-roundish' sample comprises 
	some compact features but also more extended relics in distant clusters. We speculate that compact features in clusters 
	are spurious being, most likely, the result of a combination of the simulation technique and the lack of energy feedback 
	in our simulations.  
	However, extended objects at high redshift may reflect realistic merger shocks that are not detected in NVSS 
	since they have a small LAS, they appear almost roundish, and most importantly, their host clusters have not yet 
	been discovered. The relic candidates found in CIZA\,J0107$+$54 are an example of `small-roundish' double relics in NVSS. 
	The MUSIC-2 mock relic sample suggests that many more similar, but not yet identified, objects should be present in NVSS. 
	\\
	
	%
	\item The mock `elongated' sample nicely reproduces the correlation between radio power and LLS. We argue that this correlation 
	is equivalent to the finding that the average surface brightness of all relics is located in a fairly narrow range. 
	This small range may originate from the fact that deeper observations usually reveal more regions with low surface brightness 
	thus increasing the relic extension and balancing the brightening. 
	\\
	
	\item Consistent with previous studies, we found that relics in the NVSS sample tend to be tangentially oriented with respect 
	to the axis joining the relic and cluster centres. This suggests that the vast majority of structures found in NVSS are most 
	likely produced in head-on cluster collisions or close. We confirm this result using our simulations. In particular, elongated 
	structures both in NVSS and the mock relic sample are typically clustered at $(s,\phi)=(\lesssim0.5,\gtrsim60^{\circ})$ in 
	the shape-orientation plane. For the `elongated' and NVSS samples, the median orientation angle turns out to be $\phi\simeq69^{\circ}$. 
	As expected, this signal is essentially lost when considering more roundish objects also present 
	in our simulations.  
	\\

	%
	\item We found that the mock `elongated' relic sample spans similar LLS and projected distance ($D_{\rm proj}$) values 
	than NVSS. However, in average, `elongated' mock relics tend to have somewhat smaller sizes and to be 
	located closer to the cluster centre in comparison to observations 
	(i.e., $\langle {\rm LLS}\rangle=0.7,\,0.82\,$Mpc and $\langle D_{\rm proj}\rangle=0.71,\,1.07\,$Mpc for the `elongated' 
	and NVSS samples, respectively). 
	Moreover, these discrepancies become larger when considering more 
	distant relics at $z\gtrsim0.3$. A possible explanation to this might be given by the lack of supernova and AGN feedback in 
	our simulations.  
	Energy feedback would heat the ICM and increase the sound speed in the central parts of the simulated galaxy clusters. 
	Nonetheless, further work is needed to decide on this issue. 
	\\
	
	%
	\item We found that the well-known radio power versus X-ray cluster luminosity correlation is to a large extent determined by the 
	flux detection threshold both in radio and X-ray. Similarly, the related radio power versus cluster mass 
	correlation recently proposed is artificially strengthened by the flux limits: at higher redshifts, bright relics in massive 
	clusters are preferentially detected. Since scatter in these correlations is large, our simulations suggest that 
	the {\it peak} radio luminosity actually scales with mass but in a less-steep way than found for 
	Malmquist bias-affected samples. 
	\\
	  
	%
	\item We estimated a flux-dependent completeness of the NVSS relic sample of about 40\% at $10\,$mJy. We note that this 
	might be partly spurious since the simulation may overpredict low-luminosity relics. However, we argue that the 
	derived efficiency is plausible given the fact that $10\,$mJy relics are difficult to discover in NVSS. 
	Furthermore, if we apply the {\it ROSAT} X-ray flux limit to the `elongated' mock relic sample --which is basically the limit 
	shown by NVSS clusters-- we roughly reproduce the observed cluster number counts. If it is true that the discovery 
	of radio relics is currently limited by the number of available X-ray clusters used to identify them, 
	this would provide additional support to our conclusion on the completeness of the NVSS sample. 
	\end{itemize}
	
	%
	Overall the mock `elongated' gischt sample reproduces the NVSS-derived relic properties reasonably well, especially 
	regarding shape, orientation, flux, radio luminosity and cluster X-ray luminosity distributions, when taking into account all known 
	biases of the samples. 
	This suggests that the morphology and abundance of merger shocks in a cosmological framework 
	is, in principle, enough to explain radio relics. These findings may rule out the scenario where merger shocks 
	are only partly illuminated by pre-existing radio plasma. If such a plasma was necessary to explain {\it all} radio relics, 
	it would have to be distributed rather homogeneously in all galaxy clusters.

	Clearly, upcoming deeper radio surveys, such as LOFAR-Tier-1, EMU or Apertif, will significantly enhance the sample of 
	known radio relics. Exploiting these data with the aid of more advanced theoretical models and simulations will help us 
	to better understand the elusive nature of these objects.

\section*{Acknowledgements}

The authors thank the anonymous referee for his/her constructive comments. 
The MUSIC simulations have been performed in the MareNostrum supercomputer at the
Barcelona Supercomputing Centre thanks to the computing time awarded by the Red Espa\~nola de Supercomputaci\'on. 
SEN, JG and MH acknowledge support by the Deutsche Forschungsgemeinschaft under grant NU 332/2-1 
and research group FOR 1254.
GY acknowledges support from MINECO/FEDER under grants AYA2012-31101 and AYA2015-63810-P.

\bibliography{references}

\begin{thebibliography}{}
\makeatletter
\relax
\def\mn@urlcharsother{\let\do\@makeother \do\$\do\&\do\#\do\^\do\_\do\%\do\~}
\def\mn@doi{\begingroup\mn@urlcharsother \@ifnextchar [ {\mn@doi@}
  {\mn@doi@[]}}
\def\mn@doi@[#1]#2{\def\@tempa{#1}\ifx\@tempa\@empty \href
  {http://dx.doi.org/#2} {doi:#2}\else \href {http://dx.doi.org/#2} {#1}\fi
  \endgroup}
\def\mn@eprint#1#2{\mn@eprint@#1:#2::\@nil}
\def\mn@eprint@arXiv#1{\href {http://arxiv.org/abs/#1} {{\tt arXiv:#1}}}
\def\mn@eprint@dblp#1{\href {http://dblp.uni-trier.de/rec/bibtex/#1.xml}
  {dblp:#1}}
\def\mn@eprint@#1:#2:#3:#4\@nil{\def\@tempa {#1}\def\@tempb {#2}\def\@tempc
  {#3}\ifx \@tempc \@empty \let \@tempc \@tempb \let \@tempb \@tempa \fi \ifx
  \@tempb \@empty \def\@tempb {arXiv}\fi \@ifundefined
  {mn@eprint@\@tempb}{\@tempb:\@tempc}{\expandafter \expandafter \csname
  mn@eprint@\@tempb\endcsname \expandafter{\@tempc}}}

\bibitem[\protect\citeauthoryear{{Akamatsu} et~al.,}{{Akamatsu}
  et~al.}{2015}]{2015A&A...582A..87A}
{Akamatsu} H.,  et~al., 2015, \mn@doi [\aap] {10.1051/0004-6361/201425209},
  \href {http://adsabs.harvard.edu/abs/2015A%26A...582A..87A} {582, A87}

\bibitem[\protect\citeauthoryear{{Bagchi}, {Durret}, {Neto}  \&
  {Paul}}{{Bagchi} et~al.}{2006}]{2006Sci...314..791B}
{Bagchi} J.,  {Durret} F.,  {Neto} G.~B.~L.,   {Paul} S.,  2006, \mn@doi
  [Science] {10.1126/science.1131189}, \href
  {http://adsabs.harvard.edu/abs/2006Sci...314..791B} {314, 791}

\bibitem[\protect\citeauthoryear{{Bagchi} et~al.,}{{Bagchi}
  et~al.}{2011}]{2011ApJ...736L...8B}
{Bagchi} J.,  et~al., 2011, \mn@doi [\apjl] {10.1088/2041-8205/736/1/L8}, \href
  {http://adsabs.harvard.edu/abs/2011ApJ...736L...8B} {736, L8}

\bibitem[\protect\citeauthoryear{{Basu}, {Vazza}, {Erler}  \& {Sommer}}{{Basu}
  et~al.}{2016}]{2016A&A...591A.142B}
{Basu} K.,  {Vazza} F.,  {Erler} J.,   {Sommer} M.,  2016, \mn@doi [\aap]
  {10.1051/0004-6361/201527726}, \href
  {http://adsabs.harvard.edu/abs/2016A%26A...591A.142B} {591, A142}

\bibitem[\protect\citeauthoryear{{Becker}, {White}  \& {Helfand}}{{Becker}
  et~al.}{1995}]{Becker95}
{Becker} R.~H.,  {White} R.~L.,   {Helfand} D.~J.,  1995, \mn@doi [\apj]
  {10.1086/176166}, \href {http://adsabs.harvard.edu/abs/1995ApJ...450..559B}
  {450, 559}

\bibitem[\protect\citeauthoryear{{Bertin} \& {Arnouts}}{{Bertin} \&
  {Arnouts}}{1996}]{1996A&AS..117..393B}
{Bertin} E.,  {Arnouts} S.,  1996, \mn@doi [\aaps] {10.1051/aas:1996164}, \href
  {http://adsabs.harvard.edu/abs/1996A%26AS..117..393B} {117, 393}

\bibitem[\protect\citeauthoryear{{Blandford} \& {Eichler}}{{Blandford} \&
  {Eichler}}{1987}]{1987PhR...154....1B}
{Blandford} R.,  {Eichler} D.,  1987, \mn@doi [\physrep]
  {10.1016/0370-1573(87)90134-7}, \href
  {http://adsabs.harvard.edu/abs/1987PhR...154....1B} {154, 1}

\bibitem[\protect\citeauthoryear{{B{\"o}hringer} et~al.,}{{B{\"o}hringer}
  et~al.}{2000}]{2000ApJS..129..435B}
{B{\"o}hringer} H.,  et~al., 2000, \mn@doi [\apjs] {10.1086/313427}, \href
  {http://adsabs.harvard.edu/abs/2000ApJS..129..435B} {129, 435}

\bibitem[\protect\citeauthoryear{{B{\"o}hringer}, {Chon}  \&
  {Collins}}{{B{\"o}hringer} et~al.}{2014}]{2014A&A...570A..31B}
{B{\"o}hringer} H.,  {Chon} G.,   {Collins} C.~A.,  2014, \mn@doi [\aap]
  {10.1051/0004-6361/201323155}, \href
  {http://adsabs.harvard.edu/abs/2014A%26A...570A..31B} {570, A31}

\bibitem[\protect\citeauthoryear{{Bonafede}, {Giovannini}, {Feretti}, {Govoni}
  \& {Murgia}}{{Bonafede} et~al.}{2009a}]{2009A&A...494..429B}
{Bonafede} A.,  {Giovannini} G.,  {Feretti} L.,  {Govoni} F.,   {Murgia} M.,
  2009a, \mn@doi [\aap] {10.1051/0004-6361:200810588}, \href
  {http://adsabs.harvard.edu/abs/2009A%26A...494..429B} {494, 429}

\bibitem[\protect\citeauthoryear{{Bonafede} et~al.,}{{Bonafede}
  et~al.}{2009b}]{2009A&A...503..707B}
{Bonafede} A.,  et~al., 2009b, \mn@doi [\aap] {10.1051/0004-6361/200912520},
  \href {http://adsabs.harvard.edu/abs/2009A%26A...503..707B} {503, 707}

\bibitem[\protect\citeauthoryear{{Bonafede}, {Feretti}, {Murgia}, {Govoni},
  {Giovannini}, {Dallacasa}, {Dolag}  \& {Taylor}}{{Bonafede}
  et~al.}{2010}]{Bonafede10}
{Bonafede} A.,  {Feretti} L.,  {Murgia} M.,  {Govoni} F.,  {Giovannini} G.,
  {Dallacasa} D.,  {Dolag} K.,   {Taylor} G.~B.,  2010, \mn@doi [\aap]
  {10.1051/0004-6361/200913696}, \href
  {http://adsabs.harvard.edu/abs/2010A%26A...513A..30B} {513, A30+}

\bibitem[\protect\citeauthoryear{{Bonafede} et~al.,}{{Bonafede}
  et~al.}{2012}]{2012MNRAS.426...40B}
{Bonafede} A.,  et~al., 2012, \mn@doi [\mnras]
  {10.1111/j.1365-2966.2012.21570.x}, \href
  {http://adsabs.harvard.edu/abs/2012MNRAS.426...40B} {426, 40}

\bibitem[\protect\citeauthoryear{{Bonafede}, {Vazza}, {Br{\"u}ggen}, {Murgia},
  {Govoni}, {Feretti}, {Giovannini}  \& {Ogrean}}{{Bonafede}
  et~al.}{2013}]{2013MNRAS.433.3208B}
{Bonafede} A.,  {Vazza} F.,  {Br{\"u}ggen} M.,  {Murgia} M.,  {Govoni} F.,
  {Feretti} L.,  {Giovannini} G.,   {Ogrean} G.,  2013, \mn@doi [\mnras]
  {10.1093/mnras/stt960}, \href
  {http://adsabs.harvard.edu/abs/2013MNRAS.433.3208B} {433, 3208}

\bibitem[\protect\citeauthoryear{{Bonafede} et~al.,}{{Bonafede}
  et~al.}{2015}]{2015MNRAS.454.3391B}
{Bonafede} A.,  et~al., 2015, \mn@doi [\mnras] {10.1093/mnras/stv2065}, \href
  {http://adsabs.harvard.edu/abs/2015MNRAS.454.3391B} {454, 3391}

\bibitem[\protect\citeauthoryear{{Boschin}, {Barrena}, {Girardi}  \&
  {Spolaor}}{{Boschin} et~al.}{2008}]{2008A&A...487...33B}
{Boschin} W.,  {Barrena} R.,  {Girardi} M.,   {Spolaor} M.,  2008, \mn@doi
  [\aap] {10.1051/0004-6361:200809620}, \href
  {http://adsabs.harvard.edu/abs/2008A%26A...487...33B} {487, 33}

\bibitem[\protect\citeauthoryear{{Botteon}, {Gastaldello}, {Brunetti}  \&
  {Dallacasa}}{{Botteon} et~al.}{2016}]{2016MNRAS.460L..84B}
{Botteon} A.,  {Gastaldello} F.,  {Brunetti} G.,   {Dallacasa} D.,  2016,
  \mn@doi [\mnras] {10.1093/mnrasl/slw082}, \href
  {http://adsabs.harvard.edu/abs/2016MNRAS.460L..84B} {460, L84}

\bibitem[\protect\citeauthoryear{{Brown}, {Duesterhoeft}  \& {Rudnick}}{{Brown}
  et~al.}{2011}]{2011ApJ...727L..25B}
{Brown} S.,  {Duesterhoeft} J.,   {Rudnick} L.,  2011, \mn@doi [\apjl]
  {10.1088/2041-8205/727/1/L25}, \href
  {http://adsabs.harvard.edu/abs/2011ApJ...727L..25B} {727, L25+}

\bibitem[\protect\citeauthoryear{{Cantwell}, {Scaife}, {Oozeer}, {Wen}  \&
  {Han}}{{Cantwell} et~al.}{2016}]{2016MNRAS.458.1803C}
{Cantwell} T.~M.,  {Scaife} A.~M.~M.,  {Oozeer} N.,  {Wen} Z.~L.,   {Han}
  J.~L.,  2016, \mn@doi [\mnras] {10.1093/mnras/stw419}, \href
  {http://adsabs.harvard.edu/abs/2016MNRAS.458.1803C} {458, 1803}

\bibitem[\protect\citeauthoryear{{Caprioli}}{{Caprioli}}{2012}]{2012JCAP...07..038C}
{Caprioli} D.,  2012, \mn@doi [\jcap] {10.1088/1475-7516/2012/07/038}, \href
  {http://adsabs.harvard.edu/abs/2012JCAP...07..038C} {7, 038}

\bibitem[\protect\citeauthoryear{{Clarke} \& {Ensslin}}{{Clarke} \&
  {Ensslin}}{2006}]{2006AJ....131.2900C}
{Clarke} T.~E.,  {Ensslin} T.~A.,  2006, \mn@doi [\aj] {10.1086/504076}, \href
  {http://adsabs.harvard.edu/abs/2006AJ....131.2900C} {131, 2900}

\bibitem[\protect\citeauthoryear{{Cohen} \& {Clarke}}{{Cohen} \&
  {Clarke}}{2011}]{2011AJ....141..149C}
{Cohen} A.~S.,  {Clarke} T.~E.,  2011, \mn@doi [\aj]
  {10.1088/0004-6256/141/5/149}, \href
  {http://adsabs.harvard.edu/abs/2011AJ....141..149C} {141, 149}

\bibitem[\protect\citeauthoryear{{Condon}, {Cotton}, {Greisen}, {Yin},
  {Perley}, {Taylor}  \& {Broderick}}{{Condon}
  et~al.}{1998}]{1998AJ....115.1693C}
{Condon} J.~J.,  {Cotton} W.~D.,  {Greisen} E.~W.,  {Yin} Q.~F.,  {Perley}
  R.~A.,  {Taylor} G.~B.,   {Broderick} J.~J.,  1998, \mn@doi [\aj]
  {10.1086/300337}, \href {http://adsabs.harvard.edu/abs/1998AJ....115.1693C}
  {115, 1693}

\bibitem[\protect\citeauthoryear{{Dolag}, {Schindler}, {Govoni}  \&
  {Feretti}}{{Dolag} et~al.}{2001}]{2001A&A...378..777D}
{Dolag} K.,  {Schindler} S.,  {Govoni} F.,   {Feretti} L.,  2001, \mn@doi
  [\aap] {10.1051/0004-6361:20011219}, \href
  {http://adsabs.harvard.edu/abs/2001A%26A...378..777D} {378, 777}

\bibitem[\protect\citeauthoryear{{Donnert}, {Stroe}, {Brunetti}, {Hoang}  \&
  {Roettgering}}{{Donnert} et~al.}{2016}]{2016MNRAS.462.2014D}
{Donnert} J.~M.~F.,  {Stroe} A.,  {Brunetti} G.,  {Hoang} D.,   {Roettgering}
  H.,  2016, \mn@doi [\mnras] {10.1093/mnras/stw1792}, \href
  {http://adsabs.harvard.edu/abs/2016MNRAS.462.2014D} {462, 2014}

\bibitem[\protect\citeauthoryear{{Drury}}{{Drury}}{1983}]{1983SSRv...36...57D}
{Drury} L.,  1983, \mn@doi [\ssr] {10.1007/BF00171901}, \href
  {http://adsabs.harvard.edu/abs/1983SSRv...36...57D} {36, 57}

\bibitem[\protect\citeauthoryear{{Duffy}, {Schaye}, {Kay}  \& {Dalla
  Vecchia}}{{Duffy} et~al.}{2008}]{2008MNRAS.390L..64D}
{Duffy} A.~R.,  {Schaye} J.,  {Kay} S.~T.,   {Dalla Vecchia} C.,  2008, \mn@doi
  [\mnras] {10.1111/j.1745-3933.2008.00537.x}, \href
  {http://adsabs.harvard.edu/abs/2008MNRAS.390L..64D} {390, L64}

\bibitem[\protect\citeauthoryear{{Ebeling}, {Voges}, {Bohringer}, {Edge},
  {Huchra}  \& {Briel}}{{Ebeling} et~al.}{1996}]{1996MNRAS.281..799E}
{Ebeling} H.,  {Voges} W.,  {Bohringer} H.,  {Edge} A.~C.,  {Huchra} J.~P.,
  {Briel} U.~G.,  1996, \mnras, \href
  {http://adsabs.harvard.edu/abs/1996MNRAS.281..799E} {281, 799}

\bibitem[\protect\citeauthoryear{{Ebeling}, {Edge}, {Bohringer}, {Allen},
  {Crawford}, {Fabian}, {Voges}  \& {Huchra}}{{Ebeling}
  et~al.}{1998}]{1998MNRAS.301..881E}
{Ebeling} H.,  {Edge} A.~C.,  {Bohringer} H.,  {Allen} S.~W.,  {Crawford}
  C.~S.,  {Fabian} A.~C.,  {Voges} W.,   {Huchra} J.~P.,  1998, \mn@doi
  [\mnras] {10.1046/j.1365-8711.1998.01949.x}, \href
  {http://adsabs.harvard.edu/abs/1998MNRAS.301..881E} {301, 881}

\bibitem[\protect\citeauthoryear{{Ebeling}, {Edge}, {Allen}, {Crawford},
  {Fabian}  \& {Huchra}}{{Ebeling} et~al.}{2000}]{2000MNRAS.318..333E}
{Ebeling} H.,  {Edge} A.~C.,  {Allen} S.~W.,  {Crawford} C.~S.,  {Fabian}
  A.~C.,   {Huchra} J.~P.,  2000, \mn@doi [\mnras]
  {10.1046/j.1365-8711.2000.03549.x}, \href
  {http://adsabs.harvard.edu/abs/2000MNRAS.318..333E} {318, 333}

\bibitem[\protect\citeauthoryear{{Ebeling}, {Mullis}  \& {Tully}}{{Ebeling}
  et~al.}{2002}]{2002ApJ...580..774E}
{Ebeling} H.,  {Mullis} C.~R.,   {Tully} R.~B.,  2002, \mn@doi [\apj]
  {10.1086/343790}, \href {http://adsabs.harvard.edu/abs/2002ApJ...580..774E}
  {580, 774}

\bibitem[\protect\citeauthoryear{{Ebeling}, {Barrett}, {Donovan}, {Ma}, {Edge}
  \& {van Speybroeck}}{{Ebeling} et~al.}{2007}]{2007ApJ...661L..33E}
{Ebeling} H.,  {Barrett} E.,  {Donovan} D.,  {Ma} C.-J.,  {Edge} A.~C.,   {van
  Speybroeck} L.,  2007, \mn@doi [\apjl] {10.1086/518603}, \href
  {http://adsabs.harvard.edu/abs/2007ApJ...661L..33E} {661, L33}

\bibitem[\protect\citeauthoryear{{En{\ss}lin} \& {Gopal-Krishna}}{{En{\ss}lin}
  \& {Gopal-Krishna}}{2001}]{2001A&A...366...26E}
{En{\ss}lin} T.~A.,  {Gopal-Krishna} 2001, \mn@doi [\aap]
  {10.1051/0004-6361:20000198}, \href
  {http://adsabs.harvard.edu/abs/2001A%26A...366...26E} {366, 26}

\bibitem[\protect\citeauthoryear{{Ensslin}, {Biermann}, {Klein}  \&
  {Kohle}}{{Ensslin} et~al.}{1998}]{1998A&A...332..395E}
{Ensslin} T.~A.,  {Biermann} P.~L.,  {Klein} U.,   {Kohle} S.,  1998, \aap,
  \href {http://adsabs.harvard.edu/abs/1998A%26A...332..395E} {332, 395}

\bibitem[\protect\citeauthoryear{{Farnsworth}, {Rudnick}, {Brown}  \&
  {Brunetti}}{{Farnsworth} et~al.}{2013}]{2013ApJ...779..189F}
{Farnsworth} D.,  {Rudnick} L.,  {Brown} S.,   {Brunetti} G.,  2013, \mn@doi
  [\apj] {10.1088/0004-637X/779/2/189}, \href
  {http://adsabs.harvard.edu/abs/2013ApJ...779..189F} {779, 189}

\bibitem[\protect\citeauthoryear{{Feretti} \& {Giovannini}}{{Feretti} \&
  {Giovannini}}{1996}]{1996IAUS..175..333F}
{Feretti} L.,  {Giovannini} G.,  1996, in {Ekers} R.~D.,  {Fanti} C.,
  {Padrielli} L.,  eds,  IAU Symposium Vol. 175, Extragalactic Radio Sources.
  p.~333

\bibitem[\protect\citeauthoryear{{Feretti}, {Fusco-Femiano}, {Giovannini}  \&
  {Govoni}}{{Feretti} et~al.}{2001}]{2001A&A...373..106F}
{Feretti} L.,  {Fusco-Femiano} R.,  {Giovannini} G.,   {Govoni} F.,  2001,
  \mn@doi [\aap] {10.1051/0004-6361:20010581}, \href
  {http://adsabs.harvard.edu/abs/2001A%26A...373..106F} {373, 106}

\bibitem[\protect\citeauthoryear{{Feretti}, {Schuecker}, {B{\"o}hringer},
  {Govoni}  \& {Giovannini}}{{Feretti} et~al.}{2005}]{2005A&A...444..157F}
{Feretti} L.,  {Schuecker} P.,  {B{\"o}hringer} H.,  {Govoni} F.,
  {Giovannini} G.,  2005, \mn@doi [\aap] {10.1051/0004-6361:20052808}, \href
  {http://adsabs.harvard.edu/abs/2005A%26A...444..157F} {444, 157}

\bibitem[\protect\citeauthoryear{{Feretti}, {Bacchi}, {Slee}, {Giovannini},
  {Govoni}, {Andernach}  \& {Tsarevsky}}{{Feretti}
  et~al.}{2006}]{2006MNRAS.368..544F}
{Feretti} L.,  {Bacchi} M.,  {Slee} O.~B.,  {Giovannini} G.,  {Govoni} F.,
  {Andernach} H.,   {Tsarevsky} G.,  2006, \mn@doi [\mnras]
  {10.1111/j.1365-2966.2006.10086.x}, \href
  {http://adsabs.harvard.edu/abs/2006MNRAS.368..544F} {368, 544}

\bibitem[\protect\citeauthoryear{{Feretti}, {Giovannini}, {Govoni}  \&
  {Murgia}}{{Feretti} et~al.}{2012}]{2012A&ARv..20...54F}
{Feretti} L.,  {Giovannini} G.,  {Govoni} F.,   {Murgia} M.,  2012, \mn@doi
  [\aapr] {10.1007/s00159-012-0054-z}, \href
  {http://adsabs.harvard.edu/abs/2012A%26ARv..20...54F} {20, 54}

\bibitem[\protect\citeauthoryear{{Finoguenov}, {Sarazin}, {Nakazawa}, {Wik}  \&
  {Clarke}}{{Finoguenov} et~al.}{2010}]{2010ApJ...715.1143F}
{Finoguenov} A.,  {Sarazin} C.~L.,  {Nakazawa} K.,  {Wik} D.~R.,   {Clarke}
  T.~E.,  2010, \mn@doi [\apj] {10.1088/0004-637X/715/2/1143}, \href
  {http://adsabs.harvard.edu/abs/2010ApJ...715.1143F} {715, 1143}

\bibitem[\protect\citeauthoryear{{Giacintucci}, {Venturi}, {Bardelli},
  {Brunetti}, {Cassano}  \& {Dallacasa}}{{Giacintucci}
  et~al.}{2006}]{2006NewA...11..437G}
{Giacintucci} S.,  {Venturi} T.,  {Bardelli} S.,  {Brunetti} G.,  {Cassano} R.,
    {Dallacasa} D.,  2006, \mn@doi [\na] {10.1016/j.newast.2005.11.001}, \href
  {http://adsabs.harvard.edu/abs/2006NewA...11..437G} {11, 437}

\bibitem[\protect\citeauthoryear{{Giacintucci}, {Venturi}, {Cassano},
  {Dallacasa}  \& {Brunetti}}{{Giacintucci} et~al.}{2009}]{2009ApJ...704L..54G}
{Giacintucci} S.,  {Venturi} T.,  {Cassano} R.,  {Dallacasa} D.,   {Brunetti}
  G.,  2009, \mn@doi [\apjl] {10.1088/0004-637X/704/1/L54}, \href
  {http://adsabs.harvard.edu/abs/2009ApJ...704L..54G} {704, L54}

\bibitem[\protect\citeauthoryear{{Giovannini} \& {Feretti}}{{Giovannini} \&
  {Feretti}}{2000}]{2000NewA....5..335G}
{Giovannini} G.,  {Feretti} L.,  2000, \mn@doi [\na]
  {10.1016/S1384-1076(00)00034-8}, \href
  {http://adsabs.harvard.edu/abs/2000NewA....5..335G} {5, 335}

\bibitem[\protect\citeauthoryear{{Giovannini}, {Tordi}  \&
  {Feretti}}{{Giovannini} et~al.}{1999}]{giovannini:99}
{Giovannini} G.,  {Tordi} M.,   {Feretti} L.,  1999, \na, 4, 141

\bibitem[\protect\citeauthoryear{{Giovannini}, {Bonafede}, {Feretti}, {Govoni},
  {Murgia}, {Ferrari}  \& {Monti}}{{Giovannini}
  et~al.}{2009}]{2009A&A...507.1257G}
{Giovannini} G.,  {Bonafede} A.,  {Feretti} L.,  {Govoni} F.,  {Murgia} M.,
  {Ferrari} F.,   {Monti} G.,  2009, \mn@doi [\aap]
  {10.1051/0004-6361/200912667}, \href
  {http://adsabs.harvard.edu/abs/2009A%26A...507.1257G} {507, 1257}

\bibitem[\protect\citeauthoryear{{Giovannini}, {Bonafede}, {Feretti}, {Govoni}
  \& {Murgia}}{{Giovannini} et~al.}{2010}]{2010A&A...511L...5G}
{Giovannini} G.,  {Bonafede} A.,  {Feretti} L.,  {Govoni} F.,   {Murgia} M.,
  2010, \mn@doi [\aap] {10.1051/0004-6361/200913983}, \href
  {http://adsabs.harvard.edu/abs/2010A%26A...511L...5G} {511, L5+}

\bibitem[\protect\citeauthoryear{{Govoni}, {Feretti}, {Giovannini},
  {B{\"o}hringer}, {Reiprich}  \& {Murgia}}{{Govoni}
  et~al.}{2001}]{2001A&A...376..803G}
{Govoni} F.,  {Feretti} L.,  {Giovannini} G.,  {B{\"o}hringer} H.,  {Reiprich}
  T.~H.,   {Murgia} M.,  2001, \mn@doi [\aap] {10.1051/0004-6361:20011016},
  \href {http://adsabs.harvard.edu/abs/2001A%26A...376..803G} {376, 803}

\bibitem[\protect\citeauthoryear{{Govoni}, {Murgia}, {Giovannini}, {Vacca}  \&
  {Bonafede}}{{Govoni} et~al.}{2011}]{2011A&A...529A..69G}
{Govoni} F.,  {Murgia} M.,  {Giovannini} G.,  {Vacca} V.,   {Bonafede} A.,
  2011, \mn@doi [\aap] {10.1051/0004-6361/201016042}, \href
  {http://adsabs.harvard.edu/abs/2011A%26A...529A..69G} {529, A69}

\bibitem[\protect\citeauthoryear{{Govoni}, {Ferrari}, {Feretti}, {Vacca},
  {Murgia}, {Giovannini}, {Perley}  \& {Benoist}}{{Govoni}
  et~al.}{2012}]{2012A&A...545A..74G}
{Govoni} F.,  {Ferrari} C.,  {Feretti} L.,  {Vacca} V.,  {Murgia} M.,
  {Giovannini} G.,  {Perley} R.,   {Benoist} C.,  2012, \mn@doi [\aap]
  {10.1051/0004-6361/201219151}, \href
  {http://adsabs.harvard.edu/abs/2012A%26A...545A..74G} {545, A74}

\bibitem[\protect\citeauthoryear{{Guo}, {Sironi}  \& {Narayan}}{{Guo}
  et~al.}{2014a}]{2014ApJ...794..153G}
{Guo} X.,  {Sironi} L.,   {Narayan} R.,  2014a, \mn@doi [\apj]
  {10.1088/0004-637X/794/2/153}, \href
  {http://adsabs.harvard.edu/abs/2014ApJ...794..153G} {794, 153}

\bibitem[\protect\citeauthoryear{{Guo}, {Sironi}  \& {Narayan}}{{Guo}
  et~al.}{2014b}]{2014ApJ...797...47G}
{Guo} X.,  {Sironi} L.,   {Narayan} R.,  2014b, \mn@doi [\apj]
  {10.1088/0004-637X/797/1/47}, \href
  {http://adsabs.harvard.edu/abs/2014ApJ...797...47G} {797, 47}

\bibitem[\protect\citeauthoryear{{Hoeft} \& {Br{\"u}ggen}}{{Hoeft} \&
  {Br{\"u}ggen}}{2007}]{2007MNRAS.375...77H}
{Hoeft} M.,  {Br{\"u}ggen} M.,  2007, \mn@doi [\mnras]
  {10.1111/j.1365-2966.2006.11111.x}, \href
  {http://adsabs.harvard.edu/abs/2007MNRAS.375...77H} {375, 77}

\bibitem[\protect\citeauthoryear{{Hoeft}, {Br{\"u}ggen}, {Yepes},
  {Gottl{\"o}ber}  \& {Schwope}}{{Hoeft} et~al.}{2008}]{2008MNRAS.391.1511H}
{Hoeft} M.,  {Br{\"u}ggen} M.,  {Yepes} G.,  {Gottl{\"o}ber} S.,   {Schwope}
  A.,  2008, \mn@doi [\mnras] {10.1111/j.1365-2966.2008.13955.x}, \href
  {http://adsabs.harvard.edu/abs/2008MNRAS.391.1511H} {391, 1511}

\bibitem[\protect\citeauthoryear{{Hoeft}, {Nuza}, {Gottl{\"o}ber}, {van
  Weeren}, {R{\"o}ttgering}  \& {Br{\"u}ggen}}{{Hoeft}
  et~al.}{2011}]{2011JApA...32..509H}
{Hoeft} M.,  {Nuza} S.~E.,  {Gottl{\"o}ber} S.,  {van Weeren} R.~J.,
  {R{\"o}ttgering} H.~J.~A.,   {Br{\"u}ggen} M.,  2011, \mn@doi [Journal of
  Astrophysics and Astronomy] {10.1007/s12036-011-9127-z}, \href
  {http://adsabs.harvard.edu/abs/2011JApA...32..509H} {32, 509}

\bibitem[\protect\citeauthoryear{{Hong}, {Kang}  \& {Ryu}}{{Hong}
  et~al.}{2015}]{2015ApJ...812...49H}
{Hong} S.~E.,  {Kang} H.,   {Ryu} D.,  2015, \mn@doi [\apj]
  {10.1088/0004-637X/812/1/49}, \href
  {http://adsabs.harvard.edu/abs/2015ApJ...812...49H} {812, 49}

\bibitem[\protect\citeauthoryear{{Hopkins}}{{Hopkins}}{2013}]{2013MNRAS.428.2840H}
{Hopkins} P.~F.,  2013, \mn@doi [\mnras] {10.1093/mnras/sts210}, \href
  {http://adsabs.harvard.edu/abs/2013MNRAS.428.2840H} {428, 2840}

\bibitem[\protect\citeauthoryear{{Intema}, {Jagannathan}, {Mooley}  \&
  {Frail}}{{Intema} et~al.}{2017}]{2017A&A...598A..78I}
{Intema} H.~T.,  {Jagannathan} P.,  {Mooley} K.~P.,   {Frail} D.~A.,  2017,
  \mn@doi [\aap] {10.1051/0004-6361/201628536}, \href
  {http://adsabs.harvard.edu/abs/2017A%26A...598A..78I} {598, A78}

\bibitem[\protect\citeauthoryear{{Kale} \& {Dwarakanath}}{{Kale} \&
  {Dwarakanath}}{2012}]{2012ApJ...744...46K}
{Kale} R.,  {Dwarakanath} K.~S.,  2012, \mn@doi [\apj]
  {10.1088/0004-637X/744/1/46}, \href
  {http://adsabs.harvard.edu/abs/2012ApJ...744...46K} {744, 46}

\bibitem[\protect\citeauthoryear{{Kale} et~al.,}{{Kale}
  et~al.}{2015}]{2015A&A...579A..92K}
{Kale} R.,  et~al., 2015, \mn@doi [\aap] {10.1051/0004-6361/201525695}, \href
  {http://adsabs.harvard.edu/abs/2015A%26A...579A..92K} {579, A92}

\bibitem[\protect\citeauthoryear{{Kang} \& {Ryu}}{{Kang} \&
  {Ryu}}{2013}]{2013ApJ...764...95K}
{Kang} H.,  {Ryu} D.,  2013, \mn@doi [\apj] {10.1088/0004-637X/764/1/95}, \href
  {http://adsabs.harvard.edu/abs/2013ApJ...764...95K} {764, 95}

\bibitem[\protect\citeauthoryear{{Kang} \& {Ryu}}{{Kang} \&
  {Ryu}}{2015}]{2015ApJ...809..186K}
{Kang} H.,  {Ryu} D.,  2015, \mn@doi [\apj] {10.1088/0004-637X/809/2/186},
  \href {http://adsabs.harvard.edu/abs/2015ApJ...809..186K} {809, 186}

\bibitem[\protect\citeauthoryear{{Kang} \& {Ryu}}{{Kang} \&
  {Ryu}}{2016}]{2016ApJ...823...13K}
{Kang} H.,  {Ryu} D.,  2016, \mn@doi [\apj] {10.3847/0004-637X/823/1/13}, \href
  {http://adsabs.harvard.edu/abs/2016ApJ...823...13K} {823, 13}

\bibitem[\protect\citeauthoryear{{Kang}, {Ryu}  \& {Jones}}{{Kang}
  et~al.}{2012}]{2012ApJ...756...97K}
{Kang} H.,  {Ryu} D.,   {Jones} T.~W.,  2012, \mn@doi [\apj]
  {10.1088/0004-637X/756/1/97}, \href
  {http://adsabs.harvard.edu/abs/2012ApJ...756...97K} {756, 97}

\bibitem[\protect\citeauthoryear{{Kempner} \& {Sarazin}}{{Kempner} \&
  {Sarazin}}{2001}]{2001ApJ...548..639K}
{Kempner} J.~C.,  {Sarazin} C.~L.,  2001, \mn@doi [\apj] {10.1086/319024},
  \href {http://adsabs.harvard.edu/abs/2001ApJ...548..639K} {548, 639}

\bibitem[\protect\citeauthoryear{{Kempner}, {Blanton}, {Clarke}, {En{\ss}lin},
  {Johnston-Hollitt}  \& {Rudnick}}{{Kempner}
  et~al.}{2004}]{2004rcfg.proc..335K}
{Kempner} J.~C.,  {Blanton} E.~L.,  {Clarke} T.~E.,  {En{\ss}lin} T.~A.,
  {Johnston-Hollitt} M.,   {Rudnick} L.,  2004, in {T.~Reiprich, J.~Kempner, \&
  N.~Soker} ed., The Riddle of Cooling Flows in Galaxies and Clusters of
  galaxies. p.~335 (\mn@eprint {} {arXiv:astro-ph/0310263})

\bibitem[\protect\citeauthoryear{{Klypin}, {Kravtsov}, {Bullock}  \&
  {Primack}}{{Klypin} et~al.}{2001}]{2001ApJ...554..903K}
{Klypin} A.,  {Kravtsov} A.~V.,  {Bullock} J.~S.,   {Primack} J.~R.,  2001,
  \mn@doi [\apj] {10.1086/321400}, \href
  {http://adsabs.harvard.edu/abs/2001ApJ...554..903K} {554, 903}

\bibitem[\protect\citeauthoryear{{Knollmann} \& {Knebe}}{{Knollmann} \&
  {Knebe}}{2009}]{Knollmann09}
{Knollmann} S.~R.,  {Knebe} A.,  2009, \mn@doi [\apjs]
  {10.1088/0067-0049/182/2/608}, \href
  {http://adsabs.harvard.edu/abs/2009ApJS..182..608K} {182, 608}

\bibitem[\protect\citeauthoryear{{Komissarov} \& {Gubanov}}{{Komissarov} \&
  {Gubanov}}{1994}]{1994A&A...285...27K}
{Komissarov} S.~S.,  {Gubanov} A.~G.,  1994, \aap, \href
  {http://adsabs.harvard.edu/abs/1994A%26A...285...27K} {285}

\bibitem[\protect\citeauthoryear{{Landau} \& {Lifshitz}}{{Landau} \&
  {Lifshitz}}{1959}]{1959flme.book.....L}
{Landau} L.~D.,  {Lifshitz} E.~M.,  1959, {Fluid mechanics}

\bibitem[\protect\citeauthoryear{{Lindner} et~al.,}{{Lindner}
  et~al.}{2014}]{2014ApJ...786...49L}
{Lindner} R.~R.,  et~al., 2014, \mn@doi [\apj] {10.1088/0004-637X/786/1/49},
  \href {http://adsabs.harvard.edu/abs/2014ApJ...786...49L} {786, 49}

\bibitem[\protect\citeauthoryear{{Macario}, {Markevitch}, {Giacintucci},
  {Brunetti}, {Venturi}  \& {Murray}}{{Macario}
  et~al.}{2011}]{2011ApJ...728...82M}
{Macario} G.,  {Markevitch} M.,  {Giacintucci} S.,  {Brunetti} G.,  {Venturi}
  T.,   {Murray} S.~S.,  2011, \mn@doi [\apj] {10.1088/0004-637X/728/2/82},
  \href {http://adsabs.harvard.edu/abs/2011ApJ...728...82M} {728, 82}

\bibitem[\protect\citeauthoryear{{Middelberg} et~al.,}{{Middelberg}
  et~al.}{2008}]{2008AJ....135.1276M}
{Middelberg} E.,  et~al., 2008, \mn@doi [\aj] {10.1088/0004-6256/135/4/1276},
  \href {http://adsabs.harvard.edu/abs/2008AJ....135.1276M} {135, 1276}

\bibitem[\protect\citeauthoryear{{Miniati}}{{Miniati}}{2014}]{2014ApJ...782...21M}
{Miniati} F.,  2014, \mn@doi [\apj] {10.1088/0004-637X/782/1/21}, \href
  {http://adsabs.harvard.edu/abs/2014ApJ...782...21M} {782, 21}

\bibitem[\protect\citeauthoryear{{Miniati}, {Ryu}, {Kang}, {Jones}, {Cen}  \&
  {Ostriker}}{{Miniati} et~al.}{2000}]{2000ApJ...542..608M}
{Miniati} F.,  {Ryu} D.,  {Kang} H.,  {Jones} T.~W.,  {Cen} R.,   {Ostriker}
  J.~P.,  2000, \mn@doi [\apj] {10.1086/317027}, \href
  {http://adsabs.harvard.edu/abs/2000ApJ...542..608M} {542, 608}

\bibitem[\protect\citeauthoryear{{Nagai}, {Kravtsov}  \& {Vikhlinin}}{{Nagai}
  et~al.}{2007}]{2007ApJ...668....1N}
{Nagai} D.,  {Kravtsov} A.~V.,   {Vikhlinin} A.,  2007, \mn@doi [\apj]
  {10.1086/521328}, \href {http://adsabs.harvard.edu/abs/2007ApJ...668....1N}
  {668, 1}

\bibitem[\protect\citeauthoryear{{Nakazawa} et~al.,}{{Nakazawa}
  et~al.}{2009}]{2009PASJ...61..339N}
{Nakazawa} K.,  et~al., 2009, \pasj, \href
  {http://adsabs.harvard.edu/abs/2009PASJ...61..339N} {61, 339}

\bibitem[\protect\citeauthoryear{{Norris} et~al.,}{{Norris}
  et~al.}{2011}]{2011PASA...28..215N}
{Norris} R.~P.,  et~al., 2011, \mn@doi [\pasa] {10.1071/AS11021}, \href
  {http://adsabs.harvard.edu/abs/2011PASA...28..215N} {28, 215}

\bibitem[\protect\citeauthoryear{{Nuza}, {Hoeft}, {van Weeren}, {Gottl{\"o}ber}
   \& {Yepes}}{{Nuza} et~al.}{2012}]{2012MNRAS.420.2006N}
{Nuza} S.~E.,  {Hoeft} M.,  {van Weeren} R.~J.,  {Gottl{\"o}ber} S.,   {Yepes}
  G.,  2012, \mn@doi [\mnras] {10.1111/j.1365-2966.2011.20118.x}, \href
  {http://adsabs.harvard.edu/abs/2012MNRAS.420.2006N} {420, 2006}

\bibitem[\protect\citeauthoryear{{Ogrean}, {Br{\"u}ggen}, {van Weeren},
  {R{\"o}ttgering}, {Croston}  \& {Hoeft}}{{Ogrean}
  et~al.}{2013}]{2013MNRAS.433..812O}
{Ogrean} G.~A.,  {Br{\"u}ggen} M.,  {van Weeren} R.~J.,  {R{\"o}ttgering} H.,
  {Croston} J.~H.,   {Hoeft} M.,  2013, \mn@doi [\mnras]
  {10.1093/mnras/stt776}, \href
  {http://adsabs.harvard.edu/abs/2013MNRAS.433..812O} {433, 812}

\bibitem[\protect\citeauthoryear{{Ogrean}, {Br{\"u}ggen}, {van Weeren},
  {Burgmeier}  \& {Simionescu}}{{Ogrean} et~al.}{2014}]{2014MNRAS.443.2463O}
{Ogrean} G.~A.,  {Br{\"u}ggen} M.,  {van Weeren} R.~J.,  {Burgmeier} A.,
  {Simionescu} A.,  2014, \mn@doi [\mnras] {10.1093/mnras/stu1299}, \href
  {http://adsabs.harvard.edu/abs/2014MNRAS.443.2463O} {443, 2463}

\bibitem[\protect\citeauthoryear{{Pandey-Pommier}, {Richard}, {Combes},
  {Dwarakanath}, {Guiderdoni}, {Ferrari}, {Sirothia}  \&
  {Narasimha}}{{Pandey-Pommier} et~al.}{2013}]{2013A&A...557A.117P}
{Pandey-Pommier} M.,  {Richard} J.,  {Combes} F.,  {Dwarakanath} K.~S.,
  {Guiderdoni} B.,  {Ferrari} C.,  {Sirothia} S.,   {Narasimha} D.,  2013,
  \mn@doi [\aap] {10.1051/0004-6361/201321809}, \href
  {http://adsabs.harvard.edu/abs/2013A%26A...557A.117P} {557, A117}

\bibitem[\protect\citeauthoryear{{Pfrommer}, {Springel}, {En{\ss}lin}  \&
  {Jubelgas}}{{Pfrommer} et~al.}{2006}]{2006MNRAS.367..113P}
{Pfrommer} C.,  {Springel} V.,  {En{\ss}lin} T.~A.,   {Jubelgas} M.,  2006,
  \mn@doi [\mnras] {10.1111/j.1365-2966.2005.09953.x}, \href
  {http://adsabs.harvard.edu/abs/2006MNRAS.367..113P} {367, 113}

\bibitem[\protect\citeauthoryear{{Pinzke}, {Oh}  \& {Pfrommer}}{{Pinzke}
  et~al.}{2013}]{2013MNRAS.435.1061P}
{Pinzke} A.,  {Oh} S.~P.,   {Pfrommer} C.,  2013, \mn@doi [\mnras]
  {10.1093/mnras/stt1308}, \href
  {http://adsabs.harvard.edu/abs/2013MNRAS.435.1061P} {435, 1061}

\bibitem[\protect\citeauthoryear{{Pizzo}, {de Bruyn}, {Feretti}  \&
  {Govoni}}{{Pizzo} et~al.}{2008}]{2008A&A...481L..91P}
{Pizzo} R.~F.,  {de Bruyn} A.~G.,  {Feretti} L.,   {Govoni} F.,  2008, \mn@doi
  [\aap] {10.1051/0004-6361:20079304}, \href
  {http://adsabs.harvard.edu/abs/2008A%26A...481L..91P} {481, L91}

\bibitem[\protect\citeauthoryear{{Planck Collaboration} et~al.,}{{Planck
  Collaboration} et~al.}{2015}]{2015A&A...581A..14P}
{Planck Collaboration} et~al., 2015, \mn@doi [\aap]
  {10.1051/0004-6361/201525787}, \href
  {http://adsabs.harvard.edu/abs/2015A%26A...581A..14P} {581, A14}

\bibitem[\protect\citeauthoryear{{Prada}, {Klypin}, {Cuesta}, {Betancort-Rijo}
  \& {Primack}}{{Prada} et~al.}{2012}]{Prada12}
{Prada} F.,  {Klypin} A.~A.,  {Cuesta} A.~J.,  {Betancort-Rijo} J.~E.,
  {Primack} J.,  2012, \mn@doi [\mnras] {10.1111/j.1365-2966.2012.21007.x},
  \href {http://adsabs.harvard.edu/abs/2012MNRAS.423.3018P} {423, 3018}

\bibitem[\protect\citeauthoryear{{Randall}, {Clarke}, {Nulsen}, {Owers},
  {Sarazin}, {Forman}  \& {Murray}}{{Randall}
  et~al.}{2010}]{2010ApJ...722..825R}
{Randall} S.~W.,  {Clarke} T.~E.,  {Nulsen} P.~E.~J.,  {Owers} M.~S.,
  {Sarazin} C.~L.,  {Forman} W.~R.,   {Murray} S.~S.,  2010, \mn@doi [\apj]
  {10.1088/0004-637X/722/1/825}, \href
  {http://adsabs.harvard.edu/abs/2010ApJ...722..825R} {722, 825}

\bibitem[\protect\citeauthoryear{{Randall} et~al.,}{{Randall}
  et~al.}{2016}]{2016ApJ...823...94R}
{Randall} S.~W.,  et~al., 2016, \mn@doi [\apj] {10.3847/0004-637X/823/2/94},
  \href {http://adsabs.harvard.edu/abs/2016ApJ...823...94R} {823, 94}

\bibitem[\protect\citeauthoryear{{Rengelink}, {Tang}, {de Bruyn}, {Miley},
  {Bremer}, {Roettgering}  \& {Bremer}}{{Rengelink}
  et~al.}{1997}]{1997A&AS..124..259R}
{Rengelink} R.~B.,  {Tang} Y.,  {de Bruyn} A.~G.,  {Miley} G.~K.,  {Bremer}
  M.~N.,  {Roettgering} H.~J.~A.,   {Bremer} M.~A.~R.,  1997, \mn@doi [\aaps]
  {10.1051/aas:1997358}, \href
  {http://adsabs.harvard.edu/abs/1997A%26AS..124..259R} {124}

\bibitem[\protect\citeauthoryear{{Reynolds}}{{Reynolds}}{2008}]{2008ARA&A..46...89R}
{Reynolds} S.~P.,  2008, \mn@doi [\araa]
  {10.1146/annurev.astro.46.060407.145237}, \href
  {http://adsabs.harvard.edu/abs/2008ARA%26A..46...89R} {46, 89}

\bibitem[\protect\citeauthoryear{{Riseley}, {Scaife}, {Wise}  \&
  {Clarke}}{{Riseley} et~al.}{2017}]{2017A&A...597A..96R}
{Riseley} C.~J.,  {Scaife} A.~M.~M.,  {Wise} M.~W.,   {Clarke} A.~O.,  2017,
  \mn@doi [\aap] {10.1051/0004-6361/201629530}, \href
  {http://adsabs.harvard.edu/abs/2017A%26A...597A..96R} {597, A96}

\bibitem[\protect\citeauthoryear{{Rottgering}, {Wieringa}, {Hunstead}  \&
  {Ekers}}{{Rottgering} et~al.}{1997}]{1997MNRAS.290..577R}
{Rottgering} H.~J.~A.,  {Wieringa} M.~H.,  {Hunstead} R.~W.,   {Ekers} R.~D.,
  1997, \mnras, \href {http://adsabs.harvard.edu/abs/1997MNRAS.290..577R} {290,
  577}

\bibitem[\protect\citeauthoryear{{Ryu}, {Kang}, {Hallman}  \& {Jones}}{{Ryu}
  et~al.}{2003}]{2003ApJ...593..599R}
{Ryu} D.,  {Kang} H.,  {Hallman} E.,   {Jones} T.~W.,  2003, \mn@doi [\apj]
  {10.1086/376723}, \href {http://adsabs.harvard.edu/abs/2003ApJ...593..599R}
  {593, 599}

\bibitem[\protect\citeauthoryear{{Sarazin} et~al.,}{{Sarazin}
  et~al.}{2014}]{2014xru..confE.181S}
{Sarazin} C.,  et~al., 2014, in The X-ray Universe 2014. p.~181

\bibitem[\protect\citeauthoryear{{Schaal} et~al.,}{{Schaal}
  et~al.}{2016}]{2016MNRAS.461.4441S}
{Schaal} K.,  et~al., 2016, \mn@doi [\mnras] {10.1093/mnras/stw1587}, \href
  {http://adsabs.harvard.edu/abs/2016MNRAS.461.4441S} {461, 4441}

\bibitem[\protect\citeauthoryear{{Sembolini}, {Yepes}, {De Petris},
  {Gottl{\"o}ber}, {Lamagna}  \& {Comis}}{{Sembolini}
  et~al.}{2013}]{2013MNRAS.429..323S}
{Sembolini} F.,  {Yepes} G.,  {De Petris} M.,  {Gottl{\"o}ber} S.,  {Lamagna}
  L.,   {Comis} B.,  2013, \mn@doi [\mnras] {10.1093/mnras/sts339}, \href
  {http://adsabs.harvard.edu/abs/2013MNRAS.429..323S} {429, 323}

\bibitem[\protect\citeauthoryear{{Sembolini} et~al.,}{{Sembolini}
  et~al.}{2016}]{2016MNRAS.459.2973S}
{Sembolini} F.,  et~al., 2016, \mn@doi [\mnras] {10.1093/mnras/stw800}, \href
  {http://adsabs.harvard.edu/abs/2016MNRAS.459.2973S} {459, 2973}

\bibitem[\protect\citeauthoryear{{Shakouri}, {Johnston-Hollitt}  \&
  {Pratt}}{{Shakouri} et~al.}{2016}]{2016MNRAS.459.2525S}
{Shakouri} S.,  {Johnston-Hollitt} M.,   {Pratt} G.~W.,  2016, \mn@doi [\mnras]
  {10.1093/mnras/stw812}, \href
  {http://adsabs.harvard.edu/abs/2016MNRAS.459.2525S} {459, 2525}

\bibitem[\protect\citeauthoryear{{Shimwell}, {Markevitch}, {Brown}, {Feretti},
  {Gaensler}, {Johnston-Hollitt}, {Lage}  \& {Srinivasan}}{{Shimwell}
  et~al.}{2015}]{2015MNRAS.449.1486S}
{Shimwell} T.~W.,  {Markevitch} M.,  {Brown} S.,  {Feretti} L.,  {Gaensler}
  B.~M.,  {Johnston-Hollitt} M.,  {Lage} C.,   {Srinivasan} R.,  2015, \mn@doi
  [\mnras] {10.1093/mnras/stv334}, \href
  {http://adsabs.harvard.edu/abs/2015MNRAS.449.1486S} {449, 1486}

\bibitem[\protect\citeauthoryear{{Sif{\'o}n}, {Menanteau}, {Hughes}, {Carrasco}
   \& {Barrientos}}{{Sif{\'o}n} et~al.}{2014}]{2014A&A...562A..43S}
{Sif{\'o}n} C.,  {Menanteau} F.,  {Hughes} J.~P.,  {Carrasco} M.,
  {Barrientos} L.~F.,  2014, \mn@doi [\aap] {10.1051/0004-6361/201321638},
  \href {http://adsabs.harvard.edu/abs/2014A%26A...562A..43S} {562, A43}

\bibitem[\protect\citeauthoryear{{Skillman}, {O'Shea}, {Hallman}, {Burns}  \&
  {Norman}}{{Skillman} et~al.}{2008}]{2008ApJ...689.1063S}
{Skillman} S.~W.,  {O'Shea} B.~W.,  {Hallman} E.~J.,  {Burns} J.~O.,   {Norman}
  M.~L.,  2008, \mn@doi [\apj] {10.1086/592496}, \href
  {http://adsabs.harvard.edu/abs/2008ApJ...689.1063S} {689, 1063}

\bibitem[\protect\citeauthoryear{{Skillman}, {Hallman}, {O'Shea}, {Burns},
  {Smith}  \& {Turk}}{{Skillman} et~al.}{2011}]{2011ApJ...735...96S}
{Skillman} S.~W.,  {Hallman} E.~J.,  {O'Shea} B.~W.,  {Burns} J.~O.,  {Smith}
  B.~D.,   {Turk} M.~J.,  2011, \mn@doi [\apj] {10.1088/0004-637X/735/2/96},
  \href {http://adsabs.harvard.edu/abs/2011ApJ...735...96S} {735, 96}

\bibitem[\protect\citeauthoryear{{Skillman}, {Xu}, {Hallman}, {O'Shea},
  {Burns}, {Li}, {Collins}  \& {Norman}}{{Skillman}
  et~al.}{2013}]{2013ApJ...765...21S}
{Skillman} S.~W.,  {Xu} H.,  {Hallman} E.~J.,  {O'Shea} B.~W.,  {Burns} J.~O.,
  {Li} H.,  {Collins} D.~C.,   {Norman} M.~L.,  2013, \mn@doi [\apj]
  {10.1088/0004-637X/765/1/21}, \href
  {http://adsabs.harvard.edu/abs/2013ApJ...765...21S} {765, 21}

\bibitem[\protect\citeauthoryear{{Slee}, {Roy}, {Murgia}, {Andernach}  \&
  {Ehle}}{{Slee} et~al.}{2001}]{2001AJ....122.1172S}
{Slee} O.~B.,  {Roy} A.~L.,  {Murgia} M.,  {Andernach} H.,   {Ehle} M.,  2001,
  \mn@doi [\aj] {10.1086/322105}, \href
  {http://adsabs.harvard.edu/abs/2001AJ....122.1172S} {122, 1172}

\bibitem[\protect\citeauthoryear{{Springel}}{{Springel}}{2005}]{Springel05}
{Springel} V.,  2005, \mn@doi [\mnras] {10.1111/j.1365-2966.2005.09655.x},
  \href {http://adsabs.harvard.edu/abs/2005MNRAS.364.1105S} {364, 1105}

\bibitem[\protect\citeauthoryear{{Stasyszyn}, {Nuza}, {Dolag}, {Beck}  \&
  {Donnert}}{{Stasyszyn} et~al.}{2010}]{2010MNRAS.408..684S}
{Stasyszyn} F.,  {Nuza} S.~E.,  {Dolag} K.,  {Beck} R.,   {Donnert} J.,  2010,
  \mn@doi [\mnras] {10.1111/j.1365-2966.2010.17166.x}, \href
  {http://adsabs.harvard.edu/abs/2010MNRAS.408..684S} {408, 684}

\bibitem[\protect\citeauthoryear{{Stobie}}{{Stobie}}{1980}]{1980JBIS...33..323S}
{Stobie} R.~S.,  1980, Journal of the British Interplanetary Society, \href
  {http://adsabs.harvard.edu/abs/1980JBIS...33..323S} {33, 323}

\bibitem[\protect\citeauthoryear{{Stroe}, {van Weeren}, {Intema},
  {R{\"o}ttgering}, {Br{\"u}ggen}  \& {Hoeft}}{{Stroe}
  et~al.}{2013}]{2013A&A...555A.110S}
{Stroe} A.,  {van Weeren} R.~J.,  {Intema} H.~T.,  {R{\"o}ttgering} H.~J.~A.,
  {Br{\"u}ggen} M.,   {Hoeft} M.,  2013, \mn@doi [\aap]
  {10.1051/0004-6361/201321267}, \href
  {http://adsabs.harvard.edu/abs/2013A%26A...555A.110S} {555, A110}

\bibitem[\protect\citeauthoryear{{Stroe} et~al.,}{{Stroe}
  et~al.}{2016}]{2016MNRAS.455.2402S}
{Stroe} A.,  et~al., 2016, \mn@doi [\mnras] {10.1093/mnras/stv2472}, \href
  {http://adsabs.harvard.edu/abs/2016MNRAS.455.2402S} {455, 2402}

\bibitem[\protect\citeauthoryear{{Subrahmanyan}, {Beasley}, {Goss}, {Golap}  \&
  {Hunstead}}{{Subrahmanyan} et~al.}{2003}]{2003AJ....125.1095S}
{Subrahmanyan} R.,  {Beasley} A.~J.,  {Goss} W.~M.,  {Golap} K.,   {Hunstead}
  R.~W.,  2003, \mn@doi [\aj] {10.1086/367797}, \href
  {http://adsabs.harvard.edu/abs/2003AJ....125.1095S} {125, 1095}

\bibitem[\protect\citeauthoryear{{Vazza} \& {Br{\"u}ggen}}{{Vazza} \&
  {Br{\"u}ggen}}{2014}]{2014MNRAS.437.2291V}
{Vazza} F.,  {Br{\"u}ggen} M.,  2014, \mn@doi [\mnras] {10.1093/mnras/stt2042},
  \href {http://adsabs.harvard.edu/abs/2014MNRAS.437.2291V} {437, 2291}

\bibitem[\protect\citeauthoryear{{Vazza}, {Brunetti}  \& {Gheller}}{{Vazza}
  et~al.}{2009}]{2009MNRAS.395.1333V}
{Vazza} F.,  {Brunetti} G.,   {Gheller} C.,  2009, \mn@doi [\mnras]
  {10.1111/j.1365-2966.2009.14691.x}, \href
  {http://adsabs.harvard.edu/abs/2009MNRAS.395.1333V} {395, 1333}

\bibitem[\protect\citeauthoryear{{Vazza}, {Dolag}, {Ryu}, {Brunetti},
  {Gheller}, {Kang}  \& {Pfrommer}}{{Vazza} et~al.}{2011}]{2011MNRAS.418..960V}
{Vazza} F.,  {Dolag} K.,  {Ryu} D.,  {Brunetti} G.,  {Gheller} C.,  {Kang} H.,
   {Pfrommer} C.,  2011, \mn@doi [\mnras] {10.1111/j.1365-2966.2011.19546.x},
  \href {http://adsabs.harvard.edu/abs/2011MNRAS.418..960V} {418, 960}

\bibitem[\protect\citeauthoryear{{Vazza}, {Br{\"u}ggen}, {van Weeren},
  {Bonafede}, {Dolag}  \& {Brunetti}}{{Vazza}
  et~al.}{2012}]{2012MNRAS.421.1868V}
{Vazza} F.,  {Br{\"u}ggen} M.,  {van Weeren} R.,  {Bonafede} A.,  {Dolag} K.,
  {Brunetti} G.,  2012, \mn@doi [\mnras] {10.1111/j.1365-2966.2011.20160.x},
  \href {http://adsabs.harvard.edu/abs/2012MNRAS.421.1868V} {421, 1868}

\bibitem[\protect\citeauthoryear{{Vazza}, {Br{\"u}ggen}, {Wittor}, {Gheller},
  {Eckert}  \& {Stubbe}}{{Vazza} et~al.}{2016}]{2016MNRAS.459...70V}
{Vazza} F.,  {Br{\"u}ggen} M.,  {Wittor} D.,  {Gheller} C.,  {Eckert} D.,
  {Stubbe} M.,  2016, \mn@doi [\mnras] {10.1093/mnras/stw584}, \href
  {http://adsabs.harvard.edu/abs/2016MNRAS.459...70V} {459, 70}

\bibitem[\protect\citeauthoryear{{Venturi}, {Giacintucci}, {Dallacasa},
  {Cassano}, {Brunetti}, {Macario}  \& {Athreya}}{{Venturi}
  et~al.}{2013}]{2013A&A...551A..24V}
{Venturi} T.,  {Giacintucci} S.,  {Dallacasa} D.,  {Cassano} R.,  {Brunetti}
  G.,  {Macario} G.,   {Athreya} R.,  2013, \mn@doi [\aap]
  {10.1051/0004-6361/201219872}, \href
  {http://adsabs.harvard.edu/abs/2013A%26A...551A..24V} {551, A24}

\bibitem[\protect\citeauthoryear{{Wenger} et~al.,}{{Wenger}
  et~al.}{2000}]{2000A&AS..143....9W}
{Wenger} M.,  et~al., 2000, \mn@doi [\aaps] {10.1051/aas:2000332}, \href
  {http://adsabs.harvard.edu/abs/2000A%26AS..143....9W} {143, 9}

\bibitem[\protect\citeauthoryear{{Yuan}, {Han}  \& {Wen}}{{Yuan}
  et~al.}{2015}]{2015ApJ...813...77Y}
{Yuan} Z.~S.,  {Han} J.~L.,   {Wen} Z.~L.,  2015, \mn@doi [\apj]
  {10.1088/0004-637X/813/1/77}, \href
  {http://adsabs.harvard.edu/abs/2015ApJ...813...77Y} {813, 77}

\bibitem[\protect\citeauthoryear{{de Gasperin}, {van Weeren}, {Br{\"u}ggen},
  {Vazza}, {Bonafede}  \& {Intema}}{{de Gasperin}
  et~al.}{2014}]{2014MNRAS.444.3130D}
{de Gasperin} F.,  {van Weeren} R.~J.,  {Br{\"u}ggen} M.,  {Vazza} F.,
  {Bonafede} A.,   {Intema} H.~T.,  2014, \mn@doi [\mnras]
  {10.1093/mnras/stu1658}, \href
  {http://adsabs.harvard.edu/abs/2014MNRAS.444.3130D} {444, 3130}

\bibitem[\protect\citeauthoryear{{de Gasperin}, {Ogrean}, {van Weeren},
  {Dawson}, {Br{\"u}ggen}, {Bonafede}  \& {Simionescu}}{{de Gasperin}
  et~al.}{2015a}]{2015MNRAS.448.2197D}
{de Gasperin} F.,  {Ogrean} G.~A.,  {van Weeren} R.~J.,  {Dawson} W.~A.,
  {Br{\"u}ggen} M.,  {Bonafede} A.,   {Simionescu} A.,  2015a, \mn@doi [\mnras]
  {10.1093/mnras/stv129}, \href
  {http://adsabs.harvard.edu/abs/2015MNRAS.448.2197D} {448, 2197}

\bibitem[\protect\citeauthoryear{{de Gasperin}, {Intema}, {van Weeren},
  {Dawson}, {Golovich}, {Wittman}, {Bonafede}  \& {Br{\"u}ggen}}{{de Gasperin}
  et~al.}{2015b}]{2015MNRAS.453.3483D}
{de Gasperin} F.,  {Intema} H.~T.,  {van Weeren} R.~J.,  {Dawson} W.~A.,
  {Golovich} N.,  {Wittman} D.,  {Bonafede} A.,   {Br{\"u}ggen} M.,  2015b,
  \mn@doi [\mnras] {10.1093/mnras/stv1873}, \href
  {http://adsabs.harvard.edu/abs/2015MNRAS.453.3483D} {453, 3483}

\bibitem[\protect\citeauthoryear{{de Gasperin} et~al.,}{{de Gasperin}
  et~al.}{2017}]{2017A&A...597A..15D}
{de Gasperin} F.,  et~al., 2017, \mn@doi [\aap] {10.1051/0004-6361/201628945},
  \href {http://adsabs.harvard.edu/abs/2017A%26A...597A..15D} {597, A15}

\bibitem[\protect\citeauthoryear{{van Weeren}, {R{\"o}ttgering}, {Br{\"u}ggen}
  \& {Cohen}}{{van Weeren} et~al.}{2009a}]{2009A&A...505..991V}
{van Weeren} R.~J.,  {R{\"o}ttgering} H.~J.~A.,  {Br{\"u}ggen} M.,   {Cohen}
  A.,  2009a, \mn@doi [\aap] {10.1051/0004-6361/200912528}, \href
  {http://adsabs.harvard.edu/abs/2009A%26A...505..991V} {505, 991}

\bibitem[\protect\citeauthoryear{{van Weeren}, {R{\"o}ttgering}, {Br{\"u}ggen}
  \& {Cohen}}{{van Weeren} et~al.}{2009b}]{2009A&A...508...75V}
{van Weeren} R.~J.,  {R{\"o}ttgering} H.~J.~A.,  {Br{\"u}ggen} M.,   {Cohen}
  A.,  2009b, \mn@doi [\aap] {10.1051/0004-6361/200912501}, \href
  {http://adsabs.harvard.edu/abs/2009A%26A...508...75V} {508, 75}

\bibitem[\protect\citeauthoryear{{van Weeren}, {R{\"o}ttgering}, {Br{\"u}ggen}
  \& {Hoeft}}{{van Weeren} et~al.}{2010}]{2010Sci...330..347V}
{van Weeren} R.~J.,  {R{\"o}ttgering} H.~J.~A.,  {Br{\"u}ggen} M.,   {Hoeft}
  M.,  2010, \mn@doi [Science] {10.1126/science.1194293}, \href
  {http://adsabs.harvard.edu/abs/2010Sci...330..347V} {330, 347}

\bibitem[\protect\citeauthoryear{{van Weeren}, {R{\"o}ttgering}  \&
  {Br{\"u}ggen}}{{van Weeren} et~al.}{2011a}]{2011A&A...527A.114V}
{van Weeren} R.~J.,  {R{\"o}ttgering} H.~J.~A.,   {Br{\"u}ggen} M.,  2011a,
  \mn@doi [\aap] {10.1051/0004-6361/201015991}, \href
  {http://adsabs.harvard.edu/abs/2011A%26A...527A.114V} {527, A114+}

\bibitem[\protect\citeauthoryear{{van Weeren}, {Hoeft}, {R{\"o}ttgering},
  {Br{\"u}ggen}, {Intema}  \& {van Velzen}}{{van Weeren}
  et~al.}{2011b}]{2011A&A...528A..38V}
{van Weeren} R.~J.,  {Hoeft} M.,  {R{\"o}ttgering} H.~J.~A.,  {Br{\"u}ggen} M.,
   {Intema} H.~T.,   {van Velzen} S.,  2011b, \mn@doi [\aap]
  {10.1051/0004-6361/201016185}, \href
  {http://adsabs.harvard.edu/abs/2011A%26A...528A..38V} {528, A38+}

\bibitem[\protect\citeauthoryear{{van Weeren}, {Br{\"u}ggen}, {R{\"o}ttgering},
  {Hoeft}, {Nuza}  \& {Intema}}{{van Weeren}
  et~al.}{2011c}]{2011A&A...533A..35V}
{van Weeren} R.~J.,  {Br{\"u}ggen} M.,  {R{\"o}ttgering} H.~J.~A.,  {Hoeft} M.,
   {Nuza} S.~E.,   {Intema} H.~T.,  2011c, \mn@doi [\aap]
  {10.1051/0004-6361/201117149}, \href
  {http://adsabs.harvard.edu/abs/2011A%26A...533A..35V} {533, A35}

\bibitem[\protect\citeauthoryear{{van Weeren}, {R{\"o}ttgering}, {Intema},
  {Rudnick}, {Br{\"u}ggen}, {Hoeft}  \& {Oonk}}{{van Weeren}
  et~al.}{2012}]{2012A&A...546A.124V}
{van Weeren} R.~J.,  {R{\"o}ttgering} H.~J.~A.,  {Intema} H.~T.,  {Rudnick} L.,
   {Br{\"u}ggen} M.,  {Hoeft} M.,   {Oonk} J.~B.~R.,  2012, \mn@doi [\aap]
  {10.1051/0004-6361/201219000}, \href
  {http://adsabs.harvard.edu/abs/2012A%26A...546A.124V} {546, A124}

\bibitem[\protect\citeauthoryear{{van Weeren} et~al.,}{{van Weeren}
  et~al.}{2013}]{2013ApJ...769..101V}
{van Weeren} R.~J.,  et~al., 2013, \mn@doi [\apj]
  {10.1088/0004-637X/769/2/101}, \href
  {http://adsabs.harvard.edu/abs/2013ApJ...769..101V} {769, 101}

\bibitem[\protect\citeauthoryear{{van Weeren}, {Brunetti}, {Br{\"u}ggen},
  {Andrade-Santos}, {Ogrean}, {Williams}, {R{\"o}ttgering}  \& {Dawson}}{{van
  Weeren} et~al.}{2016}]{2016ApJ...818..204V}
{van Weeren} R.~J.,  {Brunetti} G.,  {Br{\"u}ggen} M.,  {Andrade-Santos} F.,
  {Ogrean} G.~A.,  {Williams} W.~L.,  {R{\"o}ttgering} H.~J.~A.,   {Dawson} e.,
   2016, \mn@doi [\apj] {10.3847/0004-637X/818/2/204}, \href
  {http://adsabs.harvard.edu/abs/2016ApJ...818..204V} {818, 204}

\bibitem[\protect\citeauthoryear{{van Weeren} et~al.,}{{van Weeren}
  et~al.}{2017a}]{2017NatAs...1E...5V}
{van Weeren} R.~J.,  et~al., 2017a, \mn@doi [Nature Astronomy]
  {10.1038/s41550-016-0005}, \href
  {http://adsabs.harvard.edu/abs/2017NatAs...1E...5V} {1, 0005}

\bibitem[\protect\citeauthoryear{{van Weeren} et~al.,}{{van Weeren}
  et~al.}{2017b}]{2017ApJ...835..197V}
{van Weeren} R.~J.,  et~al., 2017b, \mn@doi [\apj]
  {10.3847/1538-4357/835/2/197}, \href
  {http://adsabs.harvard.edu/abs/2017ApJ...835..197V} {835, 197}

\makeatother
\end{thebibliography}

\end{document}